\documentclass[11pt]{article}
\usepackage{amsmath,amssymb,amsfonts}
\usepackage[english]{babel}
\def\nn{\nonumber} \def\bd{\begin{document}} \def\ed{\end{document}}
\def\ds{\documentstyle}
\let\bm=\bibitem
\newcommand{\be}{\begin{equation}}
\newcommand{\ee}{\end{equation}}
\newcommand{\bea}{\setlength\arraycolsep{2pt} \begin{eqnarray}}
\newcommand{\eea}{\end{eqnarray}}
\newcommand{\hoch}[1]{$\, ^{#1}$}
\def\p{\partial}
\title{\large {\bf A note on field equations
in generalized theories of gravity}}
\date{}

\author{Jun-Jin Peng$^{1,2}$\footnote{corresponding author:
pengjjph@163.com}
\\\ \\
\small \sl $^1$School of Physics and Electronic Science,
\small \sl Guizhou Normal University,\\
\small Guiyang, Guizhou 550001, People's Republic of China; \\
\small \sl  $^{2}$Guizhou Provincial Key Laboratory of Radio Astronomy and
Data Processing, \\
\small \sl Guizhou Normal University, \\
\small Guiyang, Guizhou 550001, People's Republic of China
}

\begin{document}

\maketitle
\vspace{-5pt}

\begin{center}
\textbf{Abstract}
\end{center}

In the work [\emph{Phys. Rev. D} \textbf{84}, 124041 (2011)],
to obtain a simple and economic formulation of field equations
for generalised theories of gravity described by the Lagrangian
$\sqrt{-g}L\big(g^{\alpha\beta},R_{\mu\nu\rho\sigma}\big)$,
the key equality $\big(\partial L/\partial
g^{\mu\nu}\big)_{R_{\alpha\beta\kappa\omega}}
=2P_{\mu}^{~\lambda\rho\sigma}R_{\nu\lambda\rho\sigma}$
was derived. In this note, it is demonstrated that
such an equality can be directly derived from an off-shell
Noether current associated with an arbitrary vector field.
As byproducts, a generalized Bianchi identity related to
the divergence for the expression of field equations,
together with the Noether potential, is obtained.
On the basis of the above, we further propose a systematic
procedure to derive the equations of motion from
the Noether current, and then this procedure is extended
to more general higher-order gravities endowed with the
Lagrangian encompassing additional terms of the
covariant derivatives of the Riemann tensor. To our
knowledge, both the detailed expressions for field
equations and the Noether potential associated with such
theories are first given at a general level. All the
results reveal that using the Noether current to determine
field equations establishes a straightforward connection
between the symmetry of the Lagrangian and the equations
of motion and such a remedy even can avoid calculating
the derivative of the Lagrangian density with respect
to the metric.

%\noindent \textbf{PACS}: 04.20.-q, 04.50.Kd, 04.70.Bw

%%%%%%%%%%%%%%%%%%%%%%%%%%%%%%%%%%%%%%%%%%%%%%%%%%%%%%%%%%%%%%%%%%%%%%%%%
\voffset=-.90pt
\vspace{10pt}
\newpage

%%%%%%%%%%%%%%%%%%%%%%%%%%%%%%%%%%
\section{Introduction}\label{one}
%%%%%%%%%%%%%%%%%%%%%%%%%%%%%%%%%%

Higher-order gravity is a significant branch among various
modified gravity theories. It has attracted considerable
attention. Specially, in the work \cite{Pady}, Padmanabhan
investigated some aspects on field equations in generalized
gravities with the Lagrangian that is a functional of the
metric and the Riemann tensor, which is of great significance
to understand such theories. Inspired with this work, here we
attempt to perform further investigation on field equations
for a wide range of gravity theories endowed with
diffeomorphism invariance from a different perspective.
For the sake of achieving this, let us first give a brief
review on the main results obtained in \cite{Pady}. Along
the lines of this work, we straightforwardly start with
the general Lagrangian without covariant derivatives of
the Riemann tensor
\be
\sqrt{-g}L=\sqrt{-g}L\big(g^{\alpha\beta},
R_{\mu\nu\rho\sigma}\big)
\, .\label{GenLag}
\ee
Here the Riemann curvature tensor $R_{\mu\nu\rho\sigma}$ is defined
through $\big(\nabla_\mu\nabla_\nu-\nabla_\nu\nabla_\mu\big)u_\rho
=R_{\mu\nu\rho\sigma}u^\sigma$ with the help of an arbitrary
smooth vector field $u_\rho$. The variation of the Lagrangian
(\ref{GenLag}) with respect to the metric tensor $g_{\mu\nu}$
gives rise to
\be
\delta\big(\sqrt{-g}L\big)=\sqrt{-g}E_{\mu\nu} \delta g^{\mu\nu}
+\sqrt{-g}\nabla_\mu \Theta^\mu(\delta g)
\, . \label{Lagvari}
\ee
Like usual,  by assumption, the variation of the
coordinates vanishes throughout the present paper.
Such an assumption implies that the variation of
the field commutes with the partial derivative,
but not the covariant derivative. In
Eq. (\ref{Lagvari}), for convenience, by introducing the
symmetric second-rank tensor $A_{\mu\nu}=A_{(\mu\nu)}$
(it is denoted by $P_{\mu\nu}$ in \cite{Pady}),
the fourth-rank one $P^{\mu\nu\rho\sigma}$ and the
second-rank one $\mathcal{R}_{\mu\nu}$ whose symmetry
deserves to be determined, defined respectively as
\be
A_{\mu\nu}=\left(\frac{\partial L}{\partial
g^{\mu\nu}}\right)_{R_{\alpha\beta\rho\sigma}}
\, , \quad
P^{\mu\nu\rho\sigma}
=\left(\frac{\partial L}{\partial R_{\mu\nu\rho\sigma}}
\right)_{g^{\alpha\beta}}, \quad
\mathcal{R}_{\mu\nu}
=P_{\mu}^{~\lambda\rho\sigma}R_{\nu\lambda\rho\sigma}
\, , \label{PcalRdef}
\ee
the expression $E_{\mu\nu}$ for the field equations,
which stands for a symmetric second-rank tensor,
is read off as
\be
E_{\mu\nu} =A_{\mu\nu}-\mathcal{R}_{(\mu\nu)}
 -2\nabla^\rho\nabla^\sigma P_{\rho(\mu\nu)\sigma}
 -\frac{1}{2}Lg_{\mu\nu}
\, , \label{MotEq1}
\ee
while the surface term
$\Theta^\mu(\delta g)$ in Eq. (\ref{Lagvari}) is written as
\be
\Theta^\mu(\delta g)
=2P^{\mu\nu\rho\sigma}\nabla_\sigma\delta g_{\rho\nu}
-2\delta g_{\nu\rho} \nabla_\sigma P^{\mu\nu\rho\sigma}
\, ,\label{Boudterm}
\ee
according to Eq. (\ref{HatPdelR}).
By calculating the Lie derivative of the Lagrangian density $L$
in two different ways and then comparing the results, Padmanabhan
established a significant connection between $A_{\mu\nu}$ and
$\mathcal{R}_{\mu\nu}$, being of the form \cite{Pady}
\be
A_{\mu\nu}=2\mathcal{R}_{\mu\nu}
\, . \label{AcalRrel}
\ee
From Eq. (\ref{AcalRrel}), it is easy to see
that the symmetry of $A_{\mu\nu}$ results in
that $\mathcal{R}_{\mu\nu}$ is symmetric,
which further leads to that the second-rank tensor
$\nabla^\rho\nabla^\sigma P_{\rho\mu\nu\sigma}$ is
symmetric as well (for this see Eq. (\ref{IdenP2})).
At last, according to the fact that the derivatives
of the Lagrangian density with respect to both the
metric and the Riemann tensor are not independent
and they relate to each other in the manner described
by Eq. (\ref{AcalRrel}), through eliminating the
symmetric tensor $A_{\mu\nu}$ in Eq. (\ref{MotEq1}),
the expression $E_{\mu\nu}$ for the equations of
motion is further rewritten as the following simple
and economic form
\be
E_{\mu\nu} =\mathcal{R}_{\mu\nu}-\frac{1}{2}Lg_{\mu\nu}
 -2\nabla^\rho\nabla^\sigma P_{\rho\mu\nu\sigma}
\, , \label{MotEq2}
\ee
involving the $P^{\mu\nu\rho\sigma}$ tensor as well
as its second-order covariant derivative.
This expression has a wide range of applications in
the evaluation for the
field equations of the higher-order gravity theories
equipped with the Lagrangian belonging to
Eq. (\ref{GenLag}) in the literature.

In the present work, we are going to concentrate on
utilizing an off-shell Noether current to give an
alternative derivation of the relation (\ref{AcalRrel}),
further yielding Eq. (\ref{MotEq2}) for field equations.
Along the way, it will be shown that the expression
for field equations is divergence-free, and the Noether
potential associated with the conserved current
will be produced in a natural manner. Inspired with
this, a method based on the Noether current
will be developed to derive the field equations
of generalized gravity theories with diffeomorphism
invariance. What is more,
to our knowledge, an expression analogous to
Eq. (\ref{MotEq2}) for the field equations is
still absent up till now within the framework
of more general higher-order modified gravities
endowed with the Lagrangian proposed in
\cite{IyerWald} (for it see Eq. (\ref{LagCovR})),
which depends not only upon both the
metric and the Riemann curvature tensor but also on
the various-order covariant derivatives of the
latter. In order to fill the gap, we shall implement the
developed method to derive the explicit expressions
for the equations of motion corresponding to such
gravity theories at a general level. Besides,
the identity (\ref{AcalRrel}) will be generalized to
these theories, and the Noether
potentials will be given.

%%%%%%%%%%%%%%%%%%%%%%%%%%%%%%%%%%%%%%%%%%%%%%%%%%%%%%%%%%%%%%%%%%%%%%%%
\section{A method based on conserved current to derive
field equations}\label{two}
%%%%%%%%%%%%%%%%%%%%%%%%%%%%%%%%%%%%%%%%%%%%%%%%%%%%%%%%%%%%%%%%%%%%%%%%
Within what follows, we shall perform an explicit and
complete derivation for the identity (\ref{AcalRrel})
in another manner. On the basis of this, a method
generating field equations for pure gravity theories
will be proposed. For such a purpose, we merely
need to impose the conditions that the Lagrangian
for the gravity theory incorporates diffeomorphism
invariance and the rank-4 tensor $P^{\mu\nu\rho\sigma}$
exhibits the algebraic symmetries
\be
P^{\mu\nu\rho\sigma}=-P^{\nu\mu\rho\sigma}
=-P^{\mu\nu\sigma\rho}=P^{\rho\sigma\mu\nu}
\, . \label{PindeSym}
\ee
Unlike in the paper \cite{Pady}, it is not essential
for the $P^{\mu\nu\rho\sigma}$ tensor to satisfy
additionally the Bianchi-type identity
$P^{\mu[\nu\rho\sigma]}=0$ in the present work
(for a discussion on how to set up the algebraic
symmetries of $P^{\mu\nu\rho\sigma}$ see Appendix
\ref{appendC}). As a matter of fact, adding the term
$kP^{\mu[\nu\rho\sigma]}$ (here $k$ denotes any
constant parameter) to the tensor
$P^{\mu\nu\rho\sigma}$, does not
alter the expression for the equations of motion,
as well as the surface term \cite{PWG23}.
In terms of Eq. (\ref{PindeSym}), we make use of the
definition and the algebraic symmetries of the Riemann
tensor to acquire \cite{Pady,PWG23}
\be
\nabla^{[\rho} \nabla^{\sigma]}P_{\mu\nu\rho\sigma}=
\mathcal{R}_{[\mu\nu]}
\, .  \label{DivRP2t}
\ee
On the basis of this, with the help of the identity
$\nabla^\rho\nabla^\sigma P_{\mu[\nu\rho\sigma]}=0$
acquired by using both the equations
$P^{[\mu\nu\rho\sigma]}=P^{\mu[\nu\rho\sigma]}$ and
$\partial_\rho\partial_\sigma
\big(\sqrt{-g}P^{[\mu\nu\rho\sigma]}\big)=0$ rather
than the one $P^{\mu[\nu\rho\sigma]}=0$ employed
in \cite{Pady}, we gain the following identity
\cite{Pady,PWG23}
\be
2\nabla^{[\rho} \nabla^{\sigma]}P_{\rho\mu\nu\sigma}
=2\nabla^{\rho} \nabla^{\sigma}P_{\rho[\mu\nu]\sigma}
=-\nabla^\rho\nabla^\sigma P_{\mu\nu\rho\sigma}
=-\mathcal{R}_{[\mu\nu]}
\, . \label{IdenP2}
\ee
Before any further process, merely for the sake
of simplifying the calculations, we make use of
Eq. (\ref{IdenP2}) to get rid of the round
brackets in the expression (\ref{MotEq1}),
yielding
\be
E_{\mu\nu} =A_{\mu\nu}-\frac{1}{2}Lg_{\mu\nu}
-\mathcal{R}_{\mu\nu}
-2\nabla^\rho\nabla^\sigma P_{\rho\mu\nu\sigma}
\, . \label{MotEq3}
\ee
Here we point out that it is unknown at this
moment whether each of the last two terms in the
right hand side of Eq. (\ref{MotEq3}) is separately
symmetric although their linear combination
$\big(\mathcal{R}_{\mu\nu}
+2\nabla^\rho\nabla^\sigma P_{\rho\mu\nu\sigma}\big)$
is symmetric indeed. However, the symmetry of
both the terms will be completely determined below.

Under diffeomorphism characterized by the coordinate
transformation $x^\mu\rightarrow x^\mu-\zeta^\mu$,
where $\zeta^\mu$ represents an arbitrary smooth vector,
the variation operator in Eq. (\ref{Lagvari}) behaves
like the Lie derivative along the diffeomorphism
generator $\zeta^\mu$. As a result,
$\delta\big(\sqrt{-g}L\big)\rightarrow \mathcal{L}_\zeta
\big(\sqrt{-g}L\big)=\sqrt{-g}\nabla_\mu(\zeta^\mu L)$,
$\delta g_{\mu\nu}\rightarrow\mathcal{L}_\zeta g_{\mu\nu}
=2\nabla_{(\mu}\zeta_{\nu)}$, and
$\Theta^\mu(\delta g)\rightarrow
\Theta^\mu(\delta\rightarrow\mathcal{L}_\zeta)$.
Such expressions aid Eq. (\ref{Lagvari}) to turn into
\be
\nabla_\mu J^\mu=2\zeta_\nu \nabla_\mu E^{\mu\nu}
\, . \label{JErel}
\ee
In the above equation, the vector field $J^\mu$,
which will be demonstrated below to be
an off-shell conserved Noether current, is
presented by \cite{TPad10,RievLL}
\be
J^\mu=2E^{\mu\nu}\zeta_\nu+\zeta^\mu L
-\Theta^\mu(\delta\rightarrow\mathcal{L}_\zeta)
\, . \label{Jdef}
\ee
Apparently, the identity (\ref{JErel}) completely arises
from the diffeomorphism invariance of the gravity theory.
In fact, it can be thought of as the foundation
for the derivations of the field equations and the
Noether potential, as well as the conservation
equation of the former. Hence, it plays a crucial role
in our analysis below. To perform further analysis,
a prominent task is to explore the explicit structure
of $\Theta^\mu(\delta\rightarrow\mathcal{L}_\zeta)
=\Theta^\mu(\delta g)|_{\delta
g_{\mu\nu}\rightarrow 2\nabla_{(\mu}\zeta_{\nu)}}$,
taking the following form
\bea
\Theta^\mu(\delta\rightarrow\mathcal{L}_\zeta)
&=&2P^{\mu\nu\rho\sigma}\nabla_\sigma\nabla_\rho \zeta_\nu
+2P^{\mu\nu\rho\sigma}\nabla_\sigma\nabla_\nu \zeta_\rho \nn \\
&&-2\big(\nabla_\nu\zeta_\rho\big) \nabla_\sigma P^{\mu\nu\rho\sigma}
-2\big(\nabla_\rho\zeta_\nu\big) \nabla_\sigma P^{\mu\nu\rho\sigma}
\, .\label{TheLie}
\eea
In the above equation, both the vectors
$P^{\mu\nu\rho\sigma}\nabla_\sigma\nabla_\rho \zeta_\nu$
and $P^{\mu\nu\rho\sigma}\nabla_\sigma\nabla_\nu \zeta_\rho$
can be expressed as the following forms
\bea
P^{\mu\nu\rho\sigma}\nabla_\sigma\nabla_\rho \zeta_\nu&=&
-P^{\mu\nu\rho\sigma}\nabla_{[\rho}\nabla_{\sigma]}\zeta_\nu
=\frac{1}{2}\mathcal{R}^{\mu\nu}\zeta_\nu  \, , \nn \\
P^{\mu\nu\rho\sigma}\nabla_\sigma\nabla_\nu \zeta_\rho&=&
\nabla_\nu\big(P^{\mu\rho\sigma\nu}\nabla_\rho\zeta_\sigma
-\zeta_\rho\nabla_\sigma P^{\mu\nu\rho\sigma}\big)
-\zeta_\nu\nabla_{\rho} \nabla_{\sigma}P^{\rho\mu\nu\sigma}
\, ,\label{Thet12}
\eea
respectively, while the two vectors
$\big(\nabla_\nu\zeta_\rho\big) \nabla_\sigma
P^{\mu\nu\rho\sigma}$
and
$\big(\nabla_\rho\zeta_\nu\big) \nabla_\sigma
P^{\mu\nu\rho\sigma}$
can be rewritten respectively as
\bea
\big(\nabla_\nu\zeta_\rho\big) \nabla_\sigma P^{\mu\nu\rho\sigma}&=&
\nabla_\nu\big(\zeta_\rho\nabla_\sigma P^{\mu\nu\rho\sigma}\big)
+\zeta_\nu\nabla_{\rho} \nabla_{\sigma}P^{\rho\mu\nu\sigma}\, , \nn \\
\big(\nabla_\rho\zeta_\nu\big) \nabla_\sigma P^{\mu\nu\rho\sigma}&=&
\nabla_\nu\big(P^{\mu\sigma\rho\nu}\nabla_\rho\zeta_\sigma\big)
-\frac{1}{2}\mathcal{R}^{\mu\nu}\zeta_\nu
\, .\label{Thet34}
\eea
According to Eqs. (\ref{Thet12}) and (\ref{Thet34}), one observes
that each term in $\Theta^\mu(\delta\rightarrow\mathcal{L}_\zeta)$ is
expressed as the combination of two parts. One is proportional
to the vector $\zeta^\mu$, and the other is a total divergence term.
Substituting Eqs. (\ref{Thet12}) and (\ref{Thet34}) into
Eq. (\ref{TheLie}) leads to
\be
\Theta^\mu(\delta\rightarrow\mathcal{L}_\zeta)
=2\zeta_\nu\big(\mathcal{R}^{\mu\nu}
-2\nabla_{\rho}\nabla_{\sigma}P^{\rho\mu\nu\sigma}\big)
-\nabla_\nu K^{\mu\nu}
\, , \label{TheLie2}
\ee
in which the anti-symmetric tensor $K^{\mu\nu}$ is given by
\be
K^{\mu\nu}=2P^{\mu\nu\rho\sigma}\nabla_{\rho}\zeta_{\sigma}
+4\zeta_\rho\nabla_\sigma P^{\mu\nu\rho\sigma}
-6P^{\mu[\nu\rho\sigma]}\nabla_\rho\zeta_\sigma
\, . \label{Kmnudef}
\ee
It will be shown below that $K^{\mu\nu}$ is just the
Noether potential, which differs from the usual one
in the literature by comprising an extra second-rank
anti-symmetric tensor related to $P^{\mu[\nu\rho\sigma]}$
since we drop the constraint $P^{\mu[\nu\rho\sigma]}=0$
throughout the present work. Apparently,
$\Theta^\mu(\delta\rightarrow\mathcal{L}_\zeta)$
inherits the decomposition to its each component.
On the basis of Eq. (\ref{Jdef}), replacing
$E^{\mu\nu}$ and
$\Theta^\mu(\delta\rightarrow\mathcal{L}_\zeta)$
with the ones given by Eqs. (\ref{MotEq3})
and (\ref{TheLie2}) respectively, we write down
\be
J^\mu=2\big(A^{\mu\nu}-2\mathcal{R}^{\mu\nu}\big)\zeta_\nu
+\nabla_\nu K^{\mu\nu}
\, . \label{Jdef2}
\ee
Naturally, $J^\mu$ has the same structure as
$\Theta^\mu(\delta\rightarrow\mathcal{L}_\zeta)$.
It is easy to verify that the divergence of
$\nabla_\nu K^{\mu\nu}$ disappears, namely,
$\nabla_\mu\nabla_\nu K^{\mu\nu}=0$.

Furthermore, the replacement of $J^\mu$ in
Eq. (\ref{JErel}) with the one in
Eq. (\ref{Jdef2}) gives rise to
\be
\big(\nabla_\mu\zeta_\nu\big)
\big(A^{\mu\nu}-2\mathcal{R}^{\mu\nu}\big)
+\zeta_\nu\big[\nabla_\mu\big(A^{\mu\nu}-2\mathcal{R}^{\mu\nu}\big)
-\nabla_\mu E^{\mu\nu}\big]=0
\, . \label{JErel2}
\ee
In order to guarantee that Eq. (\ref{JErel2}) holds for an arbitrary
vector $\zeta_\nu$, the following conditions
\bea
A^{\mu\nu}-2\mathcal{R}^{\mu\nu}&=&0 \,  \nn \\
\nabla_\mu\big(A^{\mu\nu}-2\mathcal{R}^{\mu\nu}\big)
-\nabla_\mu E^{\mu\nu}&=&0
\, \label{AREdet}
\eea
have to be satisfied. Consequently, the first
equation in Eq. (\ref{AREdet}) straightforwardly
leads to Eq. (\ref{AcalRrel}), namely,
$A^{\mu\nu}=2\mathcal{R}^{\mu\nu}$. The symmetry of
$A^{\mu\nu}$ determines that the second-rank tensor
$\mathcal{R}^{\mu\nu}$ is symmetric as well,
subsequently leading to that the tensor
$\nabla^{\rho}\nabla^{\sigma}P_{\rho\mu\nu\sigma}$
is also symmetric with the help of Eq. (\ref{IdenP2}).
All of them are our desired
results. On the other hand, as a byproduct, the
second equation in Eq. (\ref{AREdet}) gives rise to
a generalized Bianchi identity for the expression of
field equations
\be
\nabla_\mu E^{\mu\nu}=
\nabla_\mu\big(A^{\mu\nu}-2\mathcal{R}^{\mu\nu}\big)
\, , \label{NewBid}
\ee
reproducing the main result in \cite{KolBid}. Apparently,
$A^{\mu\nu}=2\mathcal{R}^{\mu\nu}$ results in
$\nabla_\mu E^{\mu\nu}=0$ and $J^{\mu}=\nabla_\nu K^{\mu\nu}$.
The disappearance of the divergence for the latter
further demonstrates that $J^{\mu}$ is a conserved
current without the requirement that the metric has to
satisfy the equations of motion, while $K^{\mu\nu}$ is
the corresponding potential. In
particular, under the decomposition
$P^{\mu\nu\rho\sigma}=\tilde{P}^{\mu\nu\rho\sigma}
+P^{\mu[\nu\rho\sigma]}$, the potential $K^{\mu\nu}$
behaves like
$K^{\mu\nu}=\tilde{K}^{\mu\nu}-
4\nabla_\rho\big(\zeta_\sigma P^{[\mu\nu\rho\sigma]}\big)$
with the two-form
$\tilde{K}^{\mu\nu}=2\tilde{P}^{\mu\nu\rho\sigma}
\nabla_{\rho}\zeta_{\sigma}
+4\zeta_\rho\nabla_\sigma \tilde{P}^{\mu\nu\rho\sigma}$,
giving rise to an unchanged current $J^{\mu}$, namely,
$J^{\mu}=\nabla_\nu K^{\mu\nu}
=\nabla_\nu \tilde{K}^{\mu\nu}$.
According to this, together with the fact that the 2-form
potential is generally determined up to a divergence
term for an arbitrary 3-form, both $K^{\mu\nu}$ and
$\tilde{K}^{\mu\nu}$ are equivalent. However, the latter
achieves a dominant position in the literature.

We switch to making several remarks on the aforementioned
procedure for deriving the identity (\ref{AcalRrel}),
as well as the field equations.
Firstly, we derive such an equality only under the
requirements that the theory is diffeomorphism invariant
and the fourth-rank tensor $P^{\mu\nu\rho\sigma}$ is
armed with the index symmetries given by
Eq. (\ref{PindeSym}). Secondly, the identity
(\ref{JErel}), completely originated from the diffeomorphism
invariance of the theory, constitutes the foundation
of the derivation for field equations. Therefore, our
procedure builds a direct connection between the
symmetry of the Lagrangian and field equations. This provides
a concrete example to show that symmetry of the theory determines
the field equations associated with it. Thirdly,
both the equality (\ref{AcalRrel}) and the generalized
Bianchi identity (\ref{NewBid}) associated with
the expression of field equations can be derived
at the same level. Fourthly, our procedure to
derive the expression for the equations of motion
takes the advantage of the surface term, which
embraces all the crucial information for the field
equations and the Noether potential. Actually, this
procedure establishes the one-to-one correspondence
between the boundary term and the expression for the
equations of motion on the bulk.
One key feature
for it is captured by calculating the Lie
derivative of the surface term with regard to
an arbitrary smooth vector field.

As usual, the expression for the equations of motion is
adopted to formulate the conserved current. However,
the above procedure provides a way to utilize the latter
to reveal some significant information of the former,
including the derivation for field equations and Bianchi-type
identity. As what has been shown before, such a strategy
could provide convenience for the evaluation of field
equations. In fact, it also allows the extension to
other modified gravity theories admitting diffeomorphism
invariance apart from the ones described by the
Lagrangian (\ref{GenLag}) because of its simple and
common starting-point. This is able to be demonstrated
in a more generic manner in what follows.

Specifically, let us take into consideration of a general
diffeomorphism-invariant Lagrangian $\sqrt{-g}L$ with the
Lagrangian density
\be
L=L(g^{\alpha\beta},
R_{\mu\nu\rho\sigma},\nabla_{\lambda_1}R_{\mu\nu\rho\sigma},
\cdot\cdot\cdot,\nabla_{\lambda_1}\cdot\cdot\cdot
\nabla_{\lambda_m}R_{\mu\nu\rho\sigma};
B_{\mu_1\cdot\cdot\cdot\mu_j},\cdot\cdot\cdot)
\, , \label{GenLagDens}
\ee
where the rank-$j$ tensor $B_{\mu_1\cdot\cdot\cdot\mu_j}
=B_{\mu_1\cdot\cdot\cdot\mu_j}
(g^{\alpha\beta},R_{\mu\nu\rho\sigma},
\nabla_\lambda R_{\mu\nu\rho\sigma},
\nabla_\gamma\nabla_\lambda R_{\mu\nu\rho\sigma},\cdot\cdot\cdot)$
is allowed to depend upon the metric $g^{\mu\nu}$, the Riemann tensor
$R_{\mu\nu\rho\sigma}$ together with the covariant derivatives of
the Riemann tensor
such as $\nabla_\lambda R_{\mu\nu\rho\sigma}$,
$\nabla_\gamma\nabla_\lambda R_{\mu\nu\rho\sigma}$,
and so on. In such a case, the variation of the Lagrangian
still takes the form given by Eq. (\ref{Lagvari}).
Under the assumption that the surface term
$\Theta^\mu(\delta\rightarrow\mathcal{L}_\zeta)$ is able
to be generally decomposed into the following form
\be
\Theta^\mu(\delta\rightarrow\mathcal{L}_\zeta)
=2X^{\mu\nu}\zeta_\nu-\nabla_\nu \check{K}^{\mu\nu}
\, , \label{TheLiegen}
\ee
where the only requirement is that the second-rank tensor
$\check{K}^{\mu\nu}$ is anti-symmetric,
the identity (\ref{JErel}), which is the
starting point of the derivation, then becomes
\be
\big(\nabla_\mu\zeta_\nu\big)
\big(2E^{\mu\nu}+g^{\mu\nu}L-2X^{\mu\nu}\big)
+\zeta_\nu\big[\nabla_\mu\big(2E^{\mu\nu}
+g^{\mu\nu}L-2X^{\mu\nu}\big)
-2\nabla_\mu E^{\mu\nu}\big]=0
\, . \label{JErelgen}
\ee
Within Eqs. (\ref{TheLiegen}) and (\ref{JErelgen}),
$X^{\mu\nu}$ denotes a second-rank tensor that
dominates the equations of motion, while
the anti-symmetric tensor $\check{K}^{\mu\nu}$
actually stands for the Noether potential.
Due to the fact that Eq. (\ref{JErelgen}) holds for
any vector field $\zeta_\nu$, one immediately obtains
the expression for field equations
\be
E_{\mu\nu}=X_{\mu\nu}-\frac{1}{2}Lg_{\mu\nu}
\, , \label{EoMgen}
\ee
together with the Bianchi-type identity
\be
2\nabla_\mu E^{\mu\nu}
=\nabla_\mu\big(2E^{\mu\nu}+g^{\mu\nu}L-2X^{\mu\nu}\big)
=0
\, \label{BiaIdE}
\ee
or $\nabla^\nu L=2\nabla_\mu X^{\mu\nu}$. Substituting
Eq. (\ref{EoMgen}) into Eq. (\ref{TheLiegen}), one obtains
\be
2E^{\mu\nu}\zeta_\nu+\zeta^\mu L
-\Theta^\mu(\delta\rightarrow\mathcal{L}_\zeta)
=\nabla_\nu \check{K}^{\mu\nu}=J^\mu
\, . \label{CCurrThie}
\ee
The identity $\nabla_\mu\nabla_\nu \check{K}^{\mu\nu}=0$
leads to $\nabla_\mu J^{\mu}=0$, confirming that
$J^{\mu}$ is a conserved current and the two-form
$\check{K}^{\mu\nu}$ is indeed the potential
associated to it.

The symmetry of the expression $E_{\mu\nu}$ for field
equations further determines that $X_{\mu\nu}$ is symmetric
as well, leading to an identity
\be
X_{[\mu\nu]}=\frac{1}{2}(X_{\mu\nu}-X_{\nu\mu})=0
\, . \label{IdenXmn}
\ee
Besides, generally, the expression for the
equations of motion derived straightforwardly out
of the variation for the Lagrangian $\sqrt{-g}L$
is able to be written as the form
\be
E_{\mu\nu}=\frac{\partial L}{\partial
g^{\mu\nu}}-\frac{1}{2}Lg_{\mu\nu}+Y_{\mu\nu}
\, , \label{EEominY}
\ee
in which the second-rank symmetric tensor $Y_{\mu\nu}$
stands for all the contributions from the variation of
the Lagrangian with respect to the other variables but
the metric $g^{\mu\nu}$, such as $R_{\mu\nu\rho\sigma}$,
$\nabla_\alpha R_{\mu\nu\rho\sigma}$,
$\nabla_\alpha\nabla_\beta R_{\mu\nu\rho\sigma}$,
and so on. From the comparison between
Eqs. (\ref{EoMgen}) and (\ref{EEominY}), one obtains
the identity
\be
\frac{\partial L}{\partial g^{\mu\nu}}
=X_{\mu\nu}-Y_{\mu\nu}
\, , \label{IdPLPgXY}
\ee
which establishes the relationship between
$\partial L/\partial g^{\mu\nu}$
and the derivatives of the Lagrangian density
with respect to the other variables.
For instance, as an application of Eq. (\ref{EoMgen})
to the Lagrangian (\ref{GenLag}), Eq. (\ref{TheLie2})
shows that $X_{\mu\nu}=\mathcal{R}_{\mu\nu}
-2\nabla^{\rho}\nabla^{\sigma}P_{\rho\mu\nu\sigma}$.
Due to this, the identity (\ref{IdenXmn}) is specific
to $X_{[\mu\nu]}=2\mathcal{R}_{[\mu\nu]}=0$, yielding
$\mathcal{R}_{\mu\nu}=\mathcal{R}_{\nu\mu}$, and
Eq. (\ref{IdPLPgXY}) turns into
$A_{\mu\nu}=2\mathcal{R}_{\mu\nu}$ with the help of
$Y_{\mu\nu}=-\mathcal{R}_{\mu\nu}
-2\nabla^{\rho}\nabla^{\sigma}P_{\rho\mu\nu\sigma}$.

Apparently, Eq. (\ref{EoMgen}) demonstrates that
the field equations of motion can completely arise
from the surface term generated out of the variation of
the Lagrangian as long as this surface term can be
put into the form (\ref{TheLiegen}) after the variation
operator in it is replaced by the Lie derivative along
an arbitrary smooth vector.
Since the term $\big(\partial L/\partial
g^{\mu\nu}\big)_{\cdot\cdot\cdot}\delta g^{\mu\nu}$
appearing in the variation of the Lagrangian is only
proportional to the variation of the metric
$\delta g^{\mu\nu}$ rather than its derivatives
so that it makes no contribution to the surface term,
the rank-2 tensor $\big(\partial L/\partial
g^{\mu\nu}\big)_{\cdot\cdot\cdot}$ does not enter
into the one $X^{\mu\nu}$ that thoroughly
comes from the surface term. As a consequence,
using Eq. (\ref{TheLiegen}) to derive the expression for
the equations of motion has the merit to avoid
evaluating the derivative of the Lagrangian density
with respect to the metric, which renders it of
possibility to merely adopt the partial derivatives
of the Lagrangian density with respect to the
other variables except for the metric $g^{\mu\nu}$,
together with their derivatives, to express the
field equations. In this regard, it is unnecessary
to specially derive Eq. (\ref{IdPLPgXY}) in order
to get the expression (\ref{EoMgen}) for field
equations like in the work \cite{Pady}. What is
more, determining field equations through
Eq. (\ref{TheLiegen}) also has the advantage of
producing its Bianchi-type identity (\ref{BiaIdE})
and the Noether potential $\check{K}^{\mu\nu}$ for
the Lagrangian in a convenient manner.

For convenience, we summarize the main idea of the
above procedure to derive the field equations and
the Noether potential. First, one obtains the surface
term $\Theta^\mu$ by calculating the variation of
the Lagrangian $\sqrt{-g}L$. Next, after substituting
the variation operator $\delta$ by the Lie derivative with
respect to an arbitrary smooth vector $\zeta^\mu$, namely,
$\mathcal{L}_\zeta$, one expresses the vector
$\Theta^\mu(\delta\rightarrow\mathcal{L}_\zeta)$
as the form displayed by Eq. (\ref{TheLiegen}).
Finally, one extracts the Noether potential
$\check{K}^{\mu\nu}$ from Eq. (\ref{TheLiegen}) and obtains
the field equations with the help of Eq. (\ref{EoMgen}).

As an illustrative example, we utilize Eq. (\ref{MotEq2})
to evaluate the equations of motion, as well as the Noether
potential, for the Lagrangian with the density that is
just the so-called Pontryagin topological one
\be
L_{\text{Pon}}=\frac{1}{2}\epsilon_{\mu\nu\rho\sigma}
R^{\rho\sigma}_{~~~\alpha\beta}
R^{\alpha\beta\mu\nu}
\, , \label{LPontr}
\ee
where $\epsilon_{\mu\nu\rho\sigma}$ is the Levi-Civita
tensor in four dimensions. In such a situation, by
utilizing Eq. (\ref{Pcase1}), one is able to write down
the derivative of the Lagrangian density with respect
to the Riemann tensor $R_{\mu\nu\rho\sigma}$,
taking the form
\be
P^{\mu\nu\rho\sigma}_{\text{Pon}}=
\frac{\partial L_{\text{Pon}}}{\partial
R_{\mu\nu\rho\sigma}}
=\frac{1}{2}\epsilon^{\alpha\beta\mu\nu}
R_{\alpha\beta}^{~~~\rho\sigma}
+\frac{1}{2}\epsilon^{\alpha\beta\rho\sigma}
R_{\alpha\beta}^{~~~\mu\nu}
\, . \label{PPondef}
\ee
It is not difficult to verify that
$g_{\nu\rho}P^{\mu\nu\rho\sigma}_{\text{Pon}}=0
=g_{\mu\sigma}P^{\mu\nu\rho\sigma}_{\text{Pon}}$
and $P^{\mu[\nu\rho\sigma]}_{\text{Pon}}
=P^{[\mu\nu\rho\sigma]}_{\text{Pon}}
=\frac{1}{6}R\epsilon^{\mu\nu\rho\sigma}\neq0$. Besides,
if the fourth-rank tensor $\partial L_{\text{Pon}}/\partial
R_{\mu\nu\rho\sigma}$ is additionally demanded to obey the
Bianchi-type identity $P^{\mu[\nu\rho\sigma]}_{\text{Pon}}
=0$, due to Eq. (\ref{Pcase2}), it can be redefined as
\be
\tilde{P}^{\mu\nu\rho\sigma}_{\text{Pon}}
=P^{\mu\nu\rho\sigma}_{\text{Pon}}
- P^{\mu[\nu\rho\sigma]}_{\text{Pon}}
=\frac{1}{2}\epsilon^{\alpha\beta\mu\nu}
R_{\alpha\beta}^{~~~\rho\sigma}
+\frac{1}{2}\epsilon^{\alpha\beta\rho\sigma}
R_{\alpha\beta}^{~~~\mu\nu}
-\frac{1}{6}R\epsilon^{\mu\nu\rho\sigma}
\, . \label{tildPpondef}
\ee
Subsequently, replacing the tensor
$P^{\mu\nu\rho\sigma}$ in the quantities
$\mathcal{R}^{\mu\nu}$ and
$\nabla_\rho\nabla_\sigma
P^{\rho\mu\nu\sigma}$ with the one
$P^{\mu\nu\rho\sigma}_{\text{Pon}}$
gives rise to
\bea
P^{\mu\lambda\rho\sigma}_{\text{Pon}}
R^{\nu}_{~\lambda\rho\sigma}&=&
\epsilon^{\alpha\beta\mu\lambda}
R_{\alpha\beta}^{~~~\rho\sigma}
R^{\nu}_{~\lambda\rho\sigma}
+\epsilon^{\rho\sigma\mu\alpha}
R_{\rho\sigma}^{~~~\nu\beta}
R_{\alpha\beta}  \, , \nn \\
2\nabla_\rho\nabla_\sigma
P^{\rho\mu\nu\sigma}_{\text{Pon}}&=&
\epsilon^{\rho\sigma\mu\alpha}
R_{\rho\sigma}^{~~~\upsilon\beta}
R_{\alpha\beta}
\, . \label{PRcov2Ppon}
\eea
By making use of $\epsilon^{[\alpha\beta\mu\lambda}
R^{\nu]}_{~~\lambda\rho\sigma}=0$
and $\epsilon^{[\rho\sigma\mu\alpha}
R_{\rho\sigma}^{~~~\nu]\beta}=0$, one can check that
$P^{\mu\lambda\rho\sigma}_{\text{Pon}}
R^{\nu}_{~\lambda\rho\sigma}=
P^{\nu\lambda\rho\sigma}_{\text{Pon}}
R^{\mu}_{~\lambda\rho\sigma}$, as well as
$\nabla_\rho\nabla_\sigma
P^{\rho[\mu\nu]\sigma}_{\text{Pon}}=0$.
What is more, due to the fact that
\be
P^{\mu\lambda\rho\sigma}_{\text{Pon}}
R^{\nu}_{~\lambda\rho\sigma}=
\tilde{P}^{\mu\lambda\rho\sigma}_{\text{Pon}}
R^{\nu}_{~\lambda\rho\sigma}
+P^{\mu[\lambda\rho\sigma]}_{\text{Pon}}
R^{\nu}_{~[\lambda\rho\sigma]}
=\tilde{P}^{\mu\lambda\rho\sigma}_{\text{Pon}}
R^{\nu}_{~\lambda\rho\sigma}
\,  \label{PtildPpon}
\ee
and $2\nabla_\rho\nabla_\sigma
P^{\rho\mu\nu\sigma}_{\text{Pon}}
=2\nabla_\rho\nabla_\sigma
\tilde{P}^{\rho\mu\nu\sigma}_{\text{Pon}}$,
one observes that
$P^{\mu[\nu\rho\sigma]}_{\text{Pon}}$ makes no
contribution to the field equations with the
help of Eq. (\ref{MotEq2}). As a consequence of
Eq. (\ref{PRcov2Ppon}), the
field equations corresponding to the Lagrangian
$\sqrt{-g}L_{\text{Pon}}$ is given by
\be
E^{\mu\nu}_{\text{Pon}}=
P^{\mu\lambda\rho\sigma}_{\text{Pon}}
R^{\nu}_{~\lambda\rho\sigma}
-2\nabla_\rho\nabla_\sigma
P^{\rho\mu\nu\sigma}_{\text{Pon}}
-\frac{1}{4}g^{\mu\nu}
\epsilon^{\gamma\lambda\rho\sigma}
R_{\rho\sigma\alpha\beta}
R_{\gamma\lambda}^{~~~\alpha\beta}
\equiv0 \, . \label{EoMforLpon}
\ee
In order to arrive at the last equality in the above
equation, the identity
$g^{\nu[\mu}\epsilon^{\gamma\lambda\rho\sigma]}=0$
has been used. Eq. (\ref{EoMforLpon}) demonstrates
that the Pontryagin topological density makes
no contribution to the field equations. In spite of
this, it could contribute to the Noether potential. As
a matter of fact, according to Eq. (\ref{Kmnudef}),
one has the following potential
\be
K^{\mu\nu}_{\text{Pon}}=
\epsilon^{\alpha\beta\mu\nu}
\big(R_{\alpha\beta\rho\sigma}\nabla^\rho\zeta^\sigma
-4\zeta^\sigma
\nabla_\alpha R_{\beta\sigma}
-R\nabla_\alpha\zeta_\beta\big)
+\epsilon_{\alpha\beta\rho\sigma}
R^{\alpha\beta\mu\nu}
\nabla^\rho\zeta^\sigma
\, , \label{KmnPon}
\ee
or equivalently,
\bea
\epsilon_{\mu\nu\rho\sigma}K^{\rho\sigma}_{\text{Pon}}
&=&8\big(2\zeta^\sigma
\nabla_{[\mu} R_{\nu]\sigma}
-R_{\mu\nu\rho\sigma}\nabla^\rho\zeta^\sigma
+R_{\sigma[\mu}\nabla^\sigma\zeta_{\nu]}
-R_{\sigma[\mu}\nabla_{\nu]}\zeta^\sigma\big) \nn \\
&=&8\big[2\nabla_{[\mu}
\big(R_{\nu]}^{\sigma}\zeta_\sigma\big)
-R_{\mu\nu\rho\sigma}\nabla^\rho\zeta^\sigma
+R^\sigma_{\mu}\nabla_{(\sigma}\zeta_{\nu)}
-R^\sigma_{\nu}\nabla_{(\sigma}\zeta_{\mu)}\big]
\, . \label{dualKmnPon}
\eea
Apart from this, the Noether potential defined in
terms of the rank-4 tensor
$\tilde{P}^{\mu\nu\rho\sigma}_{\text{Pon}}$ takes
the form
\be
\tilde{K}^{\mu\nu}_{\text{Pon}}=
K^{\mu\nu}_{\text{Pon}}
+\frac{2}{3}\epsilon^{\mu\nu\rho\sigma}
\nabla_\rho(R\zeta_\sigma)
\, . \label{tildKpon}
\ee
Both the potentials $K^{\mu\nu}_{\text{Pon}}$ and
$\tilde{K}^{\mu\nu}_{\text{Pon}}$ are equivalent
since $\nabla_\nu K^{\mu\nu}_{\text{Pon}}
=\nabla_\nu \tilde{K}^{\mu\nu}_{\text{Pon}}$.
In order to demonstrate the above method more clearly,
apart from the example about the Lagrangian (\ref{LPontr}),
we also apply this method to derive equations
of motion and Noether potentials associated to the well-known
quadratic and cubic gravities within Appendix \ref{appendB0}.

%%%%%%%%%%%%%%%%%%%%%%%%%%%%%%%%%%%%%%%%%%%%%%%%%%%%%%%%%%%%%%%%%%%%%%%%
\section{The extension to the Lagrangian depending on
the Riemann tensor and its covariant derivatives}\label{three}
%%%%%%%%%%%%%%%%%%%%%%%%%%%%%%%%%%%%%%%%%%%%%%%%%%%%%%%%%%%%%%%%%%%%%%%%

\subsection{The general formalism for field equations and
Noether potential}\label{three1}

In the present section, as an illustrative and
significant application of the procedure to derive
the equations of motion proposed in the previous section,
we further take into consideration of a higher-order
generalized theory of pure gravity with the Lagrangian
that relies on the metric and the Riemann
curvature tensors, together with $i$th-order
$(i=1,2,\cdot\cdot\cdot,m)$ covariant derivatives
of the latter, being of the form \cite{IyerWald}
\be
\sqrt{-g}\hat{L}=\sqrt{-g}\hat{L}(g^{\alpha\beta},
R_{\mu\nu\rho\sigma},\nabla_{\alpha_1}R_{\mu\nu\rho\sigma},
\cdot\cdot\cdot,\nabla_{(\alpha_1}\cdot\cdot\cdot
\nabla_{\alpha_m)}R_{\mu\nu\rho\sigma})
\, , \label{LagCovR}
\ee
whose variation with regard to all the variables
$g^{\mu\nu}$, $R_{\mu\nu\rho\sigma}$, and
$\nabla_{(\alpha_1}\cdot\cdot\cdot\nabla_{\alpha_i)}
R_{\mu\nu\rho\sigma}$s brings forth
\bea
\delta\big(\sqrt{-g}\hat{L}\big)
&=&\sqrt{-g}\Big(\frac{\partial\hat{L}}{\partial g^{\mu\nu}}
-\frac{1}{2}\hat{L} g_{\mu\nu} \Big)\delta g^{\mu\nu}
+\sqrt{-g}\frac{\partial \hat{L}}{\partial R_{\mu\nu\rho\sigma}}
\delta R_{\mu\nu\rho\sigma} \nn \\
&&+\sqrt{-g}\sum^{m}_{i=1}
Z_{(i)}^{\alpha_1\cdot\cdot\cdot\alpha_i\mu\nu\rho\sigma}
\delta \nabla_{(\alpha_1}
\cdot\cdot\cdot\nabla_{\alpha_i)}R_{\mu\nu\rho\sigma}
\, , \label{LagCvariA}
\eea
with the tensors
$Z_{(i)}^{\alpha_1\cdot\cdot\cdot\alpha_i\mu\nu\rho\sigma}$s
$(i=1,\cdot\cdot\cdot,m)$ defined through
\be
Z_{(i)}^{\alpha_1\cdot\cdot\cdot\alpha_i\mu\nu\rho\sigma}
=Z_{(i)}^{(\alpha_1\cdot\cdot\cdot\alpha_i)\mu\nu\rho\sigma}
=\frac{\partial \hat{L}}
{\partial\nabla_{(\alpha_1}
\cdot\cdot\cdot\nabla_{\alpha_i)}R_{\mu\nu\rho\sigma}}
\, . \label{Zdef}
\ee
Here and in what follows the last four indices
$\mu$, $\nu$, $\rho$, and $\sigma$ in
$Z_{(i)}^{\alpha_1\cdot\cdot\cdot\alpha_i\mu\nu\rho\sigma}$
are required to have the same index symmetries as the
ones in the Riemann tensor $R^{\mu\nu\rho\sigma}$, that is,
\be
Z_{(i)}^{\alpha_1\cdot\cdot\cdot\alpha_i\mu\nu\rho\sigma}
=Z_{(i)}^{\alpha_1\cdot\cdot\cdot\alpha_i\alpha\beta\gamma\lambda}
\Delta^{\mu\nu\rho\sigma}_{\alpha\beta\gamma\lambda}
\, , \label{Zdef2}
\ee
with the tensor $\Delta^{\mu\nu\rho\sigma}_{\alpha\beta\gamma\lambda}$
given by Eq. (\ref{Del4index}). However, the identity
$Z_{(i)}^{\alpha_1\cdot\cdot\cdot\alpha_i\mu[\nu\rho\sigma]}=0$
is unnecessary. Besides, without particular requirement,
the integer $i$ runs from one up to $m$.

Moreover, with the help of the rank-4 tensor
$\hat{P}^{\mu\nu\rho\sigma}$ that accommodates
all the index symmetries in Eq. (\ref{PindeSym})
and is defined via \cite{IyerWald}
\be
\hat{P}^{\mu\nu\rho\sigma}=\frac{\partial\hat{L}}{\partial
R_{\mu\nu\rho\sigma}}
+\sum^{m}_{i=1} (-1)^i\nabla_{(\alpha_1}\cdot\cdot\cdot
\nabla_{\alpha_i)}
Z_{(i)}^{\alpha_1\cdot\cdot\cdot\alpha_i\mu\nu\rho\sigma}
\, , \label{hatPdef}
\ee
the equation (\ref{LagCvariA}) for the variation of
the Lagrangian can be reformulated as
\bea
\delta\big(\sqrt{-g}\hat{L}\big)
&=& \sqrt{-g}\left[\left(\frac{\partial \hat{L}}{\partial g^{\mu\nu}}
-\frac{1}{2}\hat{L} g_{\mu\nu} +B_{\mu\nu}\right)\delta g^{\mu\nu}
+\hat{P}^{\mu\nu\rho\sigma}\delta R_{\mu\nu\rho\sigma}
+\nabla_\mu \hat{\Theta}_{(2)}^\mu(\delta g)\right]\nn \\
&=&\sqrt{-g}\hat{E}_{\mu\nu} \delta g^{\mu\nu}
+\sqrt{-g}\nabla_\mu\hat{\Theta}^\mu(\delta g)
\, . \label{LagCvari}
\eea
Within Eq. (\ref{LagCvari}), the second-rank symmetric
tensor $B_{\mu\nu}$ arises from all the terms proportional
to $\delta \Gamma^\lambda_{\alpha\beta}$, which are
acquired in the procedure to extract terms proportional
to $\delta R_{\mu\nu\rho\sigma}$ out of the ones
$Z_{(i)}^{\alpha_1\cdot\cdot\cdot\alpha_i\mu\nu\rho\sigma}
\delta \nabla_{(\alpha_1}
\cdot\cdot\cdot\nabla_{\alpha_i)}R_{\mu\nu\rho\sigma}$s
so that the collection of such terms constitutes the
$\hat{P}^{\mu\nu\rho\sigma}\delta R_{\mu\nu\rho\sigma}$
term (for details see \cite{IyerWald} or equations
from (\ref{Upsilikdef}) to (\ref{ZidelRiem})
in Appendix \ref{appendA}).
It disappears when the Lagrangian is independent of
the variables $\nabla_{(\alpha_1}
\cdot\cdot\cdot\nabla_{\alpha_i)}R_{\mu\nu\rho\sigma}$s.
For instance, in the situation where $i=1$, in order to
separate out a term proportional to
$\delta R_{\mu\nu\rho\sigma}$ from
$Z_{(1)}^{\lambda\mu\nu\rho\sigma}
\delta \nabla_{\lambda}R_{\mu\nu\rho\sigma}$, one is able
to achieve this by expressing such a term as
\bea
Z_{(1)}^{\lambda\mu\nu\rho\sigma}
\delta \nabla_{\lambda}R_{\mu\nu\rho\sigma}
&=& Z_{(1)}^{\lambda\mu\nu\rho\sigma}
\nabla_{\lambda}\delta R_{\mu\nu\rho\sigma}
-4Z_{(1)}^{\alpha\beta\nu\rho\sigma}
R_{\lambda\nu\rho\sigma}
\delta \Gamma^\lambda_{\alpha\beta} \nn \\
&=&\nabla_{\mu}\big(Z_{(1)}^{\mu\alpha\beta\rho\sigma}
\delta R_{\alpha\beta\rho\sigma}\big)
-\big(\nabla_{\lambda}Z_{(1)}^{\lambda\mu\nu\rho\sigma}
\big)\delta R_{\mu\nu\rho\sigma}
-4Z_{(1)}^{\alpha\beta\nu\rho\sigma}
R_{\lambda\nu\rho\sigma}
\delta \Gamma^\lambda_{\alpha\beta}
\, . \quad \label{Z1delRiem}
\eea
The term $4Z_{(1)}^{\alpha\beta\nu\rho\sigma}
R_{\lambda\nu\rho\sigma}
\delta \Gamma^\lambda_{\alpha\beta}$
in Eq. (\ref{Z1delRiem}) can be
further written as
\be
4Z_{(1)}^{\alpha\beta\nu\rho\sigma}
R_{\lambda\nu\rho\sigma}
\delta \Gamma^\lambda_{\alpha\beta}
= \big(\nabla_\alpha H^{\alpha\mu\nu}_{(1)}\big)
\delta g_{\mu\nu}
-\nabla_\mu\big(H^{\mu\alpha\beta}_{(1)}
\delta g_{\alpha\beta}\big)
\, , \label{Z1delGam}
\ee
with the rank-3 tensor
$H^{\alpha\mu\nu}_{(1)}=H^{\alpha\nu\mu}_{(1)}$
defined through
\be
H^{\alpha\mu\nu}_{(1)}=-2\Big(Z^{\alpha(\mu|\tau\rho\sigma|}_{(1)}
R^{\nu)}_{~~\tau\rho\sigma}+Z^{(\mu|\alpha\tau\rho\sigma|}_{(1)}
R^{\nu)}_{~~\tau\rho\sigma}-Z^{(\mu\nu)\tau\rho\sigma}_{(1)}
R^\alpha_{~\tau\rho\sigma}\Big)
\, . \label{Htendef}
\ee
As a consequence of Eqs. (\ref{Z1delRiem}) and
(\ref{Z1delGam}), the term
$Z_{(1)}^{\lambda\mu\nu\rho\sigma}
\delta \nabla_{\lambda}R_{\mu\nu\rho\sigma}$ is
recast into
\bea
Z_{(1)}^{\lambda\mu\nu\rho\sigma}
\delta \nabla_{\lambda}R_{\mu\nu\rho\sigma}
&=&\nabla_{\mu}\big(Z_{(1)}^{\mu\alpha\beta\rho\sigma}
\delta R_{\alpha\beta\rho\sigma}
+H^{\mu\alpha\beta}_{(1)}
\delta g_{\alpha\beta}\big)
-\big(\nabla_{\lambda}Z_{(1)}^{\lambda\mu\nu\rho\sigma}
\big)\delta R_{\mu\nu\rho\sigma} \nn \\
&&-\big(\nabla_\alpha H^{\alpha\mu\nu}_{(1)}\big)
\delta g_{\mu\nu}
\, . \quad \label{Z1delRiem2}
\eea
From Eq. (\ref{Z1delRiem2}), it is easy to see that the
term $4Z_{(1)}^{\alpha\beta\nu\rho\sigma}
R_{\lambda\nu\rho\sigma}
\delta \Gamma^\lambda_{\alpha\beta}$ contributes to
the tensor $B_{\mu\nu}$ by
$\nabla_\alpha H^{\alpha\mu\nu}_{(1)}$. As is shown
in \cite{IyerWald}, the concrete expression for
$B_{\mu\nu}$ is absent there although a procedure on
how to yield it was sketched out. For convenience of
computing field equations, in the present
work, $B_{\mu\nu}$ is given explicitly by
Eq. (\ref{Bgendef}) with the help of
Eq. (\ref{ZidelRiem}) in Appendix \ref{appendA}.

What is more, the surface term
$\hat{\Theta}^\mu(\delta g)$ in Eq. (\ref{LagCvari}),
given by
\be
\hat{\Theta}^\mu(\delta g)=\hat{\Theta}_{(1)}^\mu(\delta g)
+\hat{\Theta}_{(2)}^\mu(\delta g)
\, , \label{HatThet}
\ee
is divided into two components for convenience. One of
them is the surface term $\hat{\Theta}_{(1)}^\mu(\delta g)$
stemming from the $\hat{P}^{\mu\nu\rho\sigma}
\delta R_{\mu\nu\rho\sigma}$ term. To figure out its
concrete expression, we make use of the Palatini identity
\be
\delta R^\mu_{~\nu\rho\sigma}=
\nabla_\rho \delta \Gamma^\mu_{\nu\sigma}
-\nabla_\sigma \delta \Gamma^\mu_{\nu\rho}
\,  \label{PalaIden}
\ee
to write down
\be
\hat{P}^{\mu\nu\rho\sigma}\delta R_{\mu\nu\rho\sigma}
=-\big(\hat{P}_{\mu}^{~\lambda\rho\sigma}R_{\nu\lambda\rho\sigma}
+2\nabla^\rho\nabla^\sigma \hat{P}_{\rho\mu\nu\sigma}\big)
\delta g^{\mu\nu}+\nabla_\mu \hat{\Theta}_{(1)}^\mu(\delta g)
\, , \label{HatPdelR}
\ee
where the surface term $\hat{\Theta}_{(1)}^\mu(\delta g)$
is read off as
\be
\hat{\Theta}_{(1)}^\mu(\delta g)
=2\hat{P}^{\mu\nu\rho\sigma}\nabla_\sigma\delta g_{\rho\nu}
-2\delta g_{\nu\rho} \nabla_\sigma \hat{P}^{\mu\nu\rho\sigma}
\, .\label{hatThe1}
\ee
On the basis of Eqs. (\ref{Upsiijgen}), (\ref{Upsili1def}),
(\ref{ZidelRiem}), and (\ref{Thei2gendef}),
the other surface term $\hat{\Theta}_{(2)}^\mu(\delta g)$,
which results from the terms proportional
to $\delta \Gamma^\lambda_{\alpha\beta}$ together
with the ones proportional to
$\delta \nabla\cdot\cdot\cdot\nabla R_{\mu\nu\rho\sigma}$
(for two explicit examples see Eqs. (\ref{Z1delRiem2})
and (\ref{Z2delCov2Riem2}), corresponding to the terms
$Z_{(1)}^{\lambda\mu\nu\rho\sigma}
\delta \nabla_{\lambda}R_{\mu\nu\rho\sigma}$ and
$Z^{\alpha\beta\mu\nu\rho\sigma}_{(2)}\delta
\nabla_{(\alpha}\nabla_{\beta)}R_{\mu\nu\rho\sigma}$,
respectively), is generally
expressed by
\be
\hat{\Theta}_{(2)}^\mu(\delta g)
=\sum^{m}_{i=1} \hat{\Theta}_{(i2)}^\mu(\delta g)
\, ,\label{The2genF2}
\ee
or Eq. (\ref{The2genF}) within Appendix \ref{appendA}.
In Eq. (\ref{The2genF2}), the term
$\hat{\Theta}_{(i2)}^\mu(\delta g)$, representing the
contribution from the variable $\nabla_{(\alpha_1}
\cdot\cdot\cdot\nabla_{\alpha_i)}R_{\mu\nu\rho\sigma}$
for the Lagrangian, can be determined by
Eq. (\ref{Thei2gendef}), taking the form
\be
\hat{\Theta}_{(i2)}^\mu(\delta g)
=\sum^i_{k=1}(-1)^{k-1}\Big(\nabla_{\lambda_1}
\cdot\cdot\cdot\nabla_{\lambda_{k-1}}
Z_{(i)}^{\mu\lambda_1\cdot\cdot\cdot\lambda_{i-1}
\alpha\beta\rho\sigma}\Big)\delta
\nabla_{\lambda_{k}}
\cdot\cdot\cdot\nabla_{\lambda_{i-1}}
R_{\alpha\beta\rho\sigma}
+H^{\mu\alpha\beta}_{(i)}\delta g_{\alpha\beta}
\, ,  \label{Thei2genF}
\ee
with the third-rank tensor $H^{\mu\alpha\beta}_{(i)}=
H^{\mu\beta\alpha}_{(i)}$ defined by
\be
H^{\mu\alpha\beta}_{(i)}=\frac{1}{2}
\Big(\hat{U}^{(\alpha|\mu|\beta)}_{(i)}
+\hat{U}^{(\alpha\beta)\mu}_{(i)}
-\hat{U}^{\mu(\alpha\beta)}_{(i)}
+2\tilde{U}^{(\alpha\beta)\mu}_{(i)}
-\tilde{U}^{\mu\alpha\beta}_{(i)}\Big)
\, . \label{HigLdef}
\ee
In the above equation, both the rank-three tensors
$\hat{U}^{\lambda\mu\nu}_{(i)}$ and
$\tilde{U}^{\lambda\mu\nu}_{(i)}$, built
from terms proportional to
$Z_{(i)}^{\lambda_1\cdot\cdot\cdot
\lambda_i\mu\nu\rho\sigma}$ or it with its
derivatives, are given respectively by
\bea
\hat{U}^{\mu\alpha\beta}_{(i)}&=&
4\sum^i_{k=1}(-1)^k\Big(\nabla_{\lambda_1}
\cdot\cdot\cdot\nabla_{\lambda_{k-1}}
Z_{(i)}^{\lambda_1\cdot\cdot\cdot
\lambda_{k-1}\alpha\lambda_{k+1}\cdot\cdot\cdot
\lambda_{i}\beta\nu\rho\sigma}\Big)
\nabla_{\lambda_{k+1}}\cdot\cdot\cdot\nabla_{\lambda_i}
R^\mu_{~\nu\rho\sigma}
\nn \\
\tilde{U}^{\mu\alpha\beta}_{(i)}&=&
\sum^{i-1}_{k=1}\sum^i_{j=k+1}(-1)^k
\Big(\nabla_{\lambda_1}
\cdot\cdot\cdot\nabla_{\lambda_{k-1}}
Z_{(i)}^{\lambda_1\cdot\cdot\cdot\lambda_{k-1}
\alpha\lambda_{k+1}\cdot\cdot\cdot
\lambda_{j-1}\beta\lambda_{j+1}
\cdot\cdot\cdot\lambda_{i}
\tau\nu\rho\sigma}\Big) \nn \\
&&\times \nabla_{\lambda_{k+1}}\cdot\cdot\cdot
\nabla_{\lambda_{j-1}}\nabla^\mu
\nabla_{\lambda_{j+1}}\cdot\cdot\cdot
\nabla_{\lambda_{i}}R_{\tau\nu\rho\sigma}
\, . \label{Ui12def}
\eea
From Eqs. (\ref{Thei2genF}) and (\ref{HigLdef}),
one observes that the tensor
$Z_{(i)}^{\lambda_1\cdot\cdot\cdot
\lambda_i\mu[\nu\rho\sigma]}$ makes no contribution
to the surface term
$\hat{\Theta}_{(2)}^\mu(\delta g)$ and such a term
vanishes when the Lagrangian does not encompass
terms consisting of the derivative of the Riemann
tensor.

By the aid of Eq. (\ref{HatPdelR}), the expression for
the equations of motion $\hat{E}_{\mu\nu}=\hat{E}_{(\mu\nu)}$
in Eq. (\ref{LagCvari}) is read off as
\bea
\hat{E}_{\mu\nu} &=&\frac{\partial \hat{L}}{\partial g^{\mu\nu}}
+B_{\mu\nu}
-\hat{P}_{(\mu}^{~~\lambda\rho\sigma}R_{\nu)\lambda\rho\sigma}
-2\nabla^\rho\nabla^\sigma \hat{P}_{\rho(\mu\nu)\sigma}
-\frac{1}{2}\hat{L} g_{\mu\nu} \nn \\
&=&\frac{\partial \hat{L}}{\partial g^{\mu\nu}}
+\sum_{i=1}^m\nabla^\lambda H_{(i)\lambda\mu\nu}
-\hat{P}_{\mu}^{~~\lambda\rho\sigma}R_{\nu\lambda\rho\sigma}
-2\nabla^\rho\nabla^\sigma \hat{P}_{\rho\mu\nu\sigma}
-\frac{1}{2}\hat{L} g_{\mu\nu}
\, . \label{hatEmn1}
\eea
In the second equality of the above equation, the round
brackets in the terms
$\hat{P}_{(\mu}^{~~\lambda\rho\sigma}R_{\nu)\lambda\rho\sigma}$
and $\nabla^\rho\nabla^\sigma \hat{P}_{\rho(\mu\nu)\sigma}$
are removed with the help of Eq. (\ref{IdenP2}).

Within the remainder of the present subsection, apart from the
expression (\ref{hatEmn1}) for $\hat{E}_{\mu\nu}$, it
will be demonstrated that $\hat{E}_{\mu\nu}$ can be
alternatively expressed as a simple and economic form
in terms of the tensor
$\partial \hat{L}/\partial R_{\mu\nu\rho\sigma}$ and the ones
$Z_{(i)}^{\alpha_1\cdot\cdot\cdot\alpha_i\mu\nu\rho\sigma}$s
$(i=1,\cdot\cdot\cdot,m)$, together with their derivatives,
in a manner similar to $E_{\mu\nu}$. By straightforward
analogy with Eq. (\ref{TheLie2}), the replacement
of the variation operator $\delta$ in the surface term
$\hat{\Theta}_{(1)}^\mu(\delta g)$ by the Lie derivative
with respect to the arbitrary smooth vector $\zeta_\nu$
brings about
\bea
\hat{\Theta}_{(1)}^\mu(\delta\rightarrow\mathcal{L}_\zeta)&=&
\Theta^\mu(\delta\rightarrow\mathcal{L}_\zeta)
\big|_{\text{Eq. (\ref{TheLie2})}}
\left(P\rightarrow\hat{P}\right)\nn \\
&=&2\zeta_\nu\left(\hat{P}^{\mu\lambda\rho\sigma}
R^{\nu}_{~\lambda\rho\sigma}
-2\nabla_{\rho}\nabla_{\sigma}
\hat{P}^{\rho\mu\nu\sigma}\right)
-\nabla_\nu \hat{K}_{(1)}^{\mu\nu}
\, , \label{hatTheLie1}
\eea
where the anti-symmetric tensor
$\hat{K}_{(1)}^{\mu\nu}$ is defined by
$\hat{K}_{(1)}^{\mu\nu}=K^{\mu\nu}
|_{P\rightarrow\hat{P}}$. Apart from this,
it has been demonstrated in great detail
within Appendix \ref{appendA} that the
surface term $\hat{\Theta}_{(2)}^\mu(\delta g)$
under diffeomorphism generated by the arbitrary
smooth vector $\zeta^\mu$ can
be generally expressed as the following form
\be
\hat{\Theta}_{(2)}^\mu(\delta\rightarrow\mathcal{L}_\zeta)
=2\zeta_\nu W^{\mu\nu}-\nabla_\nu \hat{K}_{(2)}^{\mu\nu}
\, , \label{LiehatThe2}
\ee
where $W^{\mu\nu}$ is given by Eq. (\ref{Wgendef}) or
(\ref{Wgendef2}) and the anti-symmetric tensor
$\hat{K}_{(2)}^{\mu\nu}=-\zeta_\lambda T^{\mu\nu\lambda}$
with $T^{\mu\nu\lambda}=T^{[\mu\nu]\lambda}$
presented by Eq. (\ref{ThiOTdef}) or (\ref{TigenLag})
according to Eq. (\ref{HatThgenL}). It is worth pointing
out that both the tensors $W^{\mu\nu}$
and $T^{\mu\nu\lambda}$ depend upon the rank-$(i+4)$
tensors $Z_{(i)}^{\lambda_1\cdot\cdot\cdot
\lambda_{i}\mu\nu\rho\sigma}$s
$(i=1,\cdot\cdot\cdot,m)$ together with their covariant
derivatives. As a consequence of Eqs. (\ref{hatTheLie1})
and (\ref{LiehatThe2}), under diffeomorphism generated
by $\zeta^\mu$, the total surface term
$\hat{\Theta}^\mu(\delta g)$ stemming from the variation
of the Lagrangian $\sqrt{-g}\hat{L}$ is transformed into
\bea
\hat{\Theta}^\mu(\delta\rightarrow\mathcal{L}_\zeta)
&=&\hat{\Theta}_{(1)}^\mu(\delta\rightarrow\mathcal{L}_\zeta)
+\hat{\Theta}_{(2)}^\mu(\delta\rightarrow\mathcal{L}_\zeta) \nn \\
&=&2\zeta_\nu\left(\hat{P}^{\mu\lambda\rho\sigma}
R^{\nu}_{~\lambda\rho\sigma}-2\nabla_{\rho}\nabla_{\sigma}
\hat{P}^{\rho\mu\nu\sigma}+W^{\mu\nu} \right)
-\nabla_\nu\hat{K}^{\mu\nu}
\, , \label{hatTheLie}
\eea
where the anti-symmetric tensor $\hat{K}^{\mu\nu}
=\hat{K}_{(1)}^{\mu\nu}+\hat{K}_{(2)}^{\mu\nu}$.
Within Eq. (\ref{hatTheLie}), like before, the surface term
$\hat{\Theta}^\mu(\delta\rightarrow\mathcal{L}_\zeta)$
is responsible for the determination of the Noether
potential and the equations of motion.

In fact, on the basis of Eq. (\ref{TheLiegen}), the
anti-symmetric tensor $\hat{K}^{\mu\nu}$ in Eq. (\ref{hatTheLie})
is the off-shell Noether potential for the Lagrangian
(\ref{LagCovR}), given explicitly by
\bea
\hat{K}^{\mu\nu}&=&2\hat{P}^{\mu\nu\rho\sigma}
\nabla_{\rho}\zeta_{\sigma}
+4\zeta_\rho\nabla_\sigma \hat{P}^{\mu\nu\rho\sigma}
-6\hat{P}^{\mu[\nu\rho\sigma]}\nabla_\rho\zeta_\sigma
-\zeta_\lambda T^{\mu\nu\lambda} \nn \\
&=&\zeta_\rho\sum_{i=1}^m\left(\hat{U}^{\rho[\mu\nu]}_{(i)}
+\hat{U}^{[\mu|\rho|\nu]}_{(i)}
+\hat{U}^{[\mu\nu]\rho}_{(i)}
+\tilde{U}^{[\mu|\rho|\nu]}_{(i)}
+\tilde{U}^{[\mu\nu]\rho}_{(i)}\right) \nn \\
&&+2\hat{P}^{\mu\nu\rho\sigma}
\nabla_{\rho}\zeta_{\sigma}
+4\zeta_\rho\nabla_\sigma \hat{P}^{\mu\nu\rho\sigma}
-6\hat{P}^{\mu[\nu\rho\sigma]}\nabla_\rho\zeta_\sigma
\, . \label{NoePLagCov}
\eea
Moreover, according to Eq. (\ref{EoMgen}), an
alternative expression for field equations
$\hat{E}_{\mu\nu}$ can be read off from
Eq. (\ref{hatTheLie}), having the form
\footnote{Here the expression $\hat{E}_{\mu\nu}$
gives rise to the field equations for pure gravity
theories, that is, $\hat{E}_{\mu\nu}=0$. Nevertheless,
in the presence of matter fields, the contributions
from these fields have to be included. For instance,
we consider the generic Lagrangian
$\sqrt{-g}\big(\hat{L}+L_{\text{M}}\big)$, where the
scalar $L_{\text{M}}$ stands for the Lagrangian density
associated to the matter fields. If it is assumed that
$L_{\text{M}}$ does not involve the derivatives of the
metric tensor, the energy-momentum tensor for the matter
fields can be defined as
$T^{\text{(M)}}_{\mu\nu}=-\frac{1}{\sqrt{-g}}
\frac{\partial(\sqrt{-g}{L}_{\text{M}})}{\partial{g}^{\mu\nu}}$
like usual. Hence, the equations of motion are given by
the conventional form $\hat{E}_{\mu\nu}=T^{\text{(M)}}_{\mu\nu}$.}

\be
\hat{E}_{\mu\nu}=W_{\mu\nu}+
\hat{P}_{\mu}^{~\lambda\rho\sigma}
R_{\nu\lambda\rho\sigma}-2\nabla^{\rho}\nabla^{\sigma}
\hat{P}_{\rho\mu\nu\sigma}
-\frac{1}{2}\hat{L}g_{\mu\nu}
\, . \label{EoMhatL}
\ee
Apparently, $\hat{E}_{\mu\nu}$ is independent of the rank-2
symmetric tensor $\partial \hat{L}/\partial g^{\mu\nu}$, and
it will be shown in Eq. (\ref{TPhCovE}) that it is
divergence-free. This is also proved through a direct
calculation on the divergence of the expression
for field equations in Appendix \ref{appendD}.
In contrast with the expression
(\ref{MotEq2}) for the field equations
corresponding to the Lagrangian (\ref{GenLag}),
the derivatives of the Riemann tensor included as
variables of the Lagrangian (\ref{LagCovR}) lead
to the existence of an additional term $W_{\mu\nu}$
in the expression $\hat{E}_{\mu\nu}$ for the
equations of motion. More specifically,
substituting Eqs. (\ref{hatPdef})
and (\ref{SymWmnidef}) into the above equation
brings $\hat{E}^{\mu\nu}$ to the following form
\be
\hat{E}^{\mu\nu}=
\hat{E}^{(\mu\nu)}_{Z}+P^{(\mu|\tau\rho\sigma|}_{(0)}
R^{\nu)}_{~~\tau\rho\sigma}
-2\nabla_{\rho}\nabla_{\sigma}
P^{\rho(\mu\nu)\sigma}_{(0)}
-\frac{1}{2}\hat{L}g^{\mu\nu}
\, . \label{EoMhatL2}
\ee
Within Eq. (\ref{EoMhatL2}), for convenience, the
fourth-rank tensor $P^{\mu\nu\rho\sigma}_{(0)}=
\partial \hat{L}/\partial R_{\mu\nu\rho\sigma}$
represents the derivative of the Lagrangian with
respect to the Riemann tensor, while the second-rank
tensor
$\hat{E}^{\mu\nu}_Z$, which consists of all
the terms relevant to the tensors
$Z_{(i)}^{\lambda_1\cdot\cdot\cdot
\lambda_i\mu\nu\rho\sigma}$s in $\hat{E}^{\mu\nu}$ and
is defined in terms of the expression
$\hat{E}^{(\mu\nu)}_Z=W^{(\mu\nu)}+
\left(\hat{P}^{(\mu|\tau\rho\sigma|}
-P^{(\mu|\tau\rho\sigma|}_{(0)}\right)R^{\nu)}_{~~\tau\rho\sigma}
-2\nabla_{\rho}\nabla_{\sigma}
\big(\hat{P}^{\rho(\mu\nu)\sigma}
-P^{\rho(\mu\nu)\sigma}_{(0)}\big)$,
has the explicit form
\bea
\hat{E}_Z^{\mu\nu}&=&
\sum^{m}_{i=1} (-1)^{i}\Big[
\Big(\nabla_{\lambda_1}\cdot\cdot\cdot
\nabla_{\lambda_i}
Z_{(i)}^{\lambda_1\cdot\cdot\cdot\lambda_i\mu\tau\rho\sigma}
\Big)R^{\nu}_{~\tau\rho\sigma}
-2\nabla_{\rho}\nabla_{\sigma}\nabla_{\lambda_1}
\cdot\cdot\cdot\nabla_{\lambda_i}
Z_{(i)}^{\lambda_1\cdot\cdot\cdot\lambda_i\rho\mu\nu\sigma}
\Big]\nn \\
&&+\frac{1}{2}\sum^{m}_{i=1}
\sum^i_{k=1}(-1)^{k-1}\Big(\nabla_{\lambda_1}
\cdot\cdot\cdot\nabla_{\lambda_{k-1}}
Z_{(i)}^{\lambda_1\cdot\cdot\cdot\lambda_{i-1}
\mu\alpha\beta\rho\sigma}\Big)
\nabla^{\nu}\nabla_{\lambda_{k}}
\cdot\cdot\cdot\nabla_{\lambda_{i-1}}
R_{\alpha\beta\rho\sigma} \nn \\
&& +\frac{1}{2}\sum^{m}_{i=1}
\Big(\nabla_\lambda\hat{U}^{\mu\nu\lambda}_{(i)}
-\nabla_\lambda\hat{U}^{\lambda\mu\nu}_{(i)}\Big)
+\frac{1}{2}\sum^{m}_{i=1}
\Big(\nabla_\lambda\tilde{U}^{\mu\nu\lambda}_{(i)}
-\nabla_\lambda\tilde{U}^{\lambda\mu\nu}_{(i)}\Big)
\, , \label{EoMhatLZ}
\eea
where both the rank-three tensors
$\hat{U}^{\lambda\mu\nu}_{(i)}$ and
$\tilde{U}^{\lambda\mu\nu}_{(i)}$ that depend upon terms
proportional to $Z_{(i)}^{\lambda_1\cdot\cdot\cdot
\lambda_i\mu\nu\rho\sigma}$ or it with its
derivatives are given by Eq. (\ref{Ui12def}).
In particular, it is easy to prove that the
transformation
$Z_{(i)}^{\lambda_1\cdot\cdot\cdot
\lambda_{i}\mu\nu\rho\sigma}\rightarrow
Z_{(i)}^{\lambda_1\cdot\cdot\cdot
\lambda_{i}\mu\nu\rho\sigma}
-kZ_{(i)}^{\lambda_1\cdot\cdot\cdot
\lambda_{i}\mu[\nu\rho\sigma]}$, in which
$k$ denotes an arbitrary constant parameter,
leaves $\hat{E}_Z^{\mu\nu}$ unaltered.
This is attributed to the fact that the quantity
$\left(\nabla_{\alpha_1}\cdot\cdot\cdot
\nabla_{\alpha_j}Z_{(i)}^{\lambda_1\cdot\cdot\cdot
\lambda_i\mu[\tau\rho\sigma]}\right)
R^{\nu}_{~\tau\rho\sigma}$ disappears identically
by virtue of the Bianchi identity
$R_{\mu[\nu\rho\sigma]}=0$ for the Riemann tensor
and there exists the identity
$\nabla_{\rho}\nabla_{\sigma}\nabla_{\lambda_1}
\cdot\cdot\cdot\nabla_{\lambda_i}
Z_{(i)}^{\lambda_1\cdot\cdot\cdot
\lambda_i\rho[(\mu\nu)\sigma]}\equiv0$, together
with the conclusion that both the tensors
$\hat{U}^{\mu\alpha\beta}_{(i)}$ and
$\tilde{U}^{\mu\alpha\beta}_{(i)}$ remain
unchanged under such a transformation.
As a result, the expression $\hat{E}^{\mu\nu}$
for field equations keep unaltered as well.

Remarkably, unlike in the situation
where the Lagrangian (\ref{GenLag}) depends merely
on the metric and the Riemann tensor, here
$\hat{E}_{\mu\nu}=\hat{E}_{\nu\mu}$ only
gives rise to the equality
\be
2\hat{P}_{[\mu}^{~~\tau\rho\sigma}
R_{\nu]\tau\rho\sigma}=-W_{[\mu\nu]}
\, ,\label{PRsymm}
\ee
rather than the one $P^{[\mu|\tau\rho\sigma|}_{(0)}
R^{\nu]}_{~~\tau\rho\sigma}=0$. Concretely,
with the help of Eq. (\ref{AnSymWmni}), the
identity (\ref{PRsymm}) is further transformed into
\bea
P^{[\mu|\tau\rho\sigma|}_{(0)}
R^{\nu]}_{~~\tau\rho\sigma}
&=&\frac{1}{4}\sum^m_{i=1}
\sum^i_{k=1}(-1)^{k}\Big(\nabla_{\lambda_1}
\cdot\cdot\cdot\nabla_{\lambda_{k-1}}
Z_{(i)}^{\lambda_1\cdot\cdot\cdot\lambda_{i-1}
[\mu|\alpha\beta\rho\sigma|}\Big)
\nabla^{\nu]}\nabla_{\lambda_{k}}
\cdot\cdot\cdot\nabla_{\lambda_{i-1}}
R_{\alpha\beta\rho\sigma} \nn \\
&&+\sum^{m}_{i=1} (-1)^{i-1}
\Big(\nabla_{\lambda_1}\cdot\cdot\cdot
\nabla_{\lambda_i}
Z_{(i)}^{\lambda_1\cdot\cdot\cdot\lambda_i[\mu|\tau\rho\sigma|}
\Big)R^{\nu]}_{~~\tau\rho\sigma}\nn \\
&&-\frac{1}{4}\sum^m_{i=1}
\nabla_\lambda\hat{U}^{[\mu|\lambda|\nu]}_{(i)}
-\frac{1}{4}\sum^m_{i=1}
\nabla_\lambda\tilde{U}^{[\mu\nu]\lambda}_{(i)}
\, . \label{PRsymm2}
\eea
Since each term in the right-hand side of the above
equality is proportional to the
$Z_{(i)}^{\lambda_1\cdot\cdot\cdot
\lambda_i\mu\nu\rho\sigma}$ tensor or it with its
derivatives, it disappears obviously in the absence
of all $Z_{(i)}^{\lambda_1\cdot\cdot\cdot
\lambda_i\mu\nu\rho\sigma}$s, returning naturally
to $\mathcal{R}_{[\mu\nu]}=0$ for the Lagrangian
(\ref{GenLag}). In addition, comparing
Eq. (\ref{hatEmn1}) with Eq. (\ref{EoMhatL}),
one acquires another identity for the derivative
of the Lagrangian density $\hat{L}$ with respect
to the inverse metric $g^{\mu\nu}$, that is,
\bea
\frac{\partial \hat{L}}{\partial g^{\mu\nu}}
&=&2\hat{P}_{\mu}^{~\lambda\rho\sigma}
R_{\nu\lambda\rho\sigma}
+W_{\mu\nu}-B_{\mu\nu} \nn \\
&=&2\hat{P}_{(\mu}^{~~\lambda\rho\sigma}
R_{\nu)\lambda\rho\sigma}
+W_{(\mu\nu)}-B_{\mu\nu}
\, . \label{ParLgW}
\eea
The above equation can be regarded as the
generalization of Eq. (\ref{AcalRrel}) and it
also confirms the identity (\ref{IdPLPgXY}).
Moreover, by the aid of Eqs. (\ref{Bgendef}) and
(\ref{SymWmnidef}), it is reexpressed as
\bea
\hat{A}^{\mu\nu}
&=&\frac{1}{2}\sum^m_{i=1}
\sum^i_{k=1}(-1)^{k-1}\Big(\nabla_{\lambda_1}
\cdot\cdot\cdot\nabla_{\lambda_{k-1}}
Z_{(i)}^{\lambda_1\cdot\cdot\cdot\lambda_{i-1}
(\mu|\alpha\beta\rho\sigma|}\Big)
\nabla^{\nu)}\nabla_{\lambda_{k}}
\cdot\cdot\cdot\nabla_{\lambda_{i-1}}
R_{\alpha\beta\rho\sigma} \nn \\
&&-\frac{1}{2}\sum^m_{i=1}
\nabla_\lambda\hat{U}^{(\mu|\lambda|\nu)}_{(i)}
-\frac{1}{2}\sum^m_{i=1}
\nabla_\lambda\tilde{U}^{(\mu\nu)\lambda}_{(i)}
+2\hat{P}^{(\mu|\tau\rho\sigma|}_{(0)}
R^{\nu)}_{~~\tau\rho\sigma}
\nn \\
&&+2\sum^{m}_{i=1} (-1)^{i}
\Big(\nabla_{\lambda_1}\cdot\cdot\cdot
\nabla_{\lambda_i}
Z_{(i)}^{\lambda_1\cdot\cdot\cdot\lambda_i(\mu|\tau\rho\sigma|}
\Big)R^{\nu)}_{~~\tau\rho\sigma}
\, , \label{ParLgW2}
\eea
where $\hat{A}^{\mu\nu}=g^{\mu\rho}g^{\nu\sigma}
\big(\partial \hat{L}/\partial g^{\rho\sigma}\big)
=\hat{A}^{\nu\mu}$. Apparently, the symmetry of
$\partial \hat{L}/\partial g^{\mu\nu}$ and $B_{\mu\nu}$
in Eq. (\ref{ParLgW}) yields the identity (\ref{PRsymm})
as well. In particular, when $\hat{L}=L$, Eq. (\ref{ParLgW2})
turns into the equality (\ref{AcalRrel}) without the
contributions from the
$Z_{(i)}^{\lambda_1\cdot\cdot\cdot\lambda_i\mu\nu\rho\sigma}$
tensors.

\subsection{Two examples}\label{three2}

In this subsection, for the sake of illustrating
Eqs. (\ref{EoMhatL}) and (\ref{ParLgW}),
as a significant example, we consider the Lagrangian density
that depends on the inverse metric $g^{\alpha\beta}$,
the Riemann tensor $R_{\mu\nu\rho\sigma}$ and the
first-order derivative of the Riemann tensor
$\nabla_{\lambda}R_{\mu\nu\rho\sigma}$, read off as
\be
\hat{L}_{(1)} =\hat{L}_{(1)}(g^{\alpha\beta},
R_{\mu\nu\rho\sigma},\nabla_{\lambda}R_{\mu\nu\rho\sigma})
\, . \label{Lag1cov}
\ee
For such a Lagrangian, in light of Eq. (\ref{Z1delRiem2}),
the tensors
$Z_{(i)}^{\alpha_1\cdot\cdot\cdot\alpha_i\mu\nu\rho\sigma}$
and $\hat{P}^{\mu\nu\rho\sigma}$ are specific to
\be
Z^{\mu\alpha\beta\rho\sigma}_{(1)}
=\frac{\partial\hat{L}_{(1)}}{\partial
\nabla_{\mu}R_{\alpha\beta\rho\sigma}}
\, , \quad
\hat{P}^{\mu\nu\rho\sigma}_{(1)}
=\frac{\partial\hat{L}_{(1)}}{\partial
R_{\mu\nu\rho\sigma}}
-\nabla_\lambda Z^{\lambda\mu\nu\rho\sigma}_{(1)}
\, , \label{ZP1def}
\ee
respectively. By the aid of Eq. (\ref{Z1delRiem2}),
the symmetric tensor $B^{\mu\nu}$ in the expression
(\ref{hatEmn1}) for the equations of motion
takes the value
\be
B^{\mu\nu}_{(1)}=\nabla_\alpha H^{\alpha\mu\nu}_{(1)}
\, , \label{B1def}
\ee
with $H^{\alpha\mu\nu}_{(1)}=H^{\alpha(\mu\nu)}_{(1)}$
given by Eq. (\ref{Htendef}). As a consequence of
Eq. (\ref{Z1delRiem2}), the surface term
$\hat{\Theta}_{(2)}^\mu(\delta g)$ turns into
\be
\hat{\Theta}_{(12)}^\mu(\delta g)=
Z^{\mu\alpha\beta\rho\sigma}_{(1)}
\delta R_{\alpha\beta\rho\sigma}
+H^{\mu\alpha\beta}_{(1)}\delta g_{\alpha\beta}
\, , \label{HatTh12def}
\ee
corresponding to the $i=1$ case of Eq. (\ref{Thei2genF}).
According to the above equation,
$\hat{\Theta}_{(12)}^\mu(\delta\rightarrow\mathcal{L}_\zeta)$
is read off as
\bea
\hat{\Theta}_{(12)}^\mu(\delta\rightarrow\mathcal{L}_\zeta)&=&
Z^{\mu\alpha\beta\rho\sigma}_{(1)}
\mathcal{L}_\zeta R_{\alpha\beta\rho\sigma}
+2H^{\mu(\alpha\beta)}_{(1)}\nabla_{\alpha}\zeta_{\beta}
\nn\\
&=&2\zeta_\nu W^{\mu\nu}_{(1)}
+\nabla_\nu \Big(\zeta_\lambda T^{\mu\nu\lambda}_{(1)}\Big)
\, , \label{HatTLi12}
\eea
in which the rank-3 tensor $T^{\mu\alpha\beta}_{(1)}$
has the form
\bea
T^{\mu\alpha\beta}_{(1)}&=&4Z^{\mu\alpha\lambda\rho\sigma}_{(1)}
R^\beta_{~\lambda\rho\sigma}
+2H^{\mu(\alpha\beta)}_{(1)}  \nn \\
&=&4Z^{[\mu\alpha]\lambda\rho\sigma}_{(1)}
R^\beta_{~\lambda\rho\sigma}
-4Z^{[\mu|\beta\lambda\rho\sigma|}_{(1)}
R^{\alpha]}_{~~\lambda\rho\sigma}
-4Z^{\beta[\mu|\lambda\rho\sigma|}_{(1)}
R^{\alpha]}_{~~\lambda\rho\sigma}
\, . \label{Rank3Tdef}
\eea
Here $T^{\mu\alpha\beta}_{(1)}=T^{[\mu\alpha]\beta}_{(1)}$
is in accordance with the symmetry of the tensor
$T^{\mu\alpha\beta}$ exhibited in Eq. (\ref{TPhCovE}).
On the basis of $T^{\mu\alpha\beta}_{(1)}$, the rank-2
tensor $W^{\mu\nu}_{(1)}$ in Eq. (\ref{HatTLi12})
is expressed as
\be
W^{\mu\nu}_{(1)}=\frac{1}{2}Z^{\mu\alpha\beta\rho\sigma}_{(1)}
\nabla^\nu R_{\alpha\beta\rho\sigma}
-\frac{1}{2}\nabla_\lambda T^{\mu\lambda\nu}_{(1)}
=W^{(\mu\nu)}_{(1)}+W^{[\mu\nu]}_{(1)}
\, , \label{Rank2W1def}
\ee
with $W^{(\mu\nu)}_{(1)}$ and $W^{[\mu\nu]}_{(1)}$ defined
respectively through
\bea
W^{(\mu\nu)}_{(1)}&=&\frac{1}{2}
Z^{(\mu|\alpha\beta\rho\sigma|}_{(1)}
\nabla^{\nu)} R_{\alpha\beta\rho\sigma}
+2\Big(\nabla^\lambda Z^{(\mu\nu)\tau\rho\sigma}_{(1)}\Big)
R_{\lambda\tau\rho\sigma}
-2Z^{(\mu|\lambda\tau\rho\sigma|}_{(1)}
\nabla_\lambda R^{\nu)}_{~~\tau\rho\sigma} \nn \\
&&-2\Big(\nabla_\lambda
Z^{(\mu|\lambda\tau\rho\sigma|}_{(1)}\Big)
R^{\nu)}_{~~\tau\rho\sigma}
+4Z^{(\mu\nu)\tau\rho\sigma}_{(1)}
\nabla_\rho R_{\sigma\tau}
\, , \nn \\
W^{[\mu\nu]}_{(1)}&=&
\frac{1}{2}Z^{[\mu|\alpha\beta\rho\sigma|}_{(1)}
\nabla^{\nu]} R_{\alpha\beta\rho\sigma}
+2\Big(\nabla_\lambda Z^{\lambda[\mu|\tau\rho\sigma|}_{(1)}\Big)
R^{\nu]}_{~~\tau\rho\sigma}
+2Z^{\lambda[\mu|\tau\rho\sigma|}_{(1)}
\nabla_\lambda R^{\nu]}_{~~\tau\rho\sigma}
\, . \label{W1anss}
\eea
In contrast with Eq. (\ref{LiehatThe2}),
Eq. (\ref{HatTLi12}) shows that the anti-symmetric
tensor $\hat{K}_{(2)}^{\mu\nu}$ there is replaced by
the one $\hat{K}_{(12)}^{\mu\nu}=
-\zeta_\lambda T^{\mu\nu\lambda}_{(1)}$ in the
situation for the Lagrangian (\ref{Lag1cov}).
As a result, substituting Eqs. (\ref{B1def})
and (\ref{Rank2W1def}) into Eq. (\ref{ParLgW})
gives rise to
\bea
\frac{\partial \hat{L}_{(1)}}{\partial g^{\mu\nu}}
&=&2\hat{P}_{\mu}^{(1)\lambda\rho\sigma}
R_{\nu\lambda\rho\sigma}
+\frac{1}{2}Z^{(1)\alpha\beta\rho\sigma}_{\mu}
\nabla_\nu R_{\alpha\beta\rho\sigma}
-\frac{1}{2}\nabla^\lambda T^{(1)}_{\mu\lambda\nu}
-\nabla^\alpha H^{(1)}_{\alpha(\mu\nu)} \nn \\
&=&\frac{1}{2}g_{\mu\beta}
\Big(4P_{(01)}^{\beta\tau\rho\sigma}
R_{\nu\tau\rho\sigma}
+Z^{\beta\alpha\tau\rho\sigma}_{(1)}
\nabla_\nu R_{\alpha\tau\rho\sigma}
+4Z^{\alpha\beta\tau\rho\sigma}_{(1)}
\nabla_\alpha R_{\nu\tau\rho\sigma}\Big)
\, . \label{ParLgW1}
\eea
Here and in the remainder of this section
$P_{(01)}^{\beta\tau\rho\sigma}$ stands for
$\partial\hat{L}_{(1)}/\partial R_{\beta\tau\rho\sigma}$.
Due to the fact that
$\partial \hat{L}_{(1)}/\partial g^{\mu\nu}$
is a symmetric rank-2 tensor, one obtains the
following identity
\be
4P^{[\mu|\tau\rho\sigma|}_{(01)}
R^{\nu]}_{~~\tau\rho\sigma}
+Z^{[\mu|\alpha\tau\rho\sigma|}_{(1)}
\nabla^{\nu]} R_{\alpha\tau\rho\sigma}
+4Z^{\alpha[\mu|\tau\rho\sigma|}_{(1)}
\nabla_{\alpha} R^{\nu]}_{~~\tau\rho\sigma}
=0 \, . \label{Z1identi}
\ee
Actually, Eq. (\ref{ParLgW1}) is in agreement
with Eq. (A6) in \cite{DLM16}, which
was derived by following Padmanabhan's
procedure proposed in \cite{Pady}.
Besides, the expression for the equations of
motion corresponding to the Lagrangian
(\ref{Lag1cov}) is given by
\bea
\hat{E}^{\mu\nu}_{(1)}
&=&P^{(\mu|\tau\rho\sigma|}_{(01)}
R^{\nu)}_{~~\tau\rho\sigma}-2\nabla_{\rho}\nabla_{\sigma}
P^{\rho(\mu\nu)\sigma}_{(01)}
-\Big(\nabla_\lambda
Z^{\lambda(\mu|\tau\rho\sigma|}_{(1)}\Big)
R^{\nu)}_{~~\tau\rho\sigma}
+2\nabla_{\rho}\nabla_{\sigma}\nabla_\lambda
Z^{\lambda\rho(\mu\nu)\sigma}_{(1)}\nn \\
&&-2Z^{(\mu|\lambda\tau\rho\sigma|}_{(1)}
\nabla_\lambda R^{\nu)}_{~~\tau\rho\sigma}
-2\Big(\nabla_\lambda
Z^{(\mu|\lambda\tau\rho\sigma|}_{(1)}\Big)
R^{\nu)}_{~~\tau\rho\sigma}
+4Z^{(\mu\nu)\tau\rho\sigma}_{(1)}
\nabla_\rho R_{\sigma\tau}\nn \\
&&+2\Big(\nabla^\lambda Z^{(\mu\nu)\tau\rho\sigma}_{(1)}\Big)
R_{\lambda\tau\rho\sigma}
+\frac{1}{2}Z^{(\mu|\lambda\tau\rho\sigma|}_{(1)}
\nabla^{\nu)} R_{\lambda\tau\rho\sigma}
-\frac{1}{2}\hat{L}_{(1)}g^{\mu\nu}
\, , \label{PotenL1cov}
\eea
while the Noether potential $\hat{K}_{\text{(1st)}}^{\mu\nu}
=\hat{K}_{(11)}^{\mu\nu}+\hat{K}_{(12)}^{\mu\nu}$
takes the form
\bea
\hat{K}_{\text{(1st)}}^{\mu\nu}
&=&2P^{\mu\nu\rho\sigma}_{(01)}
\nabla_{\rho}\zeta_{\sigma}
+4\zeta_\rho\nabla_\sigma
P^{\mu\nu\rho\sigma}_{(01)}
-6P^{\mu[\nu\rho\sigma]}_{(01)}
\nabla_\rho\zeta_\sigma
-2\Big(\nabla_\lambda
Z^{\lambda\mu\nu\rho\sigma}_{(1)}\Big)
\nabla_{\rho}\zeta_{\sigma} \nn \\
&&-4\zeta_\rho\nabla_\sigma\nabla_\lambda
Z^{\lambda\mu\nu\rho\sigma}_{(1)}
+6\Big(\nabla_\lambda
Z^{\lambda\mu[\nu\rho\sigma]}_{(1)}\Big)
\nabla_\rho\zeta_\sigma
-4\zeta_\lambda Z^{[\mu\nu]\tau\rho\sigma}_{(1)}
R^\lambda_{~\tau\rho\sigma}\nn \\
&&+4\zeta_\lambda Z^{[\mu|\lambda\tau\rho\sigma|}_{(1)}
R^{\nu]}_{~~\tau\rho\sigma}
+4\zeta_\lambda Z^{\lambda[\mu|\tau\rho\sigma|}_{(1)}
R^{\nu]}_{~~\tau\rho\sigma}
\, . \label{PotforL1}
\eea
Without matter fields, the expression (\ref{PotenL1cov})
for the equations of motion coincides with the one
given by Eq. (2.12) in the work \cite{ERR22},
where the field equations corresponding to the
Lagrangian $\sqrt{-g}\hat{L}_{(1)}$ was also
calculated in a different way. What is more, as an example
to show the application of Eq. (\ref{PotenL1cov}), it will be
utilized to derive the equations of motion associated to
the Lagrangian $-\sqrt{-g}(\nabla_\mu R)\nabla^\mu R$
in Appendix \ref{appendB}.

Apart from the above example, let us take into account
a more complex one, namely, the Lagrangian depending
on up to the second-order derivative of the Riemann tensor
\be
\sqrt{-g}\hat{L}_{(2)}=\sqrt{-g}\hat{L}_{(2)}\big(g^{\alpha\beta},
R_{\mu\nu\rho\sigma},\nabla_{\lambda}R_{\mu\nu\rho\sigma},
\nabla_{(\alpha}\nabla_{\beta)}R_{\mu\nu\rho\sigma}\big)
\, . \label{Lag2cov}
\ee
In this case, with the help of
\be
Z^{\lambda\mu\nu\rho\sigma}_{(1)}
=\frac{\partial\hat{L}_{(2)}}{\partial
\nabla_{\lambda}R_{\mu\nu\rho\sigma}}
\, , \qquad
Z^{\alpha\beta\mu\nu\rho\sigma}_{(2)}
=Z^{(\alpha\beta)\mu\nu\rho\sigma}_{(2)}
=\frac{\partial\hat{L}_{(2)}}{\partial
\nabla_{(\alpha}\nabla_{\beta)}R_{\mu\nu\rho\sigma}}
\, , \label{Z2P2def}
\ee
varying the Lagrangian (\ref{Lag2cov}) with respect
to all the four variables gives rise to
\bea
\delta\big(\sqrt{-g}\hat{L}_{(2)}\big)
&=& \sqrt{-g}\left[\left(\frac{\partial
\hat{L}_{(2)}}{\partial g^{\mu\nu}}
-\frac{1}{2}g_{\mu\nu}\hat{L}_{(2)}\right)\delta g^{\mu\nu}
+\frac{\partial\hat{L}_{(2)}}{\partial
R_{\mu\nu\rho\sigma}}\delta R_{\mu\nu\rho\sigma}
\right. \nn \\
&&\left.
+Z^{\lambda\mu\nu\rho\sigma}_{(1)}\delta\nabla_{\lambda}
R_{\mu\nu\rho\sigma}
+Z^{\alpha\beta\mu\nu\rho\sigma}_{(2)}\delta
\nabla_{(\alpha}\nabla_{\beta)}R_{\mu\nu\rho\sigma}
\right]
\, . \label{LaC2vari0}
\eea
The last term
$Z^{\alpha\beta\mu\nu\rho\sigma}_{(2)}\delta
\nabla_{(\alpha}\nabla_{\beta)}R_{\mu\nu\rho\sigma}$
in the right hand side of Eq. (\ref{LaC2vari0})
can be expressed as
\bea
Z^{\alpha\beta\mu\nu\rho\sigma}_{(2)}\delta
\nabla_{(\alpha}\nabla_{\beta)}R_{\mu\nu\rho\sigma}
&=&Z^{\alpha\beta\mu\nu\rho\sigma}_{(2)}
\nabla_{\alpha}\big(\delta\nabla_{\beta}
R_{\mu\nu\rho\sigma}\big)
-Z^{\alpha\beta\mu\nu\rho\sigma}_{(2)}
\big(\nabla_\lambda R_{\mu\nu\rho\sigma}\big)
\delta \Gamma^\lambda_{\alpha\beta}\nn \\
&&-4Z^{\mu\alpha\beta\nu\rho\sigma}_{(2)}
\big(\nabla_\mu R_{\lambda\nu\rho\sigma}\big)
\delta \Gamma^\lambda_{\alpha\beta} \nn \\
&=&\nabla_\mu\left[
Z^{\mu\lambda\alpha\beta\rho\sigma}_{(2)}
\delta\nabla_\lambda R_{\alpha\beta\rho\sigma}
-\big(\nabla_\lambda
Z^{\lambda\mu\alpha\beta\rho\sigma}_{(2)}\big)
\delta R_{\alpha\beta\rho\sigma}\right] \nn \\
&&+g_{\lambda\gamma}U^{\lambda\alpha\beta}_{(2)}
\delta \Gamma^\gamma_{\alpha\beta}
+\big(\nabla_{\alpha}\nabla_{\beta}
Z^{\alpha\beta\mu\nu\rho\sigma}_{(2)}\big)
\delta R_{\mu\nu\rho\sigma}
\, , \label{Z2delCov2Riem}
\eea
where the rank-3 tensor
$U^{\lambda\alpha\beta}_{(2)}=U^{\lambda\beta\alpha}_{(2)}$
is given by
\be
U^{\lambda\alpha\beta}_{(2)}=
4R^\lambda_{~\nu\rho\sigma}
\nabla_\mu Z^{\mu(\alpha\beta)\nu\rho\sigma}_{(2)}
-Z^{\alpha\beta\mu\nu\rho\sigma}_{(2)}
\nabla^\lambda R_{\mu\nu\rho\sigma}
-4Z^{\mu(\alpha\beta)\nu\rho\sigma}_{(2)}
\nabla_\mu R^\lambda_{~\nu\rho\sigma}
\, .  \label{U2def}
\ee
By introducing the rank-3 tensor $H^{\lambda\alpha\beta}_{(2)}$
defined through
\be
H^{\lambda\alpha\beta}_{(2)}=
H^{\lambda(\alpha\beta)}_{(2)}
=\frac{1}{2}\Big(U^{\alpha\lambda\beta}_{(2)}
+U^{\beta\alpha\lambda}_{(2)}
-U^{\lambda\alpha\beta}_{(2)}\Big)
\, , \label{H2def}
\ee
together with the surface term
$\hat{\Theta}_{(22)}^\mu(\delta g)$
taking the form
\be
\hat{\Theta}_{(22)}^\mu(\delta g)
=Z^{\mu\lambda\alpha\beta\rho\sigma}_{(2)}
\delta\nabla_\lambda R_{\alpha\beta\rho\sigma}
-\big(\nabla_\lambda
Z^{\lambda\mu\alpha\beta\rho\sigma}_{(2)}\big)
\delta R_{\alpha\beta\rho\sigma}
+H^{\mu\alpha\beta}_{(2)}\delta g_{\alpha\beta}
\, , \label{The22def}
\ee
one is able to further express the term
$Z^{\alpha\beta\mu\nu\rho\sigma}_{(2)}\delta
\nabla_{(\alpha}\nabla_{\beta)}R_{\mu\nu\rho\sigma}$
as the linear combination of a total divergence term
and two terms proportional to
$\delta R_{\mu\nu\rho\sigma}$ and $\delta g_{\mu\nu}$
respectively, which has the specific form
\be
Z^{\alpha\beta\mu\nu\rho\sigma}_{(2)}\delta
\nabla_{(\alpha}\nabla_{\beta)}R_{\mu\nu\rho\sigma}
=\nabla_\mu\hat{\Theta}_{(22)}^\mu(\delta g)
+\big(\nabla_{\alpha}\nabla_{\beta}
Z^{\alpha\beta\mu\nu\rho\sigma}_{(2)}\big)
\delta R_{\mu\nu\rho\sigma}
-\big(\nabla_\lambda H^{\lambda\mu\nu}_{(2)}\big)
 \delta g_{\mu\nu}
\, . \label{Z2delCov2Riem2}
\ee
Subsequently, according to Eqs. (\ref{Z1delRiem2}) and
(\ref{Z2delCov2Riem2}), Eq. (\ref{LaC2vari0}) is further
expressed as
\bea
\delta\big(\sqrt{-g}\hat{L}_{(2)}\big)
&=& \sqrt{-g}\left(\frac{\partial \hat{L}_{(2)}}{\partial g^{\mu\nu}}
-\frac{1}{2}g_{\mu\nu}\hat{L}_{(2)} +B^{(1)}_{\mu\nu}
 +B^{(2)}_{\mu\nu}\right)\delta g^{\mu\nu}
+\sqrt{-g}\hat{P}^{\mu\nu\rho\sigma}_{(2)}\delta R_{\mu\nu\rho\sigma}
\nn \\
&&+\sqrt{-g}\nabla_\mu\Big( \hat{\Theta}_{(12)}^\mu(\delta g)
+\hat{\Theta}_{(22)}^\mu(\delta g)\Big)
\, . \label{LaC2vari}
\eea
Within Eq. (\ref{LaC2vari}), the rank-4 tensor
$\hat{P}^{\mu\nu\rho\sigma}_{(2)}$ is presented by
\be
\hat{P}^{\mu\nu\rho\sigma}_{(2)}=
\frac{\partial\hat{L}_{(2)}}{\partial
R_{\mu\nu\rho\sigma}}
-\nabla_\lambda Z^{\lambda\mu\nu\rho\sigma}_{(1)}
+\nabla_{\alpha}\nabla_{\beta}
Z^{\alpha\beta\mu\nu\rho\sigma}_{(2)}
\, , \label{HatP2def}
\ee
while the symmetric tensor $B_{(2)}^{\mu\nu}$ can be
read directly from Eq. (\ref{Z2delCov2Riem2}),
being of the form
\be
B_{(2)}^{\mu\nu}=\nabla_\lambda H^{\lambda\mu\nu}_{(2)}
\, . \label{B2def}
\ee
Besides, both the symmetric tensor $B_{(1)}^{\mu\nu}$
and the surface term
$\hat{\Theta}_{(12)}^\mu(\delta g)$
are still given by Eqs. (\ref{B1def}) and
(\ref{HatTh12def}), respectively, but the quantity
$Z^{\lambda\mu\nu\rho\sigma}_{(1)}$ in them is
alternatively defined in terms of the Lagrangian
density $\hat{L}_{(2)}$, given by Eq. (\ref{Z2P2def}).

In terms of Eq. (\ref{The22def}), the replacement of
$\delta$ by $\mathcal{L}_\zeta$ brings
$\hat{\Theta}_{(22)}^\mu(\delta g)$ to
the following form
\be
\hat{\Theta}_{(22)}^\mu(\delta\rightarrow\mathcal{L}_\zeta)
=2\zeta_\nu W^{\mu\nu}_{(2)}
+\nabla_\nu \Big(\zeta_\lambda T^{\mu\nu\lambda}_{(2)}\Big)
\, , \label{HatTL22}
\ee
which together with
$\hat{\Theta}_{(12)}^\mu
(\delta\rightarrow\mathcal{L}_\zeta)$
accounts for the equations of motion and the
Noether potential. In the above equation, the
tensor $T^{\mu\alpha\beta}_{(2)}$ is defined through
\bea
T^{\mu\alpha\beta}_{(2)}
&=&4Z^{\lambda[\mu\alpha]\nu\rho\sigma}_{(2)}
\nabla_\lambda R^\beta_{~\nu\rho\sigma}
-4R^\beta_{~\nu\rho\sigma}
\nabla_\lambda Z^{\lambda[\mu\alpha]\nu\rho\sigma}_{(2)}
-4R^{[\mu}_{~~\nu\rho\sigma}
\nabla_\lambda Z^{\alpha]\lambda\beta\nu\rho\sigma}_{(2)}
\nn \\
&&-4R^{[\mu}_{~~\nu\rho\sigma}
\nabla_\lambda
Z^{|\lambda\beta|\alpha]\nu\rho\sigma}_{(2)}
+2\big(\nabla^{[\mu} R_{\lambda\nu\rho\sigma}\big)
Z^{\alpha]\beta\lambda\nu\rho\sigma}_{(2)} \nn \\
&&+4\big(\nabla_\lambda R^{[\mu}_{~~\nu\rho\sigma}\big)
Z^{\alpha]\lambda\beta\nu\rho\sigma}_{(2)}
+4\big(\nabla_\lambda R^{[\mu}_{~~\nu\rho\sigma}\big)
Z^{|\lambda\beta|\alpha]\nu\rho\sigma}_{(2)}
\, . \label{T2def}
\eea
Apparently, $T^{\mu\alpha\beta}_{(2)}=-T^{\alpha\mu\beta}_{(2)}$,
confirming the first equation in Eq. (\ref{TPhCovE}) once again.
Besides, the tensor $W^{\mu\nu}_{(2)}$ in Eq. (\ref{HatTL22})
is defined by
\be
W^{\mu\nu}_{(2)}=W^{(\mu\nu)}_{(2)}+W^{[\mu\nu]}_{(2)}
\, , \label{W2def}
\ee
with $W^{(\mu\nu)}_{(2)}$ given by
\bea
W^{(\mu\nu)}_{(2)}&=&\frac{1}{2}\Big[
Z^{\lambda(\mu|\alpha\beta\rho\sigma|}_{(2)}
\nabla^{\nu)} \nabla_\lambda R_{\alpha\beta\rho\sigma}
-\Big(\nabla_\lambda
Z^{\lambda(\mu|\alpha\beta\rho\sigma|}_{(2)}\Big)
\nabla^{\nu)} R_{\alpha\beta\rho\sigma} \nn \\
&&-\big(\nabla^{(\mu}  R_{\alpha\beta\rho\sigma}\big)
\nabla_\lambda Z^{\nu)\lambda\alpha\beta\rho\sigma}_{(2)}
-\big(\nabla_\lambda\nabla^{(\mu}
R_{\alpha\beta\rho\sigma}\big)
Z^{\nu)\lambda\alpha\beta\rho\sigma}_{(2)} \nn \\
&&+\big(\nabla^\lambda Z^{\mu\nu\alpha\beta\rho\sigma}_{(2)}\big)
\nabla_{\lambda}  R_{\alpha\beta\rho\sigma}
+Z^{\mu\nu\alpha\beta\rho\sigma}_{(2)}
\nabla^\lambda\nabla_{\lambda} R_{\alpha\beta\rho\sigma} \nn \\
&&+\nabla_\lambda\hat{U}^{(\mu\nu)\lambda}_{(2)}
-\nabla_\lambda\hat{U}^{\lambda(\mu\nu)}_{(2)}
\Big]
\,  \label{W2symmdef}
\eea
and $W^{[\mu\nu]}_{(2)}$ being of the form
\bea
W^{[\mu\nu]}_{(2)}&=&\frac{1}{2}\Big[
Z^{\lambda[\mu|\alpha\beta\rho\sigma|}_{(2)}
\nabla^{\nu]} \nabla_\lambda R_{\alpha\beta\rho\sigma}
-\Big(\nabla_\lambda
Z^{\lambda[\mu|\alpha\beta\rho\sigma|}_{(2)}\Big)
\nabla^{\nu]} R_{\alpha\beta\rho\sigma} \nn \\
&&-\big(\nabla^{[\mu}  R_{\alpha\beta\rho\sigma}\big)
\nabla_\lambda Z^{\nu]\lambda\alpha\beta\rho\sigma}_{(2)}
-\big(\nabla_\lambda\nabla^{[\mu}
R_{\alpha\beta\rho\sigma}\big)
Z^{\nu]\lambda\alpha\beta\rho\sigma}_{(2)}  \nn \\
&&+4R^{[\mu}_{~~\tau\rho\sigma}
\nabla_\gamma\nabla_\lambda
Z^{|\gamma\lambda|\nu]\tau\rho\sigma}_{(2)}
-4\big(\nabla_\gamma\nabla_\lambda
R^{[\mu}_{~~\tau\rho\sigma}\big)
Z^{|\gamma\lambda|\nu]\tau\rho\sigma}_{(2)}\Big]
\, . \label{W2ansymdef}
\eea
By the aid of Eq. (\ref{Ui12def}), the rank-3 tensor
$\hat{U}^{\mu\nu\lambda}_{(2)}$ in
Eq. (\ref{W2symmdef}) has the form
\be
\hat{U}^{\mu\nu\lambda}_{(2)}=
4R^\mu_{~\tau\rho\sigma}
\nabla_\gamma Z^{\gamma\nu\lambda\tau\rho\sigma}_{(2)}
-4Z^{\gamma\nu\lambda\tau\rho\sigma}_{(2)}
\nabla_\gamma R^\mu_{~\tau\rho\sigma}
\, . \label{hatU2def}
\ee

Furthermore, by making use of Eqs. (\ref{TheLiegen}) and
(\ref{EoMgen}), one obtains the expression
$\hat{E}^{\mu\nu}_{(2)}$ for field equations
associated with the Lagrangian (\ref{Lag2cov}), being of
the form
\be
\hat{E}^{\mu\nu}_{(2)}=\hat{P}^{\sigma\rho\tau(\mu}_{(2)}
R^{\nu)}_{~~\tau\rho\sigma}-2\nabla_{\rho}\nabla_{\sigma}
\hat{P}^{\rho(\mu\nu)\sigma}_{(2)}
-\frac{1}{2}\hat{L}_{(2)}g^{\mu\nu}
+W^{(\mu\nu)}_{(1)}+W^{(\mu\nu)}_{(2)}
\, , \label{EP2cov}
\ee
as well as the off-shell Noether potential
$\hat{K}_{\text{(2nd)}}^{\mu\nu}$, presented by
\be
\hat{K}_{\text{(2nd)}}^{\mu\nu}=
2\hat{P}^{\mu\nu\rho\sigma}_{(2)}
\nabla_{\rho}\zeta_{\sigma}
+4\zeta_\rho\nabla_\sigma
\hat{P}^{\mu\nu\rho\sigma}_{(2)}
-6\hat{P}^{\mu[\nu\rho\sigma]}_{(2)}
\nabla_\rho\zeta_\sigma
-\zeta_\lambda T^{\mu\nu\lambda}_{(1)}
-\zeta_\lambda T^{\mu\nu\lambda}_{(2)}
\, . \label{NoePotL2}
\ee
What is more, according to Eq. (\ref{ParLgW}),
$\partial \hat{L}_{(2)}/\partial g^{\mu\nu}$ is read off as
\be
\frac{\partial \hat{L}_{(2)}}{\partial g^{\mu\nu}}
=g_{\mu\alpha}g_{\mu\beta}
\Big(2\hat{P}_{(2)}^{\sigma\rho\tau(\alpha}
R^{\beta)}_{~~\tau\rho\sigma}
+W_{(1)}^{(\alpha\beta)}+W_{(2)}^{(\alpha\beta)}
-\nabla_\lambda H^{\lambda\alpha\beta}_{(1)}
-\nabla_\lambda H^{\lambda\alpha\beta}_{(2)}\Big)
\, . \label{PLgW12}
\ee
The identity (\ref{PRsymm}) is specific to
\be
2\hat{P}_{(2)}^{\sigma\rho\tau[\mu}
R^{\nu]}_{~~\tau\rho\sigma}=
-W^{[\mu\nu]}_{(1)}-W^{[\mu\nu]}_{(2)}
\, . \label{PW2rel}
\ee
It is worth pointing out that the tensor
$Z^{\lambda\mu\nu\rho\sigma}_{(1)}$ appearing in
$\hat{P}^{\mu\nu\rho\sigma}_{(2)}$,
$W^{\mu\nu}_{(1)}$, $T^{\mu\nu\lambda}_{(1)}$,
and $H^{\lambda\alpha\beta}_{(1)}$ within equations
from (\ref{EP2cov}) to (\ref{PW2rel}) has to be
compatible with the Lagrangian density $\hat{L}_{(2)}$.
That is to say, this tensor is defined through
$Z^{\lambda\mu\nu\rho\sigma}_{(1)}
=\partial\hat{L}_{(2)}/\partial
\nabla_{\lambda}R_{\mu\nu\rho\sigma}$ in contrast
with the previous definition
$Z^{\lambda\mu\nu\rho\sigma}_{(1)}
=\partial\hat{L}_{(1)}/\partial
\nabla_{\lambda}R_{\mu\nu\rho\sigma}$ corresponding to
the Lagrangian $\sqrt{-g}\hat{L}_{(1)}$.
In order to demonstrate the application of
Eq. (\ref{EP2cov}), we will consider the
field equations associated with the Lagrangian
density $f\big(R^n, (\Box R)^m\big)$
in appendix \ref{appendB}.

%%%%%%%%%%%%%%%%%%%%%%%%%%%%%%%%%%%%%%%%%%%%%%%%%%%%%%%%%%%%%%%%%%%%%%%%
\section{Summary}\label{four}
%%%%%%%%%%%%%%%%%%%%%%%%%%%%%%%%%%%%%%%%%%%%%%%%%%%%%%%%%%%%%%%%%%%%%%%%

In summary, within the framework of the generalized
gravity theories described by the Lagrangian (\ref{GenLag}),
we present in full detail an alternative
derivation of the equality (\ref{AcalRrel}) obtained in \cite{Pady}
by making use of the identity (\ref{JErel}), which builds the
relation between the divergences of the off-shell Noether
current and the expression for the equations of motion under
the diffeomorphism transformation. Apart from this, the
generalized Bianchi identity (\ref{NewBid}) is produced at
the same time. Inspired with the analysis related to
the Lagrangian (\ref{GenLag}), we further propose a
method based on the off-shell Noether current to derive
field equations for gravity theories admitting diffeomorphism
invariance. As an extension, such a method is
applied to the higher-order gravity theories
described by the Lagrangian (\ref{LagCovR}), which is the
functional of the metric, the Riemann tensor and the
covariant derivatives of the latter. We acquire the Noether
potential (\ref{NoePLagCov}) and the expression
(\ref{EoMhatL}) of field equations for such gravity theories,
together with the generalization of the
equality (\ref{AcalRrel}) given by Eq. (\ref{PRsymm}).
Particularly, we take into consideration of the
Lagrangians (\ref{Lag1cov}) and (\ref{Lag2cov}). The results
demonstrate that the field equations can be derived out of
the surface term with the variation operator replaced
by the Lie derivative along an arbitrary smooth vector
and such field equations can naturally avoid the appearance
for the derivative of the Lagrangian density with
respect to the metric. As a consequence, it is unnecessary
to specially compute this quantity. Besides, using the
conserved current to determine the equations of motion
renders them connected straightforwardly with the symmetry
of the Lagrangian.

In the present work, we only consider the pure gravity theories
in the traditional Riemann geometry setting. As is well known,
within such a geometry setting,
the affine connection is chosen to be the Levi-Civita connection,
which is torsion-free and metric-compatible and is completely
determined by the metric tensor. In addition, there also exist a series
of modified gravity theories in the framework of non-Riemann geometry,
such as the well-known Einstein-Cartan theory and the teleparallel
gravities. These modified theories of gravity treat both the metric
and the affine connection as independent dynamical variables.
In general, the affine connection involves three geometric quantities,
which are curvature, torsion and non-metricity.
Naturally, for the sake of testing the universality of the method
proposed in this paper, apart from the metric and curvature,
it is of great necessity to take into
account the contributions from the torsion and the non-metricity
together with various matter fields in the future work.

\section*{Acknowledgments}

This work was supported by the National Natural Science
Foundation of China under Grant Nos. 11865006. It was
also partially supported by the Guizhou Science and
Technology Department Fund under Grant Nos. [2018]5769.

%%%%%%%%%%%%%%%%%%%%%%%%%%%%%%%%%%%%%%%%%%%%%%%%%%%%%%%%%%%
\appendix
%%%%%%%%%%%%%%%%%%%%%%%%%%%%%%%%%%%%%%%%%%%%%%%%%%%%%%%%
\section{The derivation of Eq. (\ref{LiehatThe2}) and
the expressions for the tensors $W^{\mu\nu}$ and
$T^{\mu\alpha\beta}$}\label{appendA}
%%%%%%%%%%%%%%%%%%%%%%%%%%%%%%%%%%%%%%%%%%%%%%%%%%%%

Within Sec. \ref{three}, we have expressed the
surface term $\hat{\Theta}_{(2)}^\mu(\delta g)$
in Eq. (\ref{LagCvari}) in terms of
Eq. (\ref{The2genF2}). This will be demonstrated
below. For the sake of comparison, we begin with
an alternative form of $\hat{\Theta}_{(2)}^\mu(\delta g)$
presented in the work \cite{IyerWald}, that is,
\be
\hat{\Theta}_{(2)}^\mu(\delta g)
=H^{\mu\alpha\beta}\delta g_{\alpha\beta}
+\sum^{m-1}_{i=0} \Psi_{(i)}^{\mu\nu_1\cdot\cdot\cdot\nu_{i}
\alpha\beta\rho\sigma}
\delta N^{(i)}_{\nu_1\cdot\cdot\cdot\nu_{i}
\alpha\beta\rho\sigma}
\, ,\label{The2genF}
\ee
where $N^{(i)}_{\nu_1\cdot\cdot\cdot\nu_{i}
\alpha\beta\rho\sigma}=\nabla_{\nu_1}\cdot\cdot\cdot
\nabla_{\nu_{i}}R_{\alpha\beta\rho\sigma}$ for convenience
and $\Psi_{(i)}^{\mu\nu_1\cdot\cdot\cdot\nu_{i}
\alpha\beta\rho\sigma}=\Psi_{(i)}^{\mu(\nu_1\cdot\cdot\cdot\nu_{i})
\alpha\beta\rho\sigma}$. In particular, when $i=0$,
$\Psi_{(i)}^{\mu\nu_1\cdot\cdot\cdot\nu_{i}
\alpha\beta\rho\sigma}$ and
$N^{(i)}_{\nu_1\cdot\cdot\cdot\nu_{i}\alpha\beta\rho\sigma}$
are denoted by $\Psi_{(0)}^{\mu\alpha\beta\rho\sigma}$
and $N^{(0)}_{\alpha\beta\rho\sigma}
=R_{\alpha\beta\rho\sigma}$, respectively. To our knowledge,
the concrete expressions for the tensors
$H^{\mu\alpha\beta}$ and  $\Psi_{(i)}^{\mu\nu_1\cdot\cdot
\cdot\nu_{i}\alpha\beta\rho\sigma}$s are absent
in the literature. However, in what follows, we shall
achieve this by expressing them in terms of the tensors
$Z_{(i)}^{\alpha_1\cdot\cdot\cdot\alpha_i\mu\nu\rho\sigma}$s
together with their covariant derivatives.
For this purpose, we start with the Lagrangian
$\sqrt{-g}\hat{L}_{(2)}$ in Eq. (\ref{Lag2cov}),
whose surface term
$\hat{\Theta}_{(2)}^\mu(\delta g)$ takes the following value
\be
\hat{\Theta}_{(2)}^\mu(\delta g)\big|_{\hat{L}_{(2)}}
=\Big(H^{\mu\alpha\beta}_{(1)}
+H^{\mu\alpha\beta}_{(2)}\Big)
\delta g_{\alpha\beta}
+\Big(Z^{\mu\alpha\beta\rho\sigma}_{(1)}
-\nabla_\lambda Z^{\lambda\mu\alpha\beta\rho\sigma}_{(2)}
\Big)\delta R_{\alpha\beta\rho\sigma}
+Z^{\mu\lambda\alpha\beta\rho\sigma}_{(2)}
\delta\nabla_\lambda R_{\alpha\beta\rho\sigma}
\, . \label{The2forLcov2}
\ee
By contrast with Eq. (\ref{The2genF}), one obtains
\bea
&&H^{\mu\alpha\beta}\big|_{\hat{L}_{(2)}}
=H^{\mu\alpha\beta}_{(1)}
+H^{\mu\alpha\beta}_{(2)} \, , \nn \\
&&\Psi_{(0)}^{\mu\alpha\beta\rho\sigma}
=Z^{\mu\alpha\beta\rho\sigma}_{(1)}
-\nabla_\lambda Z^{\lambda\mu\alpha\beta\rho\sigma}_{(2)}
\, , \quad
\Psi_{(1)}^{\mu\lambda\alpha\beta\rho\sigma}
=Z^{\mu\lambda\alpha\beta\rho\sigma}_{(2)}
\, . \label{PsiHforLcov2}
\eea
Here the tensor $H^{\mu\alpha\beta}_{(1)}=H^{\mu\alpha\beta}_{(1)}
\big|_{\text{Eq. (}\ref{Htendef}\text{)}}
(\hat{L}_{(1)}\rightarrow\hat{L}_{(2)})$, the one
$H^{\mu\alpha\beta}_{(2)}$ is given by Eq. (\ref{H2def}),
and the tensor $Z^{\mu\alpha\beta\rho\sigma}_{(1)}
=Z^{\mu\alpha\beta\rho\sigma}_{(1)}
\big|_{\text{Eq. (}\ref{ZP1def}\text{)}}
(\hat{L}_{(1)}\rightarrow\hat{L}_{(2)})$. What is more,
as another example, for the Lagrangian density
depending on up to the third-order derivative of the
Riemann curvature tensor
\be
\hat{L}_{(3)}=\hat{L}_{(3)}\big(g^{\alpha\beta},
R_{\mu\nu\rho\sigma},\nabla_{\lambda}R_{\mu\nu\rho\sigma},
\nabla_{(\alpha}\nabla_{\beta)}R_{\mu\nu\rho\sigma},
\nabla_{(\lambda}\nabla_{\alpha}\nabla_{\beta)}
R_{\mu\nu\rho\sigma}\big)
\, , \label{Lag3cov}
\ee
a direct calculation for the variation of the Lagrangian
$\sqrt{-g}\hat{L}_{(3)}$ gives rise to that the
surface term $\hat{\Theta}_{(2)}^\mu(\delta g)$
corresponding to such a Lagrangian is read off as
\bea
\hat{\Theta}_{(2)}^\mu(\delta g)\big|_{\hat{L}_{(3)}}
&=&H^{\mu\alpha\beta}\big|_{\hat{L}_{(3)}}
\delta g_{\alpha\beta}
+\Big(Z^{\mu\alpha\beta\rho\sigma}_{(1)}
-\nabla_\lambda Z^{\lambda\mu\alpha\beta\rho\sigma}_{(2)}
+\nabla_\gamma\nabla_\lambda
Z^{\gamma\lambda\mu\alpha\beta\rho\sigma}_{(3)}\Big)
\delta R_{\alpha\beta\rho\sigma} \nn \\
&&+\Big(Z^{\mu\lambda\alpha\beta\rho\sigma}_{(2)}
-\nabla_\gamma
Z^{\gamma\mu\lambda\alpha\beta\rho\sigma}_{(3)}\Big)
\delta\nabla_\lambda R_{\alpha\beta\rho\sigma}
+Z^{\mu\lambda\gamma\alpha\beta\rho\sigma}_{(3)}
\delta\nabla_\lambda\nabla_\gamma R_{\alpha\beta\rho\sigma}
\, . \label{The2forLcov3}
\eea
Within Eq. (\ref{The2forLcov3}), both the tensors
$Z^{\mu\alpha\beta\rho\sigma}_{(1)}$ and
$Z^{\lambda\mu\alpha\beta\rho\sigma}_{(2)}$
are required to be defined in accordance with the Lagrangian density
$\hat{L}_{(3)}$, while the rank-3 tensor
$H^{\mu\alpha\beta}\big|_{\hat{L}_{(3)}}$ can be
decomposed into
\be
H^{\mu\alpha\beta}\big|_{\hat{L}_{(3)}}
=H^{\mu\alpha\beta}_{(1)}
\big|_{\hat{L}_{(1)}\rightarrow\hat{L}_{(3)}}
+H^{\mu\alpha\beta}_{(2)}
\big|_{\hat{L}_{(2)}\rightarrow\hat{L}_{(3)}}
+H^{\mu\alpha\beta}_{(3)}
\, , \label{HforLag3cov}
\ee
with the $H^{\lambda\alpha\beta}_{(3)}$ tensor defined by
\be
H^{\lambda\alpha\beta}_{(3)}=\frac{1}{2}
\Big(U^{\alpha\lambda\beta}_{(3)}
+U^{\beta\alpha\lambda}_{(3)}
-U^{\lambda\alpha\beta}_{(3)}\Big)
\, , \label{H3def}
\ee
in which
\bea
U^{\lambda\alpha\beta}_{(3)}&=&
\big(\nabla^\lambda R_{\gamma\nu\rho\sigma}\big)
\nabla_\mu Z^{\mu\alpha\beta\gamma\nu\rho\sigma}_{(3)}
-4R^\lambda_{~\nu\rho\sigma}\nabla_\mu
\nabla_\gamma Z^{\mu\gamma(\alpha\beta)\nu\rho\sigma}_{(3)}
\nn \\
&&+4\big(\nabla_\gamma R^\lambda_{~\nu\rho\sigma}\big)
\nabla_\mu Z^{\mu\gamma(\alpha\beta)\nu\rho\sigma}_{(3)}
-Z^{\alpha\beta\gamma\mu\nu\rho\sigma}_{(3)}
\nabla^\lambda\nabla_\gamma R_{\mu\nu\rho\sigma}
\nn \\
&&-Z^{\alpha\beta\gamma\mu\nu\rho\sigma}_{(3)}
\nabla_\gamma\nabla^\lambda R_{\mu\nu\rho\sigma}
-4Z^{\mu\gamma(\alpha\beta)\nu\rho\sigma}_{(3)}
\nabla_\mu\nabla_\gamma R^\lambda_{~\nu\rho\sigma}
\, . \label{U3def}
\eea
From Eq. (\ref{The2forLcov3}), we obtain
the tensors $\Psi_{(i)}^{\mu\nu_1\cdot\cdot\cdot\nu_{i}
\alpha\beta\rho\sigma}$s $(i=0,1,2)$, being of the form
\bea
&&\Psi_{(0)}^{\mu\alpha\beta\rho\sigma}
=Z^{\mu\alpha\beta\rho\sigma}_{(1)}
-\nabla_\lambda Z^{\lambda\mu\alpha\beta\rho\sigma}_{(2)}
+\nabla_\gamma\nabla_\lambda
Z^{\gamma\lambda\mu\alpha\beta\rho\sigma}_{(3)}
 \, , \nn \\
&&\Psi_{(1)}^{\mu\lambda\alpha\beta\rho\sigma}
=Z^{\mu\lambda\alpha\beta\rho\sigma}_{(2)}
-\nabla_\gamma
Z^{\gamma\mu\lambda\alpha\beta\rho\sigma}_{(3)}
\, , \nn \\
&&\Psi_{(2)}^{\mu\lambda\gamma\alpha\beta\rho\sigma}
=Z^{\mu\lambda\gamma\alpha\beta\rho\sigma}_{(3)}
\, . \label{PsiforLcov3}
\eea

Inspired with both the aforementioned examples, we go
on to take into consideration of the most general
Lagrangian (\ref{LagCovR}). In such a situation, for
convenience, we introduce a scalar
\be
\Upsilon_{(i,k)}=(-1)^{k-1}
\Big(\nabla_{\alpha_{1}}\cdot\cdot\cdot
\nabla_{\alpha_{k-1}}
Z_{(i)}^{\alpha_1\cdot\cdot\cdot\alpha_i\mu\nu\rho\sigma}
\Big) \delta \nabla_{\alpha_{k}}
\cdot\cdot\cdot\nabla_{\alpha_i}R_{\mu\nu\rho\sigma}
\, , \label{Upsilikdef}
\ee
where $k$ is allowed to run from $1$ up to $i+1$. For
instance, $\Upsilon_{(i,i+1)}=(-1)^{i}
\big(\nabla_{\alpha_{1}}\cdot\cdot\cdot
\nabla_{\alpha_{i}}
Z_{(i)}^{\alpha_1\cdot\cdot\cdot\alpha_i\mu\nu\rho\sigma}
\big) \delta R_{\mu\nu\rho\sigma}$. We
compute the term
$\Upsilon_{(i,1)}=
Z_{(i)}^{\alpha_1\cdot\cdot\cdot\alpha_i\mu\nu\rho\sigma}
\delta \nabla_{(\alpha_1}
\cdot\cdot\cdot\nabla_{\alpha_i)}R_{\mu\nu\rho\sigma}$
in Eq. (\ref{LagCvariA}), yielding
\bea
\Upsilon_{(i,1)}&=&
Z_{(i)}^{\alpha_1\cdot\cdot\cdot\alpha_i\mu\nu\rho\sigma}
\nabla_{\alpha_1} \delta\nabla_{\alpha_2}
\cdot\cdot\cdot\nabla_{\alpha_i}R_{\mu\nu\rho\sigma}
-4(\delta \Gamma^\lambda_{\alpha\beta})
Z_{(i)}^{\alpha_1\cdot\cdot\cdot\alpha_{i-1}
\alpha\beta\nu\rho\sigma}
\nabla_{\alpha_{1}}\cdot\cdot\cdot
\nabla_{\alpha_{i-1}}R_{\lambda\nu\rho\sigma}\nn\\
&&-\delta \Gamma^\lambda_{\alpha\beta}
\sum^{i}_{j=2}Z_{(i)}^{\alpha\alpha_2\cdot\cdot\cdot
\alpha_{j-1}\beta\alpha_{j+1}\cdot\cdot\cdot\alpha_{i}
\mu\nu\rho\sigma}
\nabla_{\alpha_{2}}\cdot\cdot\cdot
\nabla_{\alpha_{j-1}}\nabla_\lambda
\nabla_{\alpha_{j+1}}\cdot\cdot\cdot
\nabla_{\alpha_{i}}R_{\mu\nu\rho\sigma}
\, . \label{ZidelRiem0}
\eea
The first term in the right hand side of
Eq. (\ref{ZidelRiem0}) is expressed as
\be
Z_{(i)}^{\alpha_1\cdot\cdot\cdot\alpha_i\mu\nu\rho\sigma}
\nabla_{\alpha_1} \delta\nabla_{\alpha_2}
\cdot\cdot\cdot\nabla_{\alpha_i}R_{\mu\nu\rho\sigma}
=\nabla_{\mu} \Big(Z_{(i)}^{\mu\alpha_2\cdot\cdot\cdot
\alpha_i\tau\nu\rho\sigma}
\delta\nabla_{\alpha_2}
\cdot\cdot\cdot\nabla_{\alpha_i}R_{\tau\nu\rho\sigma}\Big)
+\Upsilon_{(i,2)}
\, . \label{ZidelRik}
\ee
According to Eqs. (\ref{ZidelRiem0}) and
(\ref{ZidelRik}), the scalar $\Upsilon_{(i,1)}$ is
related to the one $\Upsilon_{(i,2)}$ in the manner
\be
\Upsilon_{(i,1)}=\Upsilon_{(i,2)}
+\text{a divergence term }
\nabla_\mu (\bullet)^\mu
+\text{terms proportional to }
\delta \Gamma^\lambda_{\alpha\beta}
\, . \label{Upsili01}
\ee
As a matter of fact, repeating the calculations in
Eqs. (\ref{ZidelRiem0}) and (\ref{ZidelRik}) to
$\Upsilon_{(i,k)}$, we are able to figure out the
concrete relation between the scalars $\Upsilon_{(i,k)}$
and $\Upsilon_{(i,k+1)}$, which has the same structures
as the one given by Eq. (\ref{Upsili01}), read off as
\be
\Upsilon_{(i,k)}=
\Upsilon_{(i,k+1)}+\Upsilon^{\text{div}}_{(i,k)}
+\Upsilon^{\delta\Gamma}_{(i,k)} \, ,
\qquad  (k=1,\cdot\cdot\cdot,i)
\, . \label{Upsiijgen}
\ee
In Eq. (\ref{Upsiijgen}), the divergence term
$\Upsilon^{\text{div}}_{(i,k)}$ is given by
\be
\Upsilon^{\text{div}}_{(i,k)}=
\nabla_\mu\Big[(-1)^{k-1}\Big(\nabla_{\alpha_{1}}\cdot\cdot\cdot
\nabla_{\alpha_{k-1}}
Z_{(i)}^{\mu\alpha_1\cdot\cdot\cdot\alpha_{i-1}\tau\nu\rho\sigma}
\Big) \delta \nabla_{\alpha_{k}}
\cdot\cdot\cdot\nabla_{\alpha_{i-1}}R_{\tau\nu\rho\sigma}\Big]
\, , \label{Upsildiv}
\ee
while the scalar
$\Upsilon^{\delta\Gamma}_{(i,k)}$, standing for the
collection of all the terms proportional to
$\delta \Gamma^\lambda_{\alpha\beta}$, is presented by
\bea
\Upsilon^{\delta\Gamma}_{(i,k)}
&=&(-1)^{k}\left[\sum^{i}_{j=k+1}\Big(\nabla_{\alpha_{1}}
\cdot\cdot\cdot\nabla_{\alpha_{k-1}}
Z_{(i)}^{\alpha_{1}\cdot\cdot\cdot\alpha_{k-1}
\alpha\alpha_{k+1}\cdot\cdot\cdot
\alpha_{j-1}\beta\alpha_{j+1}\cdot\cdot\cdot\alpha_{i}
\mu\nu\rho\sigma}\Big) \right.\nn \\
&&\left.\times\nabla_{\alpha_{k+1}}\cdot\cdot\cdot
\nabla_{\alpha_{j-1}}\nabla_\lambda
\nabla_{\alpha_{j+1}}\cdot\cdot\cdot
\nabla_{\alpha_{i}}R_{\mu\nu\rho\sigma}\right. \nn \\
&&\left.+4\Big(\nabla_{\alpha_{1}}\cdot\cdot\cdot
\nabla_{\alpha_{k-1}}
Z_{(i)}^{\alpha_1\cdot\cdot\cdot\alpha_{i-1}
\alpha\beta\nu\rho\sigma}
\Big)\nabla_{\alpha_{k}}\cdot\cdot\cdot
\nabla_{\alpha_{i-1}}R_{\lambda\nu\rho\sigma}\right]
\delta \Gamma^\lambda_{\alpha\beta}
\, . \label{UpsilijdelG}
\eea
Summing Eq. (\ref{Upsiijgen}) with respect to $k$
from $1$ to $i$ and then eliminating
$\sum^i_{k=2}\Upsilon_{(i,k)}$ in each side of the
resulting equation, we ultimately arrive at the
following generic form
\be
\Upsilon_{(i,1)}=\Upsilon_{(i,i+1)}
+\sum^i_{k=1}\Big(\Upsilon^{\text{div}}_{(i,k)}
+\Upsilon^{\delta\Gamma}_{(i,k)}\Big)
\, . \label{Upsili1def}
\ee
One is able to check that both $\Upsilon_{(1,1)}$
and $\Upsilon_{(2,1)}$ coincide with
Eqs. (\ref{Z1delRiem}) and (\ref{Z2delCov2Riem}),
respectively. Moreover, according to Eq. (\ref{Upsili1def}),
one can write down
\bea
\Upsilon_{(i,1)}
&=&(-1)^i\Big(\nabla_{\alpha_1}\cdot\cdot\cdot
\nabla_{\alpha_i}
Z_{(i)}^{\alpha_1\cdot\cdot\cdot\alpha_i\mu\nu\rho\sigma}
\Big)\delta R_{\mu\nu\rho\sigma}
-\Big(\nabla_\mu H^{\mu\alpha\beta}_{(i)}\Big)
\delta g_{\alpha\beta}\nn \\
&&+\nabla_\mu \hat{\Theta}_{(i2)}^\mu(\delta g)
\, . \label{ZidelRiem}
\eea
Within Eq. (\ref{ZidelRiem}), the surface term
$\hat{\Theta}_{(i2)}^\mu(\delta g)$, given explicitly by
Eq. (\ref{Thei2genF}), is defined in terms of
\bea
\nabla_\mu \hat{\Theta}_{(i2)}^\mu(\delta g)
&=&\sum^i_{k=1}\Upsilon^{\text{div}}_{(i,k)}
+\nabla_\mu \Big(H^{\mu\alpha\beta}_{(i)}
\delta g_{\alpha\beta}\Big) \nn \\
&=&\nabla_\mu \left[\sum^i_{k=1}
(-1)^{k-1}\Big(\nabla_{\alpha_{1}}\cdot\cdot\cdot
\nabla_{\alpha_{k-1}}
Z_{(i)}^{\mu\alpha_1\cdot\cdot\cdot\alpha_{i-1}\tau\nu\rho\sigma}
\Big) \delta \nabla_{\alpha_{k}}
\cdot\cdot\cdot\nabla_{\alpha_{i-1}}R_{\tau\nu\rho\sigma}
\right.\nn \\
&&\left.+H^{\mu\alpha\beta}_{(i)}
\delta g_{\alpha\beta}\right]
\, , \label{Thei2gendef}
\eea
while the rank-3 tensor $H^{\mu\alpha\beta}_{(i)}$
in Eqs. (\ref{ZidelRiem}) and (\ref{Thei2gendef}),
fully coming from all the terms proportional to
$\delta \Gamma^\lambda_{\mu\nu}$,
is presented by Eq. (\ref{HigLdef}), or defined
equivalently through
\be
H^{\mu\alpha\beta}_{(i)}=\frac{1}{2}
\Big(U^{\alpha\mu\beta}_{(i)}
+U^{\beta\alpha\mu}_{(i)}
-U^{\mu\alpha\beta}_{(i)}\Big)
\, , \label{HigLcovdef}
\ee
with the rank-3 tensor
$U^{\mu\alpha\beta}_{(i)}=U^{\mu(\alpha\beta)}_{(i)}$
taking the form
\be
U^{\mu\alpha\beta}_{(i)}=\hat{U}^{\mu(\alpha\beta)}_{(i)}
+\tilde{U}^{\mu(\alpha\beta)}_{(i)}
\, . \label{genUidef}
\ee
In the above equation, $\hat{U}^{\mu\alpha\beta}_{(i)}$
and $\tilde{U}^{\mu\alpha\beta}_{(i)}=
\tilde{U}^{\mu\beta\alpha}_{(i)}$
are given by Eq. (\ref{Ui12def}). The tensor
$U^{\mu\alpha\beta}_{(i)}$ totally consists of $i(i+1)/2$
independent terms, each of which is proportional to
the rank-$(i+4)$ tensor $Z_{(i)}^{\lambda_1\cdot\cdot\cdot
\lambda_{i}\mu\nu\rho\sigma}$ or its covariant derivative, and
it covers the ones $U^{\mu\alpha\beta}_{(2)}$ and
$U^{\mu\alpha\beta}_{(3)}$, given by Eqs. (\ref{U2def})
and (\ref{U3def}), respectively.

As a consequence of Eq. (\ref{ZidelRiem}), the rank-3 tensor
$H^{\mu\alpha\beta}$ in Eq. (\ref{The2genF}) can be
defined in terms of the ones $H^{\mu\alpha\beta}_{(i)}$s,
that is,
\be
H^{\mu\alpha\beta}=\sum^m_{i=1}
H^{\mu\alpha\beta}_{(i)}
\, , \label{HgenLcovdef}
\ee
and the tensors
$\Psi_{(i)}^{\mu\nu_1\cdot\cdot\cdot\nu_{i}
\alpha\beta\rho\sigma}$s $(i=0,1,\cdot\cdot\cdot,m-1)$
are able to be expressed as the following
unified form
\be
\Psi_{(i)}^{\mu\nu_1\cdot\cdot\cdot\nu_{i}
\alpha\beta\rho\sigma}=
\sum^m_{j=i+1}(-1)^{j-i-1}\nabla_{\lambda_1}
\cdot\cdot\cdot\nabla_{\lambda_{j-i-1}}
Z_{(j)}^{\lambda_1\cdot\cdot\cdot\lambda_{j-i-1}
\mu\nu_1\cdot\cdot\cdot\nu_{i}
\alpha\beta\rho\sigma}
\, . \label{Psigendef}
\ee
Furthermore, in light of Eq. (\ref{ZidelRiem}),
the symmetric tensor $B^{\mu\nu}$ in the
expression (\ref{hatEmn1}) for field
equations is presented by
\be
B^{\mu\nu}=\nabla_\alpha H^{\alpha\mu\nu}
=\sum^m_{i=1}B^{\mu\nu}_{(i)}
=\sum^m_{i=1}\nabla_\alpha
H^{\alpha\mu\nu}_{(i)}
\, . \label{Bgendef}
\ee
By comparing both the forms for the
surface term $\hat{\Theta}_{(2)}^\mu(\delta g)$,
given by Eqs. (\ref{The2genF2}) and (\ref{The2genF}),
respectively, we point out that the former is more
convenient for analysis and calculations.

Next, we move on to take into consideration of the surface
term $\hat{\Theta}_{(2)}^\mu(\delta g)$ under the
diffeomorphism $x^\mu\rightarrow x^\mu-\zeta^\mu$.
In this situation, $\hat{\Theta}_{(2)}^\mu(\delta g)$
in Eq. (\ref{The2genF}) turns into
\bea
\hat{\Theta}_{(2)}^\mu(\delta\rightarrow\mathcal{L}_\zeta)
&=&2H^{\mu(\alpha\beta)}\nabla_\alpha\zeta_\beta
+\sum^{m-1}_{i=0} \Psi_{(i)}^{\mu\nu_1\cdot\cdot\cdot\nu_{i+4}}
\mathcal{L}_\zeta N^{(i)}_{\nu_1\cdot\cdot\cdot\nu_{i+4}} \nn \\
&=&2H^{\mu(\alpha\beta)}\nabla_\alpha\zeta_\beta
+\zeta_\nu\sum^{m-1}_{i=0}
\Psi_{(i)}^{\mu\nu_1\cdot\cdot\cdot\nu_{i+4}}
\nabla^\nu N^{(i)}_{\nu_1\cdot\cdot\cdot\nu_{i+4}} \nn \\
&&+\sum^{m-1}_{i=0} \sum^{i+4}_{j=1}
\Psi_{(i)}^{\mu\nu_1\cdot\cdot\cdot\nu_{i+4}}
N^{(i)}_{\nu_1\cdot\cdot\cdot\nu_{j-1}\lambda\nu_{j+1}
\cdot\cdot\cdot\nu_{i+4}}\nabla_{\nu_{j}}\zeta^\lambda
\, . \label{The2genFLD}
\eea
By the aid of the following tensor
\bea
T^{\mu\alpha\beta}&=&
2H^{\mu(\alpha\beta)}+g^{\lambda\beta}
\sum^{m-1}_{i=0} \sum^{i+4}_{j=1}
\Psi_{(i)}^{\mu\nu_1\cdot\cdot\cdot\nu_{j-1}\alpha\nu_{j+1}
\cdot\cdot\cdot\nu_{i+4}}
N^{(i)}_{\nu_1\cdot\cdot\cdot\nu_{j-1}\lambda\nu_{j+1}
\cdot\cdot\cdot\nu_{i+4}} \nn \\
&=&\sum^{m-1}_{i=1} \sum^{i}_{j=1}
\Psi_{(i)}^{\mu\nu_1\cdot\cdot\cdot\nu_{j-1}\alpha\nu_{j+1}
\cdot\cdot\cdot\nu_{i}\tau\nu\rho\sigma}
\nabla_{\nu_1}\cdot\cdot\cdot\nabla_{\nu_{j-1}}\nabla^\beta
\nabla_{\nu_{j+1}}\cdot\cdot\cdot
\nabla_{\nu_i}R_{\tau\nu\rho\sigma} \nn \\
&&+4\sum^{m-1}_{i=0}
\Psi_{(i)}^{\mu\nu_1\cdot\cdot\cdot\nu_{i}\alpha\nu\rho\sigma}
\nabla_{\nu_1}\cdot\cdot\cdot\nabla_{\nu_i}R^\beta_{~\nu\rho\sigma}
+2H^{\mu(\alpha\beta)}
\, , \label{ThiOTdef}
\eea
$\hat{\Theta}_{(2)}^\mu(\delta\rightarrow\mathcal{L}_\zeta)$ is recast into
\be
\hat{\Theta}_{(2)}^\mu(\delta\rightarrow\mathcal{L}_\zeta)=
2\zeta_\nu W^{\mu\nu}
+\nabla_\nu \big(\zeta_\lambda T^{\mu\nu\lambda}\big)
\, , \label{HatThgenL}
\ee
with the tensor $W^{\mu\nu}$ given by
\be
W^{\mu\nu}=\frac{1}{2}\sum^{m-1}_{i=0}
\Psi_{(i)}^{\mu\nu_1\cdot\cdot\cdot\nu_{i}\alpha\beta\rho\sigma}
\nabla^\nu \nabla_{\nu_1}\cdot\cdot\cdot\nabla_{\nu_i}
R_{\alpha\beta\rho\sigma}
-\frac{1}{2}\nabla_\lambda T^{\mu\lambda\nu}
\, . \label{Wgendef}
\ee
In what follows, we shall prove that
$T^{\mu\nu\lambda}=T^{[\mu\nu]\lambda}$. To perform
the proof, with the help of the decomposition
$\zeta_\lambda T^{\mu\nu\lambda}=
\zeta_\lambda T^{(\mu\nu)\lambda}
+\zeta_\lambda T^{[\mu\nu]\lambda}$,
the substitution of Eqs. (\ref{hatTheLie1}) and
(\ref{HatThgenL}) into Eq. (\ref{JErel}) yields
\bea
&&\big(\nabla_\mu\nabla_\nu\zeta_\lambda\big)T^{(\mu\nu)\lambda}
+\big(\nabla_\mu\zeta_\nu\big)
\Big(\Phi^{\mu\nu}+2\nabla_\lambda T^{(\lambda\mu)\nu}\Big)
\nn \\
&&+\zeta_\nu\Big(\nabla_\mu\Phi^{\mu\nu}
+\nabla_\mu\nabla_\lambda T^{(\lambda\mu)\nu}
-2\nabla_\mu \hat{E}^{\mu\nu}\Big)=0
\, , \label{EThetGen}
\eea
where $\Phi^{\mu\nu}$ is related to the expression
$\hat{E}^{\mu\nu}$ for field equations in the manner
\be
\Phi^{\mu\nu}=2\hat{E}^{\mu\nu}+g^{\mu\nu}\hat{L}
-2\hat{P}^{\mu\lambda\rho\sigma}
R^{\nu}_{~\lambda\rho\sigma}+4\nabla_{\rho}\nabla_{\sigma}
\hat{P}^{\rho\mu\nu\sigma}-2W^{\mu\nu}
\, . \label{Phidef}
\ee
The fact that Eq. (\ref{EThetGen}) holds for any vector
$\zeta_\nu$ gives rise to
\be
T^{(\mu\nu)\lambda}=0\, , \quad
\Phi^{\mu\nu}=0 \, , \quad
\nabla_\mu \hat{E}^{\mu\nu}=0
\, . \label{TPhCovE}
\ee
As a consequence, $T^{\mu\nu\lambda}=T^{[\mu\nu]\lambda}$,
supporting that
$\hat{\Theta}_{(2)}^\mu(\delta\rightarrow\mathcal{L}_\zeta)$ can be
expressed as Eq. (\ref{LiehatThe2}) with
$\hat{K}_{(2)}^{\mu\nu}=-\zeta_\lambda T^{\mu\nu\lambda}$.
Besides, $\Phi^{\mu\nu}=0$ results in the expression
(\ref{EoMhatL}) for field equations. The last identity
in Eq. (\ref{TPhCovE}) will be verified via straightforward
calculations on the divergence of $\hat{E}^{\mu\nu}$
in Appendix \ref{appendD}.

Although Eqs. (\ref{Rank3Tdef}) and (\ref{T2def}) support
that $T^{\mu\nu\lambda}=T^{[\mu\nu]\lambda}$, we shall
perform directly more calculations to confirm this
property for the third-rank tensor $T^{\mu\nu\lambda}$.
According to Eqs. (\ref{H3def}), (\ref{PsiforLcov3}),
and (\ref{ThiOTdef}), we start with computing
$T^{\mu\alpha\beta}$ for the Lagrangian
$\sqrt{-g}\hat{L}_{(3)}$ given by Eq. (\ref{Lag3cov}),
yielding
\be
T^{\mu\alpha\beta}=T^{[\mu\alpha]\beta}_{(1)}
+T^{[\mu\alpha]\beta}_{(2)}+T^{[\mu\alpha]\beta}_{(3)}
\, . \label{T3forLcov3}
\ee
Here $T^{\mu\alpha\beta}_{(1)}
=T^{\mu\alpha\beta}_{(1)}
\big|_{\text{Eq. }(\ref{Rank3Tdef})}
\big(\hat{L}_{(1)}\rightarrow\hat{L}_{(3)}\big)$ and
$T^{\mu\alpha\beta}_{(2)}
=T^{\mu\alpha\beta}_{(2)}
\big|_{\text{Eq. }(\ref{T2def})}
\big(\hat{L}_{(2)}\rightarrow\hat{L}_{(3)}\big)$, while
the tensor $T^{\mu\alpha\beta}_{(3)}$ takes the form
\bea
T^{\mu\alpha\beta}_{(3)}&=&
4R^\beta_{~\nu\rho\sigma}\nabla_\gamma
\nabla_\lambda Z^{\gamma\lambda[\mu\alpha]\nu\rho\sigma}_{(3)}
-4\big(\nabla_\gamma R^\beta_{~\nu\rho\sigma}\big)
\nabla_\lambda Z^{\gamma\lambda[\mu\alpha]\nu\rho\sigma}_{(3)}
\nn \\
&&+4Z^{\gamma\lambda[\mu\alpha]\nu\rho\sigma}_{(3)}
\nabla_\gamma\nabla_\lambda R^\beta_{~\nu\rho\sigma}
-2U^{[\mu\alpha]\beta}_{(3)}
\, . \label{T3def}
\eea
What is more, for the most general Lagrangian (\ref{LagCovR}),
the tensor $T^{\mu\alpha\beta}$ in Eq. (\ref{ThiOTdef})
can be decomposed as
$T^{\mu\alpha\beta}=T^{(\mu\alpha)\beta}
+T^{[\mu\alpha]\beta}$. In such a situation,
by the aid of
\bea
\sum^m_{i=1}\tilde{U}^{\beta\mu\alpha}_{(i)}&=&
-\sum^{m-1}_{i=1} \sum^{i}_{j=1}
\Psi_{(i)}^{\mu\nu_1\cdot\cdot\cdot\nu_{j-1}\alpha\nu_{j+1}
\cdot\cdot\cdot\nu_{i}\tau\nu\rho\sigma}
\nabla_{\nu_1}\cdot\cdot\cdot\nabla_{\nu_{j-1}}\nabla^\beta
\nabla_{\nu_{j+1}}\cdot\cdot\cdot
\nabla_{\nu_i}R_{\tau\nu\rho\sigma} \, , \nn \\
\sum^m_{i=1}\hat{U}^{\beta\mu\alpha}_{(i)}
&=&-4\sum^{m-1}_{i=0}
\Psi_{(i)}^{\nu_1\cdot\cdot\cdot\nu_{i}\mu\alpha\nu\rho\sigma}
\nabla_{\nu_1}\cdot\cdot\cdot\nabla_{\nu_i}R^\beta_{~\nu\rho\sigma}
\, , \label{Tcompident}
\eea
which can be obtained by computing the surface term
$\hat{\Theta}_{(2)}^\mu(\delta\rightarrow\mathcal{L}_\zeta)$
on the basis of Eqs. (\ref{The2genF2}) and (\ref{The2genF})
respectively and comparing the results, the calculations on the
symmetric component $T^{(\mu\alpha)\beta}$ bring forth
\be
T^{(\mu\alpha)\beta}=
-\sum^m_{i=1}\tilde{U}^{\beta(\mu\alpha)}_{(i)}
-\sum^m_{i=1}\hat{U}^{\beta(\mu\alpha)}_{(i)}
+\sum^m_{i=1}U^{\beta\mu\alpha}_{(i)}=0
\, , \label{Tsymid}
\ee
while the anti-symmetric part $T^{[\mu\alpha]\beta}$ takes the form
\bea
T^{[\mu\alpha]\beta}
&=&4\sum^{m-1}_{i=0}
\Psi_{(i)}^{\nu_1\cdot\cdot\cdot\nu_{i}[\mu\alpha]\nu\rho\sigma}
\nabla_{\nu_1}\cdot\cdot\cdot\nabla_{\nu_i}R^\beta_{~\nu\rho\sigma}
-2\sum^{m}_{i=1}U^{[\mu\alpha]\beta}_{(i)}
=T^{\mu\alpha\beta}
\, . \label{Tansymid}
\eea
Apparently, Eq. (\ref{Tsymid}) confirms that
$T^{(\mu\alpha)\beta}=0$ in a straightforward manner.
Apart from the expression (\ref{Tansymid}) for
the tensor $T^{\mu\alpha\beta}$, according to Eqs.
(\ref{HigLcovdef}), (\ref{ThiOTdef}), and (\ref{Tcompident}),
it can be alternatively expressed as
\be
T^{\mu\alpha\beta}=\sum^m_{i=1}
T^{\mu\alpha\beta}_{(i)}=\sum^m_{i=1}
\Big(2H^{\mu\alpha\beta}_{(i)}
-\hat{U}^{\beta\mu\alpha}_{(i)}
-\tilde{U}^{\beta\mu\alpha}_{(i)}\Big)
\, . \label{TforgenLag}
\ee
In terms of $\hat{U}^{\mu\alpha\beta}_{(i)}$
and $\tilde{U}^{\mu\alpha\beta}_{(i)}$
given by Eq. (\ref{Ui12def}), here the rank-3 tensor
$T^{\mu\alpha\beta}_{(i)}$ is given more explicitly by
\bea
T^{\mu\alpha\beta}_{(i)}&=&
-\hat{U}^{\beta[\mu\alpha]}_{(i)}
-2U^{[\mu\alpha]\beta}_{(i)}\nn \\
&=& -\hat{U}^{\beta[\mu\alpha]}_{(i)}
-\hat{U}^{[\mu|\beta|\alpha]}_{(i)}
-\hat{U}^{[\mu\alpha]\beta}_{(i)}
-\tilde{U}^{[\mu|\beta|\alpha]}_{(i)}
-\tilde{U}^{[\mu\alpha]\beta}_{(i)}\nn \\
&=&4\sum^i_{k=1}(-1)^{k-1}\Big(\nabla_{\lambda_1}
\cdot\cdot\cdot\nabla_{\lambda_{k-1}}
Z_{(i)}^{\lambda_1\cdot\cdot\cdot\lambda_{i-1}
[\mu\alpha]\nu\rho\sigma}\Big)
\nabla_{\lambda_{k}}\cdot\cdot\cdot\nabla_{\lambda_{i-1}}
R^\beta_{~\nu\rho\sigma}
-2U^{[\mu\alpha]\beta}_{(i)}
\, . \qquad \label{TigenLag}
\eea
Dividing $T^{\mu\alpha\beta}_{(i)}$ into the linear combination
of two parts
$T^{\mu[\alpha\beta]}_{(i)}$ and $T^{\mu(\alpha\beta)}_{(i)}$,
namely, $T^{\mu\alpha\beta}_{(i)}=T^{\mu[\alpha\beta]}_{(i)}
+T^{\mu(\alpha\beta)}_{(i)}$, one has
\bea
T^{\mu[\alpha\beta]}_{(i)}&=&
\hat{U}^{[\alpha|\mu|\beta]}_{(i)}
+\tilde{U}^{[\alpha|\mu|\beta]}_{(i)} \, , \nn \\
T^{\mu(\alpha\beta)}_{(i)}&=&
\Big(\hat{U}^{(\alpha\beta)\mu}_{(i)}
-\hat{U}^{\mu(\alpha\beta)}_{(i)}\Big)
+\Big(\tilde{U}^{(\alpha\beta)\mu}_{(i)}
-\tilde{U}^{\mu(\alpha\beta)}_{(i)}\Big)
\, . \label{Ti23decom}
\eea

By the aid of the decomposition to the tensor
$T^{\mu\alpha\beta}$ displayed by
Eq. (\ref{TforgenLag}), the rank-2 tensor
$W^{\mu\nu}$ given by Eq. (\ref{Wgendef}) can
be further rewritten as
\be
W^{\mu\nu}=\sum^m_{i=1}
W^{\mu\nu}_{(i)}
\, , \label{Wgendef2}
\ee
where $W^{\mu\nu}_{(i)}$ is defined through
\bea
W^{\mu\nu}_{(i)}&=&\frac{1}{2}
\sum^i_{k=1}(-1)^{k-1}\Big(\nabla_{\lambda_1}
\cdot\cdot\cdot\nabla_{\lambda_{k-1}}
Z_{(i)}^{\lambda_1\cdot\cdot\cdot\lambda_{i-1}
\mu\alpha\beta\rho\sigma}\Big)
\nabla^\nu\nabla_{\lambda_{k}}
\cdot\cdot\cdot\nabla_{\lambda_{i-1}}
R_{\alpha\beta\rho\sigma}
-\frac{1}{2}\nabla_\lambda
T^{\mu\lambda\nu}_{(i)}
\nn \\
&=&\frac{1}{2}
\sum^i_{k=1}(-1)^{k-1}\Big(\nabla_{\lambda_1}
\cdot\cdot\cdot\nabla_{\lambda_{k-1}}
Z_{(i)}^{\lambda_1\cdot\cdot\cdot\lambda_{i-1}
\mu\alpha\beta\rho\sigma}\Big)
\nabla^\nu\nabla_{\lambda_{k}}
\cdot\cdot\cdot\nabla_{\lambda_{i-1}}
R_{\alpha\beta\rho\sigma} \nn \\
&& -\frac{1}{2}\nabla_\lambda\hat{U}^{\nu[\lambda\mu]}_{(i)}
-\frac{1}{2}\nabla_\lambda\hat{U}^{[\lambda|\nu|\mu]}_{(i)}
-\frac{1}{2}\nabla_\lambda\hat{U}^{[\lambda\mu]\nu}_{(i)}
-\nabla_\lambda\tilde{U}^{[\lambda\mu]\nu}_{(i)}
\, . \label{Wmnidef}
\eea
For convenience to apply $W^{\mu\nu}$ in the calculations
of field equations, it is sometimes necessary to decompose
$W^{\mu\nu}_{(i)}$ into the following form
\be
W^{\mu\nu}_{(i)}=W^{(\mu\nu)}_{(i)}+W^{[\mu\nu]}_{(i)}
\, , \label{Widecom}
\ee
where the symmetric component $W^{(\mu\nu)}_{(i)}$
is given by
\bea
W^{(\mu\nu)}_{(i)}
&=&\frac{1}{2}\left[
\sum^i_{k=1}(-1)^{k-1}\Big(\nabla_{\lambda_1}
\cdot\cdot\cdot\nabla_{\lambda_{k-1}}
Z_{(i)}^{\lambda_1\cdot\cdot\cdot\lambda_{i-1}
(\mu|\alpha\beta\rho\sigma|}\Big)
\nabla^{\nu)}\nabla_{\lambda_{k}}
\cdot\cdot\cdot\nabla_{\lambda_{i-1}}
R_{\alpha\beta\rho\sigma} \right.\nn \\
&& \left.+\Big(\nabla_\lambda\hat{U}^{(\mu\nu)\lambda}_{(i)}
-\nabla_\lambda\hat{U}^{\lambda(\mu\nu)}_{(i)}\Big)
+\Big(\nabla_\lambda\tilde{U}^{(\mu\nu)\lambda}_{(i)}
-\nabla_\lambda\tilde{U}^{\lambda\mu\nu}_{(i)}\Big)\right]
\,  \label{SymWmnidef}
\eea
with the help of Eq. (\ref{Ti23decom}),
and the anti-symmetric tensor $W^{[\mu\nu]}_{(i)}$
is of the form
\bea
W^{[\mu\nu]}_{(i)}
&=&\frac{1}{2}\left[
\sum^i_{k=1}(-1)^{k-1}\Big(\nabla_{\lambda_1}
\cdot\cdot\cdot\nabla_{\lambda_{k-1}}
Z_{(i)}^{\lambda_1\cdot\cdot\cdot\lambda_{i-1}
[\mu|\alpha\beta\rho\sigma|}\Big)
\nabla^{\nu]}\nabla_{\lambda_{k}}
\cdot\cdot\cdot\nabla_{\lambda_{i-1}}
R_{\alpha\beta\rho\sigma} \right. \nn \\
&&\left.+\nabla_\lambda\hat{U}^{[\mu|\lambda|\nu]}_{(i)}
+\nabla_\lambda\tilde{U}^{[\mu\nu]\lambda}_{(i)}\right]
\, . \label{AnSymWmni}
\eea
On the basis of Eqs. (\ref{TforgenLag}) and (\ref{Wgendef2}),
the term
$\hat{\Theta}_{(i2)}^\mu(\delta\rightarrow\mathcal{L}_\zeta)$
$(i=1,2,\cdot\cdot\cdot,m)$ is read off as
\be
\hat{\Theta}_{(i2)}^\mu(\delta\rightarrow\mathcal{L}_\zeta)=
2\zeta_\nu W^{\mu\nu}_{(i)}
+\nabla_\nu \Big(\zeta_\lambda T^{\mu\nu\lambda}_{(i)}\Big)
\, . \label{Thei2Lie}
\ee

At last, it is worthwhile pointing out that all the
quantities $H^{\mu\alpha\beta}_{(i)}$,
$U^{\mu\alpha\beta}_{(i)}$, $B^{\mu\nu}_{(i)}$,
$T^{\mu\alpha\beta}_{(i)}$, $W^{\mu\nu}_{(i)}$,
and $\hat{\Theta}_{(i2)}^\mu
(\delta\rightarrow\mathcal{L}_\zeta)$
$(i=1,2,\cdot\cdot\cdot,m)$ in the aforementioned
context, given by Eqs. (\ref{HigLcovdef}),
(\ref{genUidef}), (\ref{Bgendef}), (\ref{TigenLag}),
(\ref{Wmnidef}), and (\ref{Thei2Lie}), respectively,
represent the ones corresponding to the variable
$\nabla_{(\alpha_1}\cdot\cdot\cdot
\nabla_{\alpha_i)}R_{\mu\nu\rho\sigma}$ in the general
Lagrangian (\ref{LagCovR}). All of them can be expressed
in terms of both the tensors
$\hat{U}^{\mu\alpha\beta}_{(i)}$ and
$\tilde{U}^{\mu\alpha\beta}_{(i)}$
presented by Eq. (\ref{Ui12def}). Due to this,
determining $\hat{U}^{\mu\alpha\beta}_{(i)}$ and
$\tilde{U}^{\mu\alpha\beta}_{(i)}$ plays a prominent
role in all the calculations.

%%%%%%%%%%%%%%%%%%%%%%%%%%%%%%%%%%%%%%%%%%%%%%%%%%%%%%%%
\section{Equations of motion and Noether potentials
for quadratic and cubic gravities}\label{appendB0}
%%%%%%%%%%%%%%%%%%%%%%%%%%%%%%%%%%%%%%%%%%%%%%%%%%%%

In the present appendix, as applications, we take into account
the derivation for the equations of motion and Noether potentials
associated to both the quadratic and cubic gravities that are
dependent on the metric together with the curvature terms
$R_{\mu\nu\rho\sigma}$, $R_{\mu\nu}$ and $R$. Both the two
types of higher-order gravities have garnered much attention.

As a beginning, we consider the situation for the $D$-dimensional
quadratic gravity described by the following Lagrangian
\cite{AKKLR,Sal18}
\be
\sqrt{-g}{L}_{\text{QD}}=\sqrt{-g}\left(\kappa{L}_{\text{EH}}
+{a}_{1}{L}_{\text{QD}(1)}+{a}_{2}{L}_{\text{QD}(2)}
+{a}_{3}{L}_{\text{QD}(3)}\right)
\, , \label{LagquadGr}
\ee
in which $\kappa$ and $({a}_{1},{a}_{2},{a}_{3})$ denote
generic real coefficients.
Besides, the well-known Einstein-Hilbert Lagrangian density
${L}_{\text{EH}}=R-2\Lambda$ and all the three quadratic
curvature terms ${L}_{\text{QD}(i)}$s $(i=1,2,3)$ are given by
\be
{L}_{\text{QD}(1)}=R^2 \, , \qquad
{L}_{\text{QD}(2)}=R^{\alpha\beta}R_{\alpha\beta}
\, , \qquad
{L}_{\text{QD}(3)}=R^{\mu\nu\rho\sigma}R_{\mu\nu\rho\sigma}
\, , \label{quadrcurts}
\ee
respectively. For the Lagrangian (\ref{LagquadGr}), the rank-4
$P^{\mu\nu\rho\sigma}$ tensor is defined as
\bea
P^{\mu\nu\rho\sigma}_{\text{QD}}&=&
\frac{\partial{L}_{\text{QD}}}{\partial{R}_{\mu\nu\rho\sigma}}
=\kappa\frac{\partial{L}_{\text{EH}}}{\partial{R}_{\mu\nu\rho\sigma}}
+\sum_{i=1}^{3}{a}_{i}
\frac{\partial{L}_{\text{QD}(i)}}{\partial{R}_{\mu\nu\rho\sigma}}
\nn \\
&=&\left(\kappa
+2{a}_{1}R\right){g}^{\mu[\rho}{g}^{\sigma]\nu}
+2{a}_{2}g^{\alpha\gamma}R^{\beta\lambda}
\Delta^{\mu\nu\rho\sigma}_{\alpha\beta\gamma\lambda}
+2{a}_{3}R^{\mu\nu\rho\sigma} \nn \\
&=&\left(\kappa
+2{a}_{1}R\right){g}^{\mu[\rho}{g}^{\sigma]\nu}
+2{a}_{2}g^{[\mu|[\rho}R^{\sigma]|\nu]}
+2{a}_{3}R^{\mu\nu\rho\sigma}
\, . \label{Pquadgrdef}
\eea
Here and in what follows, for convenience, we have
introduced the tensor
$\Delta^{\mu\nu\rho\sigma}_{\alpha\beta\gamma\lambda}$
which is of the form
\be
\Delta^{\mu\nu\rho\sigma}_{\alpha\beta\gamma\lambda}
=\frac{1}{8}\big(\delta^{\mu\nu}_{\alpha\beta}
\delta^{\rho\sigma}_{\gamma\lambda}
+\delta^{\rho\sigma}_{\alpha\beta}
\delta^{\mu\nu}_{\gamma\lambda}\big)
\, . \label{Del4index}
\ee
Within Eq. (\ref{Del4index}), the generalized
Kronecker delta symbol
$\delta^{\mu\nu}_{\alpha\beta}$ is presented
by $\delta^{\mu\nu}_{\alpha\beta}=
2\delta^{[\mu}_{\alpha}\delta^{\nu]}_{\beta}$.
As a consequence, the tensor
$\Delta^{\mu\nu\rho\sigma}_{\alpha\beta\gamma\lambda}$
has the algebraic symmetries
\be
\Delta^{\mu\nu\rho\sigma}_{\alpha\beta\gamma\lambda}
=\Delta^{[\mu\nu][\rho\sigma]}_{[\alpha\beta][\gamma\lambda]}
=\Delta^{[\rho\sigma][\mu\nu]}_{[\alpha\beta][\gamma\lambda]}
=\Delta^{[\mu\nu][\rho\sigma]}_{[\gamma\lambda][\alpha\beta]}
=\Delta^{[\rho\sigma][\mu\nu]}_{[\gamma\lambda][\alpha\beta]}
\, . \label{DelSymm}
\ee
Accordingly, the tensor $P^{\mu\nu\rho\sigma}_{\text{QD}}$
satisfies
$P^{\mu\nu\rho\sigma}_{\text{QD}}
=P^{[\mu\nu][\rho\sigma]}_{\text{QD}}
=P^{[\rho\sigma][\mu\nu]}_{\text{QD}}$. What is more,
it can be proved that this tensor also fulfills
\be
P^{\mu[\nu\rho\sigma]}_{\text{QD}}=0 \, ,\quad
\nabla_\sigma P^{\mu\nu\rho\sigma}_{\text{QD}}
= \frac{1}{2}(4a_1+a_2){g}^{\rho[\mu}
\nabla^{\nu]} R-(a_2+4a_3)\nabla^{[\mu}R^{\nu]\rho}
\, . \label{IdenPqd}
\ee
Replacing the tensor $P^{\mu\nu\rho\sigma}$ in Eq. (\ref{MotEq2})
with the one $P^{\mu\nu\rho\sigma}_{\text{QD}}$, we obtain the
field equations for the quadratic gravity
$E^{\mu\nu}_{\text{QD}}=0$, where the second-rank symmetric tensor
$E^{\mu\nu}_{\text{QD}}=E^{\mu\nu}\big|_{\text{Eq. }(\ref{MotEq2})}
\left(P\rightarrow{P}_{\text{QD}},L\rightarrow{L}_{\text{QD}}\right)$
has the form
\be
E^{\mu\nu}_{\text{QD}} =P^{\mu\lambda\rho\sigma}_{\text{QD}}
R^{\nu}_{~\lambda\rho\sigma}
-\frac{1}{2}g^{\mu\nu}L_{\text{QD}}
-2\nabla_\rho\nabla_\sigma P^{\rho\mu\nu\sigma}_{\text{QD}}
\, . \label{EoMforQuadGr}
\ee
Furthermore, by making use of Eq. (\ref{IdenPqd}), the field equations
can be expressed more concretely as
\bea
E^{\mu\nu}_{\text{QD}}&=&\kappa{R}^{\mu\nu}+2a_1RR^{\mu\nu}
+(2a_2+4a_3)R^{\mu\rho\nu\sigma}R_{\rho\sigma}
-4a_3R^{\mu\sigma}R_\sigma^\nu
+2a_3R^{\mu\lambda\rho\sigma}R^{\nu}_{~\lambda\rho\sigma}
 \nn  \\
&&-\frac{1}{2}g^{\mu\nu}L_{\text{QD}}
+\frac{1}{2}(4a_1+a_2)g^{\mu\nu}\Box R
+(a_2+4a_3)\Box R^{\mu\nu}
\nn  \\
&&-(2a_1+a_2+2a_3)\nabla^\mu\nabla^\nu R
=0\, ,  \label{EoMforQuadGr2}
\eea
which coincides with Eq. (2.16) in the work \cite{PWG23}
when $\kappa=1$ and $a_i=c_i$. In addition, by the aid
of Eq. (\ref{Kmnudef}), under the diffeomorphism invariance
generated by the vector field $\zeta^\mu$,
the Noether potential corresponding
to the Lagrangian (\ref{LagquadGr}) is read off as
\bea
K^{\mu\nu}_{\text{QD}}&=&2P^{\mu\nu\rho\sigma}_{\text{QD}}
\nabla_{\rho}\zeta_{\sigma}
+4\zeta_\rho\nabla_\sigma P^{\mu\nu\rho\sigma}_{\text{QD}} \nn \\
&=&2\left(\kappa
+2{a}_{1}R\right)\nabla^{[\mu}\zeta^{\nu]}
+2{a}_{2}R^{\sigma[\mu}\nabla_\sigma\zeta^{\nu]}
-2{a}_{2}R^{\sigma[\mu}\nabla^{\nu]}\zeta_\sigma\nn \\
&&+4{a}_{3}R^{\mu\nu\rho\sigma}
\nabla_{\rho}\zeta_{\sigma}
+(8a_1+2a_2)\zeta^{[\mu}
\nabla^{\nu]} R-4(a_2+4a_3)
\zeta_\sigma\nabla^{[\mu}R^{\nu]\sigma}
\, . \label{KmnQuadGr}
\eea

In the remainder of this appendix, we pay attention to
the field equations and the Noether potential for
the theory of cubic gravity in $D$-dimensions.
The Lagrangian for this theory has the generic form
\cite{CubgrBC,BCMV}
\be
\sqrt{-g}{L}_{\text{CB}}=\sqrt{-g}\left({L}_{\text{QD}}
+\sum_{i=1}^{8}{b}_{i}{L}_{\text{CB}(i)}\right)
\, , \label{LagCibGr}
\ee
where the eight scalars ${L}_{\text{CB}(i)}$s $(i=1,\cdot\cdot\cdot,8)$
are given respectively by
\bea
{L}_{\text{CB}(1)}&=&R_{\alpha\gamma\beta\lambda}
R^{\gamma\tau\lambda\omega}R_{\tau~\omega~}^{~\alpha~\beta}
 \, , \quad
{L}_{\text{CB}(2)}=R_{\alpha\beta\gamma\lambda}
R^{\gamma\lambda\tau\omega}R_{\tau\omega}^{~~~\alpha\beta}
\, , \quad
{L}_{\text{CB}(3)}=R_{\alpha\beta\gamma\lambda}
R^{\alpha\beta\gamma\tau}R^{\lambda}_{\tau} \, , \nn \\
{L}_{\text{CB}(4)}&=&R_{\alpha\beta\gamma\lambda}
R^{\alpha\beta\gamma\lambda}R \, , \quad
{L}_{\text{CB}(5)}=R_{\alpha\beta\gamma\lambda}
R^{\alpha\gamma}R^{\beta\lambda}  \, , \quad
{L}_{\text{CB}(6)}=R_{\alpha}^{\beta}
R^{\gamma}_{\beta}R_{\gamma}^{\alpha}  \, , \nn \\
{L}_{\text{CB}(7)}&=&R_{\alpha\beta}
R^{\alpha\beta}R \, , \quad
{L}_{\text{CB}(8)}=R^3
\, . \label{LCB1to8def}
\eea
For each scalar ${L}_{\text{CB}(i)}$, a fourth rank tensor
$P^{\mu\nu\rho\sigma}_{\text{CB}(i)}
=P^{[\mu\nu][\rho\sigma]}_{\text{CB}(i)}
=P^{[\rho\sigma][\mu\nu]}_{\text{CB}(i)}$ is defined through
\be
P^{\mu\nu\rho\sigma}_{\text{CB}(i)}=
\frac{\partial{L}_{\text{CB}(i)}}{\partial{R}_{\mu\nu\rho\sigma}}
\, . \label{PCBidef}
\ee
Specifically, all the tensors $P^{\mu\nu\rho\sigma}_{\text{CB}(i)}$s
$(i=1,\cdot\cdot\cdot,8)$ are presented respectively by
\bea
P^{\mu\nu\rho\sigma}_{\text{CB}(1)}&=&
3\Delta^{\mu\nu\rho\sigma}_{\alpha\gamma\beta\lambda}
R^{\gamma\tau\lambda\omega}R^{~\alpha~\beta~}_{\tau~\omega~}
=\frac{3}{2}\Delta^{\mu\nu\rho\sigma}_{\alpha\beta\gamma\lambda}\left(
2R^{\lambda\tau\beta\omega}R^{~\alpha~\gamma~}_{\tau~\omega~}
+R^{\beta\lambda\tau\omega}R_{\tau\omega}^{~~~\alpha\gamma}
\right)\, , \label{PCB1def} \\
P^{\mu\nu\rho\sigma}_{\text{CB}(2)}
&=&3\Delta^{\mu\nu\rho\sigma}_{\alpha\beta\gamma\lambda}
R^{\gamma\lambda\tau\omega}R_{\tau\omega}^{~~~\alpha\beta}
=3R^{\tau\omega\mu\nu}R_{\tau\omega}^{~~~\rho\sigma}
\, , \label{PCB2def} \\
P^{\mu\nu\rho\sigma}_{\text{CB}(3)}&=&
\Delta^{\mu\nu\rho\sigma}_{\alpha\beta\gamma\lambda}
\left(2R^{\alpha\beta\gamma\tau}R^{\lambda}_{\tau}
+g^{\lambda\eta}g^{\alpha\gamma}R^{\tau\omega\kappa\beta}
R_{\tau\omega\kappa\eta}\right)
\, , \label{PCB3def} \\
P^{\mu\nu\rho\sigma}_{\text{CB}(4)}&=&
\Delta^{\mu\nu\rho\sigma}_{\alpha\beta\gamma\lambda}
\left(2R^{\alpha\beta\gamma\lambda}R+g^{\alpha\gamma}g^{\beta\lambda}
R_{\kappa\eta\tau\omega}R^{\kappa\eta\tau\omega}\right)
\, , \label{PCB4def} \\
P^{\mu\nu\rho\sigma}_{\text{CB}(5)}&=&
R^{\mu[\rho}R^{\sigma]\nu}
+2R^{\tau\beta\omega\lambda}
R_{\tau\omega}g^{\alpha\gamma}
\Delta^{\mu\nu\rho\sigma}_{\alpha\beta\gamma\lambda}
\, , \label{PCB5def}\\
P^{\mu\nu\rho\sigma}_{\text{CB}(6)}&=&
3\Delta^{\mu\nu\rho\sigma}_{\kappa\alpha\eta\beta}g^{\kappa\eta}
R^{\beta\gamma}R_{\gamma}^{\alpha}
=3\Delta^{\mu\nu\rho\sigma}_{\alpha\beta\gamma\lambda}g^{\alpha\gamma}
R^{\lambda\tau}R_{\tau}^{\beta}\, ,\label{PCB6def} \\
P^{\mu\nu\rho\sigma}_{\text{CB}(7)}&=&
\Delta^{\mu\nu\rho\sigma}_{\alpha\beta\gamma\lambda}
g^{\alpha\gamma}\big(2RR^{\beta\lambda}
+g^{\beta\lambda}R^{\tau}_{\omega}R^{\omega}_{\tau}\big)
=2Rg^{[\mu|[\rho}R^{\sigma]|\nu]}
+R^{\tau}_{\omega}R^{\omega}_{\tau}g^{\mu[\rho}g^{\sigma]\nu}
\, , \label{PCB7def}\\
P^{\mu\nu\rho\sigma}_{\text{CB}(8)}&=&
3\Delta^{\mu\nu\rho\sigma}_{\alpha\beta\gamma\lambda}
g^{\alpha\gamma}g^{\beta\lambda}R^2
=3R^2g^{\mu[\rho}g^{\sigma]\nu}
\, . \label{PCB8def}
\eea
It is not difficult to prove that
\be
P^{\mu[\nu\rho\sigma]}_{\text{CB}(1)}
=-\frac{3}{4}R^{\alpha\beta[\mu\nu}
R^{\rho\sigma]}_{~~~\alpha\beta}
=-\frac{1}{4}P^{\mu[\nu\rho\sigma]}_{\text{CB}(2)}
\, , \label{BiaIdPCB12}
\ee
together with
$P^{\mu[\nu\rho\sigma]}_{\text{CB}(i)}=0$ $(i=3,\cdot\cdot\cdot,8)$,
and it can be also verified that
\be
P^{\mu\lambda\rho\sigma}_{\text{CB}(i)}R^{\nu}_{~\lambda\rho\sigma}
=P^{\nu\lambda\rho\sigma}_{\text{CB}(i)}R^{\mu}_{~\lambda\rho\sigma}
\qquad (i=1,\cdot\cdot\cdot,8)
\, . \label{PCBiRiemSym}
\ee
For convenience, in terms of all the tensors
$P^{\mu\nu\rho\sigma}_{\text{CB}(i)}$s,
we introduce the rank-4 tensor
\be
P^{\mu\nu\rho\sigma}_{\text{CB}}=
\sum^8_{i=1}{b}_iP^{\mu\nu\rho\sigma}_{\text{CB}(i)}
\, . \label{PCBdef}
\ee
Here $P^{\mu[\nu\rho\sigma]}_{\text{CB}}$ does not vanish
identically. In spite of this, the tensor
$P^{\mu\nu\rho\sigma}_{\text{CB}}$
can be alternatively defined as
$\tilde{P}^{\mu\nu\rho\sigma}_{\text{CB}}=
P^{\mu\nu\rho\sigma}_{\text{CB}}
-P^{\mu[\nu\rho\sigma]}_{\text{CB}}$. Thus, it is guaranteed that
$\tilde{P}^{\mu[\nu\rho\sigma]}_{\text{CB}}=0$.

With the help of $P^{\mu\nu\rho\sigma}_{\text{CB}}$ or
$\tilde{P}^{\mu\nu\rho\sigma}_{\text{CB}}$,
utilizing Eq. (\ref{MotEq2}), we obtain the field equations
for the cubic gravity $E^{\mu\nu}_{\text{CB}}=0$, where
the expression $E^{\mu\nu}_{\text{CB}}=E^{\nu\mu}_{\text{CB}}$
is of the form
\be
E^{\mu\nu}_{\text{CB}} =\bar{E}^{\mu\nu}_{\text{CB}}
+{E}^{\mu\nu}_{\text{QD}}
\, , \label{EoMforCubGr}
\ee
with $\bar{E}^{\mu\nu}_{\text{CB}}$ given by
\be
\bar{E}^{\mu\nu}_{\text{CB}} =
P^{\mu\lambda\rho\sigma}_{\text{CB}}
R^{\nu}_{~\lambda\rho\sigma}
-2\nabla_\rho\nabla_\sigma
P^{\rho\mu\nu\sigma}_{\text{CB}}
-\frac{1}{2}g^{\mu\nu}
\sum_{i=1}^{8}{b}_{i}{L}_{\text{CB}(i)}
\, . \label{EbarCBdef}
\ee
Apart from the field equations, on the basis of Eq. (\ref{Kmnudef}),
we obtain the Noether potential $K^{\mu\nu}_{\text{CB}}$
associated to the Lagrangian (\ref{LagCibGr}), expressed as
\bea
K^{\mu\nu}_{\text{CB}}&=&K^{\mu\nu}_{\text{QD}}+
2P^{\mu\nu\rho\sigma}_{\text{CB}}
\nabla_{\rho}\zeta_{\sigma}
+4\zeta_\rho\nabla_\sigma P^{\mu\nu\rho\sigma}_{\text{CB}}
+\frac{9}{2}\left({b}_1-4{b}_2\right)R^{\alpha\beta[\mu\nu}
R^{\rho\sigma]}_{~~~\alpha\beta}
\nabla_\rho\zeta_\sigma
\, , \label{KmnCubGr}
\eea
where the vector $\zeta^\mu$ represents the generator for
the diffeomorphism invariance of the gravity theory.

Within the framework of the cubic gravity theories, when
the dimension of the spacetime is set as $D=4$, there exists
the Lagrangian
\be
\sqrt{-g}\tilde{L}_{\text{CB}}=\sqrt{-g}
\left(\tilde{b}_{1}\tilde{L}_{\text{CB}(1)}
+\tilde{b}_{2}\tilde{L}_{\text{CB}(2)}
+\tilde{b}_{3}\tilde{L}_{\text{CB}(3)}
+\tilde{b}_{4}\tilde{L}_{\text{CB}(4)}
+\tilde{b}_{5}\tilde{L}_{\text{CB}(5)}\right)
\,  \label{TildLagCibGr}
\ee
that could be incorporated into these theories.
In the above equation,
all the quantities $\tilde{L}_{\text{CB}(i)}$
$(i=1,\cdot\cdot\cdot,5)$ are given by
\bea
\tilde{L}_{\text{CB}(1)}&=&\tilde{R}_{\alpha\gamma\beta\lambda}
R^{\gamma\tau\lambda\omega}R_{\tau~\omega~}^{~\alpha~\beta}
 \, , \quad
\tilde{L}_{\text{CB}(2)}=\tilde{R}_{\alpha\beta\gamma\lambda}
R^{\gamma\lambda\tau\omega}R_{\tau\omega}^{~~~\alpha\beta}
\, , \nn \\
\tilde{L}_{\text{CB}(3)}&=&\tilde{R}_{\alpha\beta\gamma\lambda}
R^{\alpha\beta\gamma\tau}R^{\lambda}_{\tau} \, , \quad
\tilde{L}_{\text{CB}(4)}=\tilde{R}_{\alpha\beta\gamma\lambda}
R^{\alpha\beta\gamma\lambda}R \, , \nn \\
\tilde{L}_{\text{CB}(5)}&=&\tilde{R}_{\alpha\beta\gamma\lambda}
R^{\alpha\gamma}R^{\beta\lambda}
\, , \label{LCB12tilde}
\eea
respectively, where
$\tilde{R}_{\mu\nu\rho\sigma}=\frac{1}{2}\epsilon_{\mu\nu\alpha\beta}
R^{\alpha\beta}_{~~~\rho\sigma}$ denotes the dual of the Riemann
curvature tensor $R_{\mu\nu\rho\sigma}$. Obviously,
$\tilde{R}_{\mu\nu\rho\sigma}=\tilde{R}_{[\mu\nu][\rho\sigma]}$.
It is worth to pointing
out that we obtain the quantity $\tilde{L}_{\text{CB}(i)}$
$(i=1,\cdot\cdot\cdot,5)$ by means of replacing one Riemann tensor
${R}_{\alpha\beta\gamma\lambda}$ in the scalar ${L}_{\text{CB}(i)}$
with the dual of the Riemann tensor $\tilde{R}_{\alpha\beta\gamma\lambda}$.
According to such a rule, there do not exist the quantities
$\tilde{L}_{\text{CB}(6)}$, $\tilde{L}_{\text{CB}(7)}$ and
$\tilde{L}_{\text{CB}(8)}$, corresponding to
${L}_{\text{CB}(6)}$, ${L}_{\text{CB}(7)}$ and
${L}_{\text{CB}(8)}$, respectively. This is
attributed to the fact that
$g^{\mu\rho}\tilde{R}_{\mu\nu\rho\sigma}
=g^{\nu\sigma}\tilde{R}_{\mu\nu\rho\sigma}=0$, leading to
$\tilde{L}_{\text{CB}(6)}=\tilde{L}_{\text{CB}(7)}
=\tilde{L}_{\text{CB}(8)}=0$.
Like before, we introduce the fourth-rank tensor
\be
\tilde{P}^{\mu\nu\rho\sigma}_{\text{CB}}=
\frac{\partial\tilde{L}_{\text{CB}}}{\partial{R}_{\mu\nu\rho\sigma}}
=\sum^{5}_{i=1}\tilde{b}_{i}\tilde{P}^{\mu\nu\rho\sigma}_{\text{CB}(i)}
\, , \label{TildPCBdef}
\ee
where the fourth-rank tensor
$\tilde{P}^{\mu\nu\rho\sigma}_{\text{CB}(1)}$
is of the form
\bea
\tilde{P}^{\mu\nu\rho\sigma}_{\text{CB}(1)}&=&\frac{1}{2}
\Delta^{\mu\nu\rho\sigma}_{\alpha\beta\gamma\lambda}
\left(g^{\gamma\varsigma}g_{\eta\vartheta}
\epsilon^{\kappa\eta\alpha\beta}
R_{\kappa\tau\varsigma\omega}
R^{\vartheta\tau\lambda\omega}
+4g^{\beta\kappa}g^{\lambda\eta}\tilde{R}^{\alpha\tau\gamma\omega}
{R}_{\kappa\tau\eta\omega}\right)
\, , \label{TildPCB1def}
\eea
the tensor $\tilde{P}^{\mu\nu\rho\sigma}_{\text{CB}(2)}$
is defined in terms of
\bea
\tilde{P}^{\mu\nu\rho\sigma}_{\text{CB}(2)}&=&
\Delta^{\mu\nu\rho\sigma}_{\alpha\beta\gamma\lambda}
\left(\tilde{R}^{\alpha\beta\tau\omega}
{R}^{\gamma\lambda}_{~~~\tau\omega}
+\tilde{R}^{\gamma\lambda\tau\omega}
{R}^{\alpha\beta}_{~~~\tau\omega}
+\tilde{R}^{\tau\omega\gamma\lambda}
{R}^{\alpha\beta}_{~~~\tau\omega}\right) \nn \\
&=&\tilde{R}^{\mu\nu\tau\omega}
{R}^{\rho\sigma}_{~~\tau\omega}
+\tilde{R}^{\rho\sigma\tau\omega}
{R}^{\mu\nu}_{~~\tau\omega}
+\frac{1}{2}\tilde{R}^{\tau\omega\mu\nu}
{R}^{\rho\sigma}_{~~\tau\omega}
+\frac{1}{2}\tilde{R}^{\tau\omega\rho\sigma}
{R}^{\mu\nu}_{~~\tau\omega}
\, , \label{TildPCB2def}
\eea
the tensor $\tilde{P}^{\mu\nu\rho\sigma}_{\text{CB}(3)}$
is given by
\be
\tilde{P}^{\mu\nu\rho\sigma}_{\text{CB}(3)}=
\Delta^{\mu\nu\rho\sigma}_{\alpha\beta\gamma\lambda}
\left(2\tilde{R}^{\alpha\beta\gamma\tau}R^{\lambda}_{\tau}
+g^{\alpha\gamma}g^{\beta\eta}R^{\tau\omega\kappa\lambda}
\tilde{R}_{\tau\omega\kappa\eta}\right)
\, , \label{TildPCB3def}
\ee
the tensor $\tilde{P}^{\mu\nu\rho\sigma}_{\text{CB}(4)}$
is expressed as
\be
\tilde{P}^{\mu\nu\rho\sigma}_{\text{CB}(4)}=
R\big(\tilde{R}^{\mu\nu\rho\sigma}
+\tilde{R}^{\rho\sigma\mu\nu}\big)
+g^{\mu[\rho}g^{\sigma]\nu}
\tilde{R}_{\kappa\eta\tau\omega}R^{\kappa\eta\tau\omega}
\, , \label{TildPCB4def}
\ee
and the tensor $\tilde{P}^{\mu\nu\rho\sigma}_{\text{CB}(5)}$
is presented by
\be
\tilde{P}^{\mu\nu\rho\sigma}_{\text{CB}(5)}=\frac{1}{2}
\Delta^{\mu\nu\rho\sigma}_{\alpha\beta\gamma\lambda}
\left(\epsilon^{\tau\omega\alpha\beta}
R^{\gamma}_{\tau}R^{\lambda}_{\omega}
+4g^{\alpha\gamma}\tilde{R}^{\tau\beta\omega\lambda}
R_{\tau\omega}\right)
\, . \label{TildPCB5def}
\ee
Calculations on $\tilde{P}^{\mu[\nu\rho\sigma]}_{\text{CB}(i)}$s
give rise to
\bea
\tilde{P}^{\mu[\nu\rho\sigma]}_{\text{CB}(1)}&=&
\frac{1}{16}\epsilon^{\mu\nu\rho\sigma}
\big(2R_{\alpha\beta}R^{\alpha\beta}
-R^{\alpha\beta\gamma\lambda}
R_{\alpha\beta\gamma\lambda}\big) 
=\frac{1}{16}\epsilon^{\mu\nu\rho\sigma}
\big(2L_{\text{QD}(2)}-L_{\text{QD}(3)}\big)
\, , \label{TildPCB1ansym} \\
\tilde{P}^{\mu[\nu\rho\sigma]}_{\text{CB}(2)}&=&
\frac{1}{12}\epsilon^{\mu\nu\rho\sigma}
\left(L_{\text{QD}(1)}-4L_{\text{QD}(2)}
+3L_{\text{QD}(3)}\right) \, , \label{TildPCB2ansym}\\
\tilde{P}^{\mu[\nu\rho\sigma]}_{\text{CB}(3)}&=&
\frac{1}{6}\epsilon^{\mu\nu\rho\sigma}
R_{\alpha\beta}R^{\alpha\beta}
=\frac{1}{6}\epsilon^{\mu\nu\rho\sigma}L_{\text{QD}(2)}
\, , \label{TildPCB3ansym}\\
\tilde{P}^{\mu[\nu\rho\sigma]}_{\text{CB}(4)}&=&
\frac{1}{6}\epsilon^{\mu\nu\rho\sigma}R^2 
=\frac{1}{6}\epsilon^{\mu\nu\rho\sigma}
L_{\text{QD}(1)}\, , \label{TildPCB4ansym}\\
\tilde{P}^{\mu[\nu\rho\sigma]}_{\text{CB}(5)}&=&
\frac{1}{24}\epsilon^{\mu\nu\rho\sigma}
\big(R^2-R_{\alpha\beta}R^{\alpha\beta}\big)
=\frac{1}{24}\epsilon^{\mu\nu\rho\sigma}
\left(L_{\text{QD}(1)}-L_{\text{QD}(2)}\right)
\, . \label{TildPCB5ansym}
\eea
By the aid of the tensor $\tilde{P}^{\mu\nu\rho\sigma}_{\text{CB}}$,
the expression for the field equations associated to the
Lagrangian (\ref{TildLagCibGr}) is read off as
\be
\tilde{E}^{\mu\nu}_{\text{CB}} =
\tilde{P}^{\mu\lambda\rho\sigma}_{\text{CB}}
R^{\nu}_{~\lambda\rho\sigma}
-2\nabla_\rho\nabla_\sigma
\tilde{P}^{\rho\mu\nu\sigma}_{\text{CB}}
-\frac{1}{2}g^{\mu\nu}
\sum^{5}_{i=1}\tilde{b}_{i}\tilde{L}_{\text{CB}(i)}
\, . \label{EtildCBdef}
\ee
Apart from this, the Noether potential
$\tilde{K}^{\mu\nu}_{\text{CB}}$ corresponding to the Lagrangian
(\ref{TildLagCibGr}) is written as
\be
\tilde{K}^{\mu\nu}_{\text{CB}}=
2\tilde{P}^{\mu\nu\rho\sigma}_{\text{CB}}
\nabla_{\rho}\zeta_{\sigma}
+4\zeta_\rho\nabla_\sigma\tilde{P}^{\mu\nu\rho\sigma}_{\text{CB}}
-6\tilde{P}^{\mu[\nu\rho\sigma]}_{\text{CB}}
\nabla_\rho\zeta_\sigma
\, . \label{KmnTildCubGr}
\ee

%%%%%%%%%%%%%%%%%%%%%%%%%%%%%%%%%%%%%%%%%%%%%%%%%%%%%%%%
\section{Field equations and Noether potentials
for the Lagrangian density $f\big(R^n, (\Box R)^m\big)$}
\label{appendB}
%%%%%%%%%%%%%%%%%%%%%%%%%%%%%%%%%%%%%%%%%%%%%%%%%%%%

In this appendix, as a more concrete example to
demonstrate the applications in computing the
field equations for higher-order derivative
gravities, we shall take into account the field equations
for the following Lagrangian density
\be
f=f\big(R^n, (\Box R)^m\big)
\, , \label{Lagf}
\ee
where the d'Alembertian operator
$\Box=\nabla^\mu \nabla_\mu$.
Since $R=g^{\mu[\rho}g^{\sigma]\nu}R_{\mu\nu\rho\sigma}$ and
$\Box R=g^{\alpha\beta}g^{\mu[\rho}g^{\sigma]\nu}
\nabla_{(\alpha}\nabla_{\beta)}R_{\mu\nu\rho\sigma}$,
all the results related to the Lagrangian (\ref{Lag2cov})
can be directly applied to the Lagrangian (\ref{Lagf}).
As a consequence, both the tensors
$Z^{\alpha\beta\mu\nu\rho\sigma}_{(2)}$ and
$\hat{P}^{\mu\nu\rho\sigma}_{(2)}$ are specific to
\be
Z^{\alpha\beta\mu\nu\rho\sigma}_f
=F_{[m]}g^{\alpha\beta}g^{\mu[\rho}g^{\sigma]\nu}\, , \quad
P^{\mu\nu\rho\sigma}_f=\big(f_{[n]}
+\Box F_{[m]}\big)g^{\mu[\rho}g^{\sigma]\nu}
\, , \label{Z2P2spf}
\ee
respectively, where
\be
f_{[n]}=nR^{n-1}\frac{\partial f}{\partial R^n} \, , \quad
F_{[m]}=m(\Box R)^{m-1}
\frac{\partial f}{\partial(\Box R)^{m}}
\, . \label{fnFm}
\ee
According to Eqs. (\ref{U2def}) and (\ref{T2def}), the
rank-3 tensors $U^{\mu\alpha\beta}_{(2)}$ and
$T^{\mu\alpha\beta}_{(2)}$ are given by
\bea
U^{\mu\alpha\beta}_f&=& 4R^{\mu(\alpha}
\nabla^{\beta)} F_{[m]}
-F_{[m]}g^{\alpha\beta}\nabla^{\mu} R
-4F_{[m]}\nabla^{(\alpha} R^{\beta)\mu}\, , \nn \\
T^{\mu\alpha\beta}_f&=&
-2F_{[m]}g^{\beta[\mu}\nabla^{\alpha]} R
\, , \label{UF2forf}
\eea
respectively, while the second-rank tensor
$W^{\mu\nu}_{(2)}$ is read off as
\be
W^{\mu\nu}_f=\frac{1}{2}
g^{\mu\nu}F_{[m]}\Box R
+\frac{1}{2} g^{\mu\nu}\big(\nabla_\lambda F_{[m]}\big)
\nabla^\lambda  R
-\big(\nabla^{(\mu} F_{[m]}\big)
\nabla^{\nu)}R
\, . \label{W2spicf}
\ee
Here $W^{[\mu\nu]}_f=0$, together with
$2P^{[\mu|\lambda\rho\sigma|}_{f}
R^{\nu]}_{~~\lambda\rho\sigma}=
-2\big(f_{[n]}+\Box F_{[m]}\big)R^{[\mu\nu]}=0$,
verifies the identity (\ref{PW2rel}).
Substituting $\hat{P}^{\mu\nu\rho\sigma}_{(2)}
=P^{\mu\nu\rho\sigma}_f$
and $W^{\mu\nu}_{(2)}=W^{\mu\nu}_f$
into $\hat{E}^{\mu\nu}_{(2)}$
given by Eq. (\ref{EP2cov}), one acquires the
expression for field equations corresponding
to the Lagrangian density (\ref{Lagf}),
being of the form
\bea
E^{\mu\nu}_{f}&=& f_{[n]} R^{\mu\nu}
-\nabla^\mu\nabla^\nu f_{[n]}+g^{\mu\nu}\Box f_{[n]}
-\frac{1}{2}fg^{\mu\nu}
+R^{\mu\nu}\Box F_{[m]}\nn \\
&& -\nabla^\mu\nabla^\nu \Box F_{[m]}
+g^{\mu\nu}\Box^2 F_{[m]}
-\big(\nabla^{(\mu} F_{[m]}\big)\nabla^{\nu)}R \nn \\
&&+\frac{1}{2}g^{\mu\nu}F_{[m]}\Box R
+\frac{1}{2} g^{\mu\nu}\big(\nabla_\lambda F_{[m]}\big)
\nabla^\lambda R
\, . \label{FieEqufLagf}
\eea
Particularly, when the variable $R^n$ in the
Lagrangian density (\ref{Lagf}) is absent and the
integer $m=1$, the expression (\ref{FieEqufLagf})
becomes the one for the field equations of the
Lagrangian $\sqrt{-g}f(\Box R)$. In such a case,
we point out that Eq. (35) in \cite{DLM16} omits
the term $-1/2f^\prime g^{ab}\Box R$
and the term $\nabla^a f^\prime \nabla^b R$ there
has to be corrected as
$\nabla^{(a} f^\prime \nabla^{b)} R$. In addition
to this, in the
situation where the Lagrangian density $f$ is
set up as the one $f_{n=1=m}=f(R,\Box R)$, the
corresponding field equations
$E_{f\mu\nu}(n=1=m)=0$ are identified with those
given by Eq. (4) in \cite{CRL19} without the
standard energy stress tensor $T_{\mu\nu}$,
as well as the field equations with $k=1$
in \cite{HJSch90}.
Apart from the above two cases, in the absence
of the variable $(\Box R)^m$,
the expression (\ref{FieEqufLagf}) with $n=1$
turns into the one for field equations of
the well-known $f(R)$ gravity. By the aid of
$2\nabla_\nu R^{\mu\nu}=\nabla^\mu R$,
$\nabla^\nu\nabla^\mu\nabla_\nu f_{[n]}=
\nabla^\mu\Box f_{[n]}+R^{\mu\nu} \nabla_\nu f_{[n]}$
and
$\nabla^\nu\nabla^\mu\nabla_\nu \Box F_{[m]}=
\nabla^\mu\Box^2 F_{[m]}+R^{\mu\nu}
\nabla_\nu \Box F_{[m]}$, the divergence of
$E^{\mu\nu}_{f}$ is given by
\be
\nabla_\nu E^{\mu\nu}_{f}=-\frac{1}{2}\nabla^\mu f
+\frac{1}{2}f_{[n]} \nabla^\mu R
+\frac{1}{2}F_{[m]}\nabla^\mu \Box R
\equiv 0, \label{DivEf}
\ee
confirming that $E^{\mu\nu}_{f}$ is indeed conserved.

As a concrete example to demonstrate the application of
Eq. (\ref{FieEqufLagf}), we take into account the
Lagrangian
\be
\sqrt{-g}f_{\text{(6th)}}=\sqrt{-g}R\Box R
\, . \label{fRBoxR}
\ee
For this Lagrangian, $f_{[1]}=\Box R$ and $F_{[1]}=R$.
Substituting them into Eq. (\ref{FieEqufLagf}) yeilds
the expression for the equations of motion
\be
E^{\mu\nu}_{\text{(6th)}}= 2R^{\mu\nu}\Box R
-2\nabla^\mu\nabla^\nu \Box R+2g^{\mu\nu}\Box^2 R
-\big(\nabla^{\mu} R\big)\nabla^{\nu}R
+\frac{1}{2} g^{\mu\nu}\big(\nabla_\sigma R\big)
\nabla^\sigma R
\, . \label{EomforRBR}
\ee
Besides, according to the decomposition
$f_{\text{(6th)}}=\nabla_\mu(R\nabla^\mu R)
-(\nabla_\mu R)(\nabla^\mu R)$, where the total
divergence term $\nabla_\mu(R\nabla^\mu R)$ is
non-dynamical, it is feasible to
compute the field equations in terms of the
Lagrangian
\be
\sqrt{-g}\tilde{f}_{\text{(6th)}}=
-\sqrt{-g}(\nabla_\mu R)\nabla^\mu R
\, . \label{Ldentildf}
\ee
Since $\sqrt{-g}\tilde{f}_{\text{(6th)}}$ is covered by the
Lagrangian (\ref{Lag1cov}), we directly adopt the
results related to the latter. Specifically, the tensors
$P^{\mu\nu\rho\sigma}_{(01)}$,
$Z^{\lambda\mu\nu\rho\sigma}_{(1)}$, and
$\hat{P}^{\mu\nu\rho\sigma}_{(1)}$ are read off as
\bea
\tilde{P}^{\mu\nu\rho\sigma}_{(01)}
&=&\frac{\partial \tilde{f}_{\text{(6th)}}}{\partial R_{\mu\nu\rho\sigma}}
=0, \quad
\tilde{Z}^{\lambda\mu\nu\rho\sigma}_{(1)}
=\frac{\partial \tilde{f}_{\text{(6th)}}}{\partial\nabla_\lambda
R_{\mu\nu\rho\sigma}}
=-2g^{\mu[\rho}g^{\sigma]\nu}\nabla^\lambda R
\, , \nn \\
\tilde{P}^{\mu\nu\rho\sigma}_{(1)}&=&
\tilde{P}^{\mu\nu\rho\sigma}_{(01)}
-\nabla_\lambda \tilde{Z}^{\lambda\mu\nu\rho\sigma}_{(1)}
=2g^{\mu[\rho}g^{\sigma]\nu}\Box R
\, , \label{ZP1tildf}
\eea
respectively. By the aid of Eq. (\ref{ZP1tildf}),
one obtains
\be
\tilde{P}^{\mu\lambda\rho\sigma}_{(1)}
R_{\nu\lambda\rho\sigma}-2\nabla_{\rho}\nabla_{\sigma}
\tilde{P}^{\rho\mu\nu\sigma}_{(1)}
=2R^{\mu\nu}\Box R
-2\nabla^\mu\nabla^\nu \Box R+2g^{\mu\nu}\Box^2 R
\, , \label{tildPcovPf}
\ee
together with the value of $W^{\mu\nu}_{(1)}$,
being of the form
\be
\tilde{W}^{\mu\nu}_{(1)}=
-\big(\nabla^{\mu} R\big)\nabla^{\nu}R
\, . \label{W1tildf}
\ee
According to Eqs. (\ref{tildPcovPf}) and (\ref{W1tildf}),
Eq. (\ref{PotenL1cov}) gives rise to the expression of
field equations
\be
\tilde{E}^{\mu\nu}_{\text{(6th)}}
=\tilde{P}^{\mu\alpha\rho\sigma}_{(1)}
R^\nu_{~\alpha\rho\sigma}-2\nabla_{\rho}\nabla_{\sigma}
\tilde{P}^{\rho\mu\nu\sigma}_{(1)}
+\tilde{W}^{\mu\nu}_{(1)}
-\frac{1}{2}g^{\mu\nu}\tilde{f}_{\text{(6th)}}
=E^{\mu\nu}_{\text{(6th)}}
\, . \label{EoMtidf6th}
\ee
This implies that treating the Lagrangian
$\sqrt{-g}R\Box R$ as one special case of
the Lagrangian (\ref{Lag1cov}) or
the one (\ref{Lag2cov}) gives rise to the
same field equations.

What is more, for the Lagrangian (\ref{Lagf}), the Noether
potential associated with a diffeomorphism generator
$\zeta^\mu$ is given by
\bea
K^{\mu\nu}_f&=&2f_{[n]}\nabla^{[\mu}\zeta^{\nu]}
+4\zeta^{[\mu}\nabla^{\nu]}f_{[n]}
+2\big(\Box F_{[m]}\big)\nabla^{[\mu}\zeta^{\nu]} \nn \\
&&+4\zeta^{[\mu}\nabla^{\nu]}\Box F_{[m]}
+2F_{[m]}\zeta^{[\mu}\nabla^{\nu]}R
\, , \label{NoPforf}
\eea
while the surface term $\Theta^\mu_f$ is read off as
\bea
\Theta^\mu_f&=&-2f_{[n]}\nabla^{[\mu}h^{\nu]}_\nu
-2h^{[\mu}_\nu \nabla^{\nu]}f_{[n]}
-2\big(\Box F_{[m]}\big)\nabla^{[\mu}h^{\nu]}_\nu
-2h^{[\mu}_\nu \nabla^{\nu]}\Box F_{[m]}
\nn \\
&&+F_{[m]} \nabla^{\mu} \delta R
-F_{[m]}h^{\mu\nu}\nabla_{\nu} R
-\big(\nabla^{\mu}F_{[m]}\big)\delta R
+\frac{1}{2}hF_{[m]}\nabla^{\mu} R
\, , \label{SurfThef}
\eea
where $h_{\mu\nu}=\delta g_{\mu\nu}$,
$h^{\mu\nu}=g^{\mu\rho}g^{\nu\sigma}h_{\rho\sigma}$,
$h=h^\nu_\nu=g^{\mu\nu}h_{\mu\nu}$, and
$\delta R=\nabla^\rho\nabla^\sigma h_{\rho\sigma}
-\Box h -h_{\rho\sigma} R^{\rho\sigma}$.
It can be verified that $\Theta^\mu_f$ under diffeomorphism
takes the form
\be
\Theta^\mu_f(\delta\rightarrow\mathcal{L}_\zeta)=
2\zeta_\nu \left(E^{\mu\nu}_{f}
+\frac{1}{2}f g^{\mu\nu}\right)-\nabla_\nu K^{\mu\nu}_f
\, . \label{ThetfLieder}
\ee

Finally, let us check Eq. (\ref{PLgW12}) to give a
further support on the expression $E^{\mu\nu}_{f}$
for field equations. According to
Eq. (\ref{H2def}), $H^{\lambda\mu\nu}_{(2)}$
turns into
\be
H^{\lambda\mu\nu}_f=2R^{\mu\nu}\nabla^\lambda F_{[m]}
+\frac{1}{2}F_{[m]}g^{\mu\nu}\nabla^\lambda R
-F_{[m]}g^{\lambda(\mu}\nabla^{\nu)} R
-2F_{[m]}\nabla^\lambda R^{\mu\nu}
\, . \label{H2forf}
\ee
Correspondingly, the tensor $B^{\mu\nu}_{(2)}$
takes the value
\be
B^{\mu\nu}_f=2R^{\mu\nu}\Box F_{[m]}
-F_{[m]}\nabla^{\mu}\nabla^\nu R
-2F_{[m]}\Box R^{\mu\nu}+W^{\mu\nu}_f
\, . \label{B2forf}
\ee
As a consequence of Eqs. (\ref{W2spicf}) and
(\ref{B2forf}), one has
\be
2P^{\mu\lambda\rho\sigma}_{f}
R^{\nu}_{~\lambda\rho\sigma}+W^{\mu\nu}_f-B^{\mu\nu}_f
=2f_{[n]}R^{\mu\nu}+F_{[m]}\nabla^{\mu}\nabla^\nu R
+2F_{[m]}\Box R^{\mu\nu}
\, . \label{P2W2B2forf}
\ee
On the other hand, under the condition that the
three tensors $g^{\mu\nu}$, $R_{\mu\nu\rho\sigma}$ and
$\nabla_{(\alpha}\nabla_{\beta)}R_{\mu\nu\rho\sigma}$
constitute all the independent variables of the Lagrangian
density $f$, a straightforward calculation
for $\partial f/\partial g^{\mu\nu}$ gives rise to
\bea
\frac{\partial f}{\partial g^{\mu\nu}}&=&
2f_{[n]}R_{\mu\nu}+F_{[m]}\nabla_{\mu}\nabla_\nu R
+2F_{[m]}\Box R_{\mu\nu} \nn \\
&=&2P_{f\mu\lambda\rho\sigma}
R_{\nu}^{~\lambda\rho\sigma}+W_{f\mu\nu}-B_{f\mu\nu}
\, . \label{PfPgforf}
\eea
The second equality in Eq. (\ref{PfPgforf})
verifies the result given by Eq. (\ref{PLgW12}).
This provides evidence to support that the
expression (\ref{FieEqufLagf}) for the equations
of motion holds true.

%%%%%%%%%%%%%%%%%%%%%%%%%%%%%%%%%%%%%%%%%%%%%%%%%%%%%%%%
\section{Two ways to define the
 $P^{\mu\nu\rho\sigma}$ tensor}\label{appendC}
%%%%%%%%%%%%%%%%%%%%%%%%%%%%%%%%%%%%%%%%%%%%%%%%%%%%

To see the algebraic symmetries for the tensor
$P^{\mu\nu\rho\sigma}$ defined in two different
ways, we consider a simple but general
example. In such a case, an arbitrary scalar $L$
is assumed to have the form
\be
L=T^{\mu\nu\rho\sigma}R_{\mu\nu\rho\sigma}
\, , \label{LforalbP}
\ee
in which the arbitrary rank-4 tensor $T^{\mu\nu\rho\sigma}$
can be treated as a quantity independent of the Riemann
curvature tensor $R_{\mu\nu\rho\sigma}$ when one
evaluates the derivative of the scalar $L$ with respect
to $R_{\mu\nu\rho\sigma}$. With the help of the
algebraic symmetries of the Riemann tensor, the scalar
$L$ can be rewritten as
\be
L=\frac{1}{2}\Big(T^{[\mu\nu][\rho\sigma]}
+T^{[\rho\sigma][\mu\nu]}\Big)R_{\mu\nu\rho\sigma}
\, . \label{LforalbP2}
\ee
Consequently, if the rank-4 tensor $P^{\mu\nu\rho\sigma}
=\partial L /\partial R_{\mu\nu\rho\sigma}$ is
merely required to possess the algebraic symmetries
given by Eq. (\ref{PindeSym}), Eq. (\ref{LforalbP2})
shows that the tensor $P^{\mu\nu\rho\sigma}$ can
be defined by
\be
P^{\mu\nu\rho\sigma}
=\frac{1}{2}\Big(T^{[\mu\nu][\rho\sigma]}
+T^{[\rho\sigma][\mu\nu]}\Big)
=T^{\alpha\beta\gamma\lambda}
\Delta^{\mu\nu\rho\sigma}_{\alpha\beta\gamma\lambda}
\, , \label{Pcase1}
\ee
with the tensor
$\Delta^{\mu\nu\rho\sigma}_{\alpha\beta\gamma\lambda}$
defined by Eq. (\ref{Del4index}). It can be proved that
\be
\delta^{\mu\nu\rho\sigma}_{\alpha\beta\gamma\lambda}=
24\Delta^{\mu[\nu\rho\sigma]}_{\alpha\beta\gamma\lambda}
=24\Delta^{[\mu\nu\rho\sigma]}_{\alpha\beta\gamma\lambda}
\, , \label{DeldelRel}
\ee
where $\delta^{\mu\nu\rho\sigma}_{\alpha\beta\gamma\lambda}
=4!\delta^{[\mu}_{\alpha}\delta^{\nu}_{\beta}
\delta^{\rho}_{\gamma}\delta^{\sigma]}_{\lambda}$.
On the other hand, by making use of
$\Delta^{\mu\nu\rho\sigma}_{\alpha\beta\gamma\lambda}
R_{\mu\nu\rho\sigma}
=R_{\alpha\beta\gamma\lambda}$ and
$\delta^{\mu\nu\rho\sigma}_{\alpha\beta\gamma\lambda}
R_{\mu\nu\rho\sigma}=
\delta^{\mu\nu\rho\sigma}_{\alpha\beta\gamma\lambda}
R_{\mu[\nu\rho\sigma]}=0$,
the scalar $L$ can be also expressed as
the following form
\be
L=\frac{1}{24}\big(
24\Delta^{\mu\nu\rho\sigma}_{\alpha\beta\gamma\lambda}
-k\delta^{\mu\nu\rho\sigma}_{\alpha\beta\gamma\lambda}\big)
T^{\alpha\beta\gamma\lambda}
R_{\mu\nu\rho\sigma}
\, , \label{LforalbP3}
\ee
where $k$ is an arbitrary constant or scalar function.
As a consequence of Eq. (\ref{LforalbP3}), in the
case where the tensor $P^{\mu\nu\rho\sigma}
=\partial L /\partial R_{\mu\nu\rho\sigma}$ is
imposed to have the algebraic symmetries given by
Eq. (\ref{PindeSym}) together with the one
$P^{\mu[\nu\rho\sigma]}=0$,
$P^{\mu\nu\rho\sigma}$ can be defined in another
manner, that is,
\bea
P^{\mu\nu\rho\sigma}
&=&\frac{1}{24}\big(
24\Delta^{\mu\nu\rho\sigma}_{\alpha\beta\gamma\lambda}
-\delta^{\mu\nu\rho\sigma}_{\alpha\beta\gamma\lambda}\big)
T^{\alpha\beta\gamma\lambda} \nn \\
&=&\frac{1}{2}T^{[\mu\nu][\rho\sigma]}
+\frac{1}{2}T^{[\rho\sigma][\mu\nu]}
-T^{[\mu\nu\rho\sigma]}
\, . \label{Pcase2}
\eea
Apparently, the tensor $P^{\mu\nu\rho\sigma}$
given by Eq. (\ref{Pcase1}) is much simpler than
the one presented by Eq. (\ref{Pcase2}).

As a matter of fact, we can even go further. Given
an arbitrary fourth-rank tensor $\Phi^{\mu\nu\rho\sigma}$,
under the transformation
\be
P^{\mu\nu\rho\sigma}\rightarrow P^{\mu\nu\rho\sigma}
+\Phi^{[\mu\nu\rho\sigma]}
\, , \label{TransP}
\ee
it can be proved that
both the surface term $\Theta^\mu(\delta g)$ and
the expression for field equations $E^{\mu\nu}$
presented by Eqs. (\ref{Boudterm}) and (\ref{MotEq2})
respectively, remain unchanged. Besides, although
the Noether potential behaves as
$K^{\mu\nu}\rightarrow K^{\mu\nu}
-4\nabla_\rho
\big(\zeta_\sigma\Phi^{[\mu\nu\rho\sigma]}\big)$,
bringing about an additional divergence term for
an arbitrary 3-form, the conserved current $J^\mu$
keeps unaltered. According to the above, under the
fundamental requirement that the rank-4 tensor
$P^{\mu\nu\rho\sigma}$ has the algebraic symmetries
$P^{\mu\nu\rho\sigma}=-P^{\nu\mu\rho\sigma}
=-P^{\mu\nu\sigma\rho}=P^{\rho\sigma\mu\nu}$, we have
great freedom to choose this tensor. That is to say,
in order to determine uniquely $P^{\mu\nu\rho\sigma}$,
extra constraints have to be introduced.

In parallel, for the Lagrangian (\ref{LagCovR}) including
the covariant derivatives of the Riemann curvature tensor,
under the transformation
\be
\hat{P}^{\mu\nu\rho\sigma}\rightarrow \hat{P}^{\mu\nu\rho\sigma}
+\Phi^{[\mu\nu\rho\sigma]}
\, , \label{TranshatP}
\ee
or the one
\be
Z_{(i)}^{\lambda_1\cdot\cdot\cdot\lambda_{i}\mu\nu\rho\sigma}
\rightarrow
Z_{(i)}^{\lambda_1\cdot\cdot\cdot\lambda_{i}\mu\nu\rho\sigma}
+\Omega_{(i+4)}^{\lambda_1\cdot\cdot\cdot\lambda_{i}[\mu\nu\rho\sigma]}
 \qquad (i=1,\cdot\cdot\cdot,m)
\, , \label{TransZi}
\ee
where the rank-$(i+4)$ tensor
$\Omega_{(i+4)}^{\lambda_1\cdot\cdot\cdot\lambda_{i}\mu\nu\rho\sigma}$s
are allowed to be arbitrary, the field equations, the surface term
and the conserved currents still remain unchanged. This is attributed
to the fact that $R^\mu_{~[\nu\rho\sigma]}=0$ and
$\nabla_\rho\nabla_\sigma\Phi^{[\rho\mu\nu\sigma]}=0$.

%%%%%%%%%%%%%%%%%%%%%%%%%%%%%%%%%%%%%%%%%%%%%%%%%%%%%%%%
\section{Another proof for $\nabla_\mu
\hat{E}^{\mu\nu}=0$}
\label{appendD}
%%%%%%%%%%%%%%%%%%%%%%%%%%%%%%%%%%%%%%%%%%%%%%%%%%%%

In this appendix, we straightforwardly prove that the expression
$\hat{E}^{\mu\nu}$ for equations of motion is
divergence-free, namely, $\nabla_\mu\hat{E}^{\mu\nu}=0$.
The computation on the divergence of $\hat{E}^{\mu\nu}$
given by Eq. (\ref{EoMhatL}) yields
\bea
\nabla_\mu\hat{E}^{\mu\nu}
&=&R^{\nu}_{~\lambda\rho\sigma}
\nabla_\mu\hat{P}^{\mu\lambda\rho\sigma}
+\hat{P}^{\mu\lambda\rho\sigma}\nabla_\mu
{R}^{\nu}_{~\lambda\rho\sigma}
+2\nabla_{[\rho}\nabla_{\mu]}
\nabla_\sigma\hat{P}^{\rho\mu\nu\sigma} \nn \\
&&+\sum_{i=1}^m\nabla_\mu W^{\mu\nu}_{(i)}
-\frac{1}{2}\nabla^\nu\hat{L}
\, . \label{DivEoMhatL}
\eea
Furthermore, substituting the divergence for the Lagrangian
density
\be
\nabla^\nu\hat{L}=
\frac{\partial\hat{L}}{\partial R_{\alpha\beta\rho\sigma}}
\nabla^{\nu}R_{\alpha\beta\rho\sigma}
+\sum^m_{i=1}
Z_{(i)}^{\lambda_1\cdot\cdot\cdot\lambda_{i}
\alpha\beta\rho\sigma}\nabla^\nu
\nabla_{\lambda_1}
\cdot\cdot\cdot\nabla_{\lambda_{i}}
R_{\alpha\beta\rho\sigma}
\, , \label{DelnuLRiem}
\ee
together with the identity
\bea
\hat{P}^{\mu\lambda\rho\sigma}\nabla_\mu
{R}^{\nu}_{~\lambda\rho\sigma}
&=&\frac{1}{2}\sum^m_{i=1}
(-1)^{i}\Big(\nabla_{\lambda_1}
\cdot\cdot\cdot\nabla_{\lambda_{i}}
Z_{(i)}^{\lambda_1\cdot\cdot\cdot\lambda_{i}
\alpha\beta\rho\sigma}\Big)
\nabla^\nu R_{\alpha\beta\rho\sigma} \nn \\
&&+\frac{1}{2}
\frac{\partial\hat{L}}{\partial R_{\alpha\beta\rho\sigma}}
\nabla^{\nu}R_{\alpha\beta\rho\sigma}
\,  \label{IdePRiem}
\eea
and the one
\bea
\nabla_{[\rho}\nabla_{\mu]}
\nabla_\sigma\hat{P}^{\rho\mu\nu\sigma}&=&
-\frac{1}{2}R^{\nu}_{~\lambda\rho\sigma}
\nabla_\mu\hat{P}^{\mu\lambda\rho\sigma}
\, , \label{IdePRiem2}
\eea
into Eq. (\ref{DivEoMhatL}), we have
\bea
\nabla_\mu\hat{E}^{\mu\nu}
&=&\sum_{i=1}^m\nabla_\mu W^{\mu\nu}_{(i)}
-\frac{1}{2}\sum^m_{i=1}
Z_{(i)}^{\lambda_1\cdot\cdot\cdot\lambda_{i}
\alpha\beta\rho\sigma}\nabla^\nu
\nabla_{\lambda_1}
\cdot\cdot\cdot\nabla_{\lambda_{i}}
R_{\alpha\beta\rho\sigma} \nn \\
&&+\frac{1}{2}\sum^m_{i=1}
(-1)^{i}\Big(\nabla_{\lambda_1}
\cdot\cdot\cdot\nabla_{\lambda_{i}}
Z_{(i)}^{\lambda_1\cdot\cdot\cdot\lambda_{i}
\alpha\beta\rho\sigma}\Big)
\nabla^\nu R_{\alpha\beta\rho\sigma}
\, . \label{DivEoMhatL2}
\eea

On the basis of Eqs. (\ref{Widecom}), (\ref{SymWmnidef})
and (\ref{AnSymWmni}), we compute the divergence of
the rank-2 tensor $W^{\mu\nu}_{(i)}$, giving rise to
\bea
\nabla_\mu W^{\mu\nu}_{(i)}&=&
\frac{1}{2}\nabla_\mu\nabla_\lambda
\left(\hat{U}^{(\mu\nu)\lambda}_{(i)}
-\hat{U}^{\lambda(\mu\nu)}_{(i)}
+\hat{U}^{[\mu|\lambda|\nu]}_{(i)}
+\tilde{U}^{(\mu\nu)\lambda}_{(i)}
-\tilde{U}^{\lambda(\mu\nu)}_{(i)}
+\tilde{U}^{[\mu|\lambda|\nu]}_{(i)}\right) \nn\\
&&+\frac{1}{2}\sum_{k=1}^i(-1)^{k-1}
\left(\nabla_{\lambda_1}
\cdot\cdot\cdot\nabla_{\lambda_{k-1}}
Z_{(i)}^{\lambda_1\cdot\cdot\cdot\lambda_{i}
\alpha\beta\rho\sigma}\right)\nabla^\nu
\nabla_{\lambda_k}
\cdot\cdot\cdot\nabla_{\lambda_{i}}
R_{\alpha\beta\rho\sigma} \nn \\
&&-\frac{1}{2}\sum_{k=1}^i(-1)^{k}
\left(\nabla_{\lambda_1}
\cdot\cdot\cdot\nabla_{\lambda_{k}}
Z_{(i)}^{\lambda_1\cdot\cdot\cdot\lambda_{i}
\alpha\beta\rho\sigma}\right)\nabla^\nu
\nabla_{\lambda_{k+1}}
\cdot\cdot\cdot\nabla_{\lambda_{i}}
R_{\alpha\beta\rho\sigma} \nn \\
&&-\frac{1}{2}R^\nu_{~\lambda\rho\sigma}
\Big(\hat{U}^{\rho\lambda\sigma}_{(i)}
+\tilde{U}^{\rho\lambda\sigma}_{(i)}\Big)
\, . \label{DivWimn0}
\eea
By making use of the following identity
\bea
&&\nabla_\mu\nabla_\lambda
\left(\hat{U}^{(\mu\nu)\lambda}_{(i)}
-\hat{U}^{\lambda(\mu\nu)}_{(i)}
+\hat{U}^{[\mu|\lambda|\nu]}_{(i)}
+\tilde{U}^{(\mu\nu)\lambda}_{(i)}
-\tilde{U}^{\lambda(\mu\nu)}_{(i)}
+\tilde{U}^{[\mu|\lambda|\nu]}_{(i)}\right) \nn \\
&&=\frac{1}{2}\big(R^\nu_{~\sigma\rho\lambda}
+R^\nu_{~\lambda\rho\sigma}
+R^\nu_{~\rho\lambda\sigma}\big)
\Big(\hat{U}^{\rho\lambda\sigma}_{(i)}
+\tilde{U}^{\rho\lambda\sigma}_{(i)}\Big) \nn \\
&&=\frac{1}{2}\big(3R^\nu_{~[\sigma\rho\lambda]}
+2R^\nu_{~\lambda\rho\sigma}\big)
\Big(\hat{U}^{\rho\lambda\sigma}_{(i)}
+\tilde{U}^{\rho\lambda\sigma}_{(i)}\Big) \nn \\
&&=R^\nu_{~\lambda\rho\sigma}
\Big(\hat{U}^{\rho\lambda\sigma}_{(i)}
+\tilde{U}^{\rho\lambda\sigma}_{(i)}\Big)
\, , \label{Div2Ui}
\eea
equation (\ref{DivWimn0}) is simplified as
\be
\nabla_\mu W^{\mu\nu}_{(i)}=
\frac{1}{2}Z_{(i)}^{\lambda_1\cdot\cdot\cdot\lambda_{i}
\alpha\beta\rho\sigma}\nabla^\nu
\nabla_{\lambda_1}
\cdot\cdot\cdot\nabla_{\lambda_{i}}
R_{\alpha\beta\rho\sigma}
-\frac{1}{2}(-1)^{i}\Big(\nabla_{\lambda_1}
\cdot\cdot\cdot\nabla_{\lambda_{i}}
Z_{(i)}^{\lambda_1\cdot\cdot\cdot\lambda_{i}
\alpha\beta\rho\sigma}\Big)
\nabla^\nu R_{\alpha\beta\rho\sigma}
\, . \label{DivWimn}
\ee
Ultimately, substituting Eq. (\ref{DivWimn}) into
Eq. (\ref{DivEoMhatL2}), we arrive at the
generalized Bianchi identity
\be
\nabla_\mu\hat{E}^{\mu\nu}\equiv0
\, . \label{DivEoMhatL3}
\ee
This apparently supports that $\hat{E}^{\mu\nu}$ is
indeed divergence-free via a direct calculation
on the divergence for the expression of the field
equations.


\begin{thebibliography}{100}
%\newcommand{\DOI}[1]{doi:\href{https://doi.org/#1}{#1}}

\bibitem{Pady}
T. Padmanabhan,
Some aspects of field equations in generalised theories of gravity,
Phys. Rev. D \textbf{84}, 124041 (2011).
%arXiv:1109.3846 [gr-qc]
%DOI: 10.1103/PhysRevD.84.124041

\bibitem{PWG23}
J.J. Peng, Y. Wang and W.J. Guo,
Conserved quantities for asymptotically AdS spacetimes in quadratic
curvature gravity in terms of a rank-4 tensor,
Phys. Rev. D \textbf{108}, 104035 (2023).
%arXiv:2305.12611 [gr-qc].
%https://doi.org/10.1103/PhysRevD.108.104035

\bibitem{TPad10}
T. Padmanabhan,
Thermodynamical aspects of gravity: new insights,
Rept. Prog. Phys. \textbf{73}, 046901 (2010).
%arXiv:0911.5004 [gr-qc]
%DOI: 10.1088/0034-4885/73/4/046901

%%%%%%%%%%%%EGB review%%%%%%%%%%%%%%%%
\bibitem{RievLL} 	
T. Padmanabhan and D. Kothawala,
Lanczos-Lovelock models of gravity,
Phys. Rept. \textbf{531}, 115 (2013).
%arXiv:1302.2151 [gr-qc]
%DOI: 10.1016/j.physrep.2013.05.007

\bibitem{KolBid}
S. Kolekar,
On the Bianchi identity in generalized theories of gravity,
Gen. Rel. Grav. \textbf{54}, 92 (2022).
%DOI: 10.1007/s10714-022-02978-5

\bibitem{IyerWald}
V. Iyer and R.M. Wald,
Some properties of the Noether charge and a proposal for
dynamical black hole entropy,
Phys. Rev. D \textbf{50}, 846 (1994).
%[arXiv:gr-qc/9403028].
%gr-qc/9403028
%DOI: 10.1103/PhysRevD.50.846

\bibitem{DLM16}
R. Dey, S. Liberati and A. Mohd,
Higher derivative gravity: field equation as the equation of state,
Phys. Rev. D \textbf{94}, 044013 (2016).
%arXiv:1605.04789 [gr-qc]
%DOI: 10.1103/PhysRevD.94.044013

\bibitem{ERR22}
J.D. Edelstein, A.S. Rivadulla and D.V. Rodr\'{i}guez,
Are there Einsteinian gravities involving covariant
derivatives of the Riemann tensor?,
J. High Energy Phys. \textbf{11}, 077 (2022).
%arXiv:2204.13567 [hep-th]

\bibitem{CRL19}
S. Carloni, J.L. Rosa and J.P.S. Lemos,
Cosmology of $f(R,\Box R)$ gravity,
Phys. Rev. D \textbf{99}, 104001 (2019).
%arXiv:1808.07316 [gr-qc]

\bibitem{HJSch90}
H.J. Schmidt,
Variational derivatives of arbitrarily high order
and multiinflation cosmological models,
Classical Quantum Gravity \textbf{7}, 1023 (1990).
%DOI: 10.1088/0264-9381/7/6/011

\bibitem{AKKLR}
L. Alvarez-Gaume, A. Kehagias, C. Kounnas, D. L\"{u}st,
and A. Riotto, Aspects of Quadratic Gravity,
Fortsch. Phys. \textbf{64}, 176 (2016).
%arXiv:1505.07657 [hep-th]

\bibitem{Sal18}
A. Salvio,
Quadratic Gravity,
Front. in Phys. \textbf{6}, 77 (2018).
%arXiv:1804.09944 [hep-th]

\bibitem{CubgrBC}
P. Bueno and P.A. Cano, Einsteinian cubic gravity,
Phys. Rev. D \textbf{94}, 104005 (2016).
%arXiv:1607.06463 [hep-th]

\bibitem{BCMV}
P. Bueno, P.A. Cano, V.S. Min and M.R. Visser,
Aspects of general higher-order gravities,
Phys. Rev. D \textbf{95}, 044010 (2017).
%arXiv:1610.08519 [hep-th]


\end{thebibliography}
\end{document}